\documentclass[11pt,a4paper]{article}

\usepackage{amsmath}
\usepackage[T1]{fontenc}

\usepackage{jheppub}
\usepackage{psfrag}
\usepackage{slashed}
\usepackage{cancel}
\usepackage{lscape}
\usepackage{caption}
\usepackage{array}
\usepackage{graphicx}
\usepackage{subcaption}
\usepackage{multirow}
\usepackage{tabularx}
\usepackage{makecell}
\usepackage[utf8]{inputenc}
\usepackage{amsmath}
\usepackage{amssymb}
\usepackage{relsize}
\usepackage{color}
\usepackage{slashed}
\usepackage{mathtools}
\usepackage{comment}
\usepackage{scalefnt}
\usepackage{siunitx}[=v2]
\renewcommand{\arraystretch}{1.3}

\definecolor{mypink}{RGB}{219, 48, 122}
\definecolor{mygreen}{rgb}{0,0.7,0}
\definecolor{raspberry}{rgb}{0.53,0.15,0.34}

\DeclareMathOperator{\Tr}{Tr}

\newcommand{\dd}{\mathop{}\!\mathrm{d}}

\def\hpl11{{\mathrm{HPL}}_{1,1}}

\newcolumntype{C}[1]{>{\hsize=#1\hsize\centering\arraybackslash}X}%

\newcolumntype{Z}{r<{\hspace{3mm}}}

\newcommand{\MSbar}{\ensuremath{\overline{\text{MS}}}}
\newcommand{\fonll}{FONLL}
\newcommand{\nlonnllpart}{NLO+NNLL\textsubscript{part}+$y_by_t$}
\newcommand{\nnnres}{$\text{N}^3\text{LL}'+\text{aN}^3\text{LO}$}

\providecommand{\href}[2]{#2}

\newcommand{\pt}{p_{\text{\relscale{0.77}T}}}

\newcommand{\qt}{{q_{\text{\relscale{0.77}T}}}}

\newcommand{\ptH}{p_{\text{\relscale{0.77}T,$H$}}}
\newcommand{\ptHjj}{p_{\text{\relscale{0.77}T,$Hjj$}}}
\newcommand{\mjj}{m_{\text{\relscale{0.77}$jj$}}}

\newcommand{\muF}{{\mu_{\text{\relscale{0.77}F}}}}
\newcommand{\muR}{{\mu_{\text{\relscale{0.77}R}}}}
\newcommand{\muB}{{\mu_{\text{\relscale{0.77}B}}}}

\newcommand{\KQ}{{K_{\text{\relscale{0.77}Q}}}}

\newcommand{\noun}[1]{{\scshape #1}}

\newcommand{\POWHEG}{\noun{Powheg}}
\newcommand{\SuSHi}{\noun{SuSHi}}

\newcommand{\minlo}{{\noun{MiNLO$^{\prime}$}}}
\newcommand{\minnlo}{{\noun{MiNNLO$_{\textrm{PS}}$}}}
\newcommand{\GENEVA}{\noun{Geneva}}

\newcommand{\PYTHIA}[1]{\noun{Pythia{#1}}}

\newcommand{\fnnnlo}{N$^3$LO}

\usepackage{xcolor}
\newcommand{\mathd}{\mathrm{d}}

\def\to{\rightarrow}

\def\mbbggs{m_{2b2\gamma}^{\star}}

\def\GeV{\mathrm{GeV}}

\newcommand{\eqn}[1]{eq.\,(\ref{#1})}

\newcommand{\fig}[1]{figure\,\ref{#1}}

\newcommand{\tab}[1]{table\,\ref{#1}}
\newcommand{\sct}[1]{section\,\ref{#1}}

\def\citere#1{\mbox{ref.\,\cite{#1}}}
\def\citeres#1{\mbox{refs.\,\cite{#1}}}

\usepackage{etoolbox}
\makeatletter
\patchcmd{\@sect}{#8}{\boldmath #8}{}{}
\let\ori@chapter\@chapter
\def\@chapter[#1]#2{\ori@chapter[\boldmath#1]{\boldmath#2}}
\makeatother

\newcommand{\bbH}{\ensuremath{b\bar{b}H}}
\newcommand{\ccH}{\ensuremath{c\bar{c}H}}
\newcommand{\bbtoH}{\ensuremath{b\bar{b}\rightarrow H}}
\newcommand{\qqtoH}{\ensuremath{q\bar{q}\rightarrow H}}
\newcommand{\ttH}{\ensuremath{t\bar{t}H}}

\newcommand{\ytsq}{\ensuremath{y_t^2}}

\newcommand{\ybsq}{\ensuremath{y_b^2}}
\newcommand{\ybyt}{\ensuremath{y_b\, y_t}}

\preprint{
\vspace{-24pt}
  \begin{flushright}
  LHCHWG-2025-11\\
  MPP-2025-116\\
  DESY-25-093\\
  TIF-UNIMI-2025-15
  \end{flushright}
}

\title{Modelling {\boldmath{$b\bar b H$}} production for the LHC at 13.6 TeV}

\author[a]{Christian Biello,}
\author[b]{Alessandro Gavardi,}
\author[b]{Rebecca von Kuk,}
\author[c,d]{Matthew A.~Lim,}
\author[e]{Stefano Manzoni,}
\author[e]{Elena Mazzeo,}
\author[f]{Javier Mazzitelli,}
\author[a,g]{Aparna Sankar,}
\author[f]{Michael Spira,}
\author[b]{Frank J. Tackmann,}
\author[a]{Marius Wiesemann,}
\author[a,g]{Giulia Zanderighi,}
\author[h]{Marco Zaro}

\affiliation[a]{Max-Planck-Institut f\"ur Physik, Boltzmannstrasse 8, 85748 Garching, Germany}
\affiliation[b]{Deutsches Elektronen-Synchrotron DESY, Notkestr. 85, 22607 Hamburg, Germany}
\affiliation[c]{Department of Physics and Astronomy, University of Sussex, Sussex House, Brighton, BN1 9RH, UK}
\affiliation[d]{Università degli Studi di Milano-Bicocca \& INFN Sezione di Milano-Bicocca, Piazza della Scienza 3, Milano 20126, Italy}
\affiliation[e]{CERN, CH-1211 Geneva 23, Switzerland}
\affiliation[f]{PSI Center for Neutron and Muon Sciences, 5232 Villigen PSI, Switzerland}
\affiliation[g]{Physik Department T31, James-Franck-Straße 1, Technische Universität München, D-85748\\Garching, Germany}
\affiliation[h]{Università degli Studi di Milano \& INFN Sezione di Milano, Via Celoria 16, 20133 Milano, Italy}

\emailAdd{biello@mpp.mpg.de, marius.wiesemann@mpp.mpg.de}

\abstract{ We present new state-of-the-art predictions for Standard
  Model Higgs boson production in association with a bottom-quark pair
  (\bbH{}). Updated cross sections are computed in accordance with the
  recommendations of the LHC Higgs Working Group, including the use of
  the PDF4LHC21 set of parton distribution functions, with a
  center-of-mass energy of 13.6 TeV. For the total inclusive cross
  section, we provide matched predictions of the massless Five-Flavour
  Scheme and the massive Four-Flavour Scheme at the fixed-order
  level. We further present recently obtained simulations matched with
  parton showers in both flavour schemes within the Standard Model,
  and also discuss them in the context of potential Beyond the
  Standard Model scenarios. In the massless scheme, we compare
  different next-to-next-to-leading order predictions matched to
  parton showers obtained through the \minnlo{} and \GENEVA{}
  generators.  In addition, the role of Four-Flavour Scheme
  predictions is studied as a background to $HH$ searches, considering
  both the top-quark and bottom-quark Yukawa contributions to $b\bar
  bH$ production. Finally, we analyse the sensitivity of the Higgs
  transverse momentum spectrum to light-quark Yukawa couplings in the
  diphoton decay channel based on \minnlo{} simulations. }

\begin{document}
\maketitle
\flushbottom

\section{Introduction}

\begin{figure}[t]
\begin{center}
    \includegraphics[height=2cm]{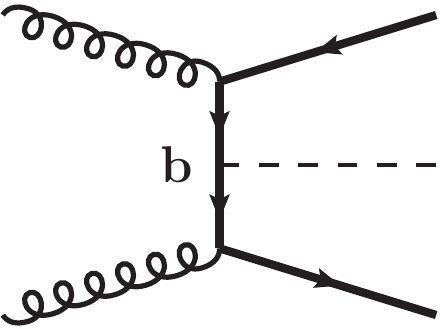}\hspace*{2cm}
    \includegraphics[height=2cm]{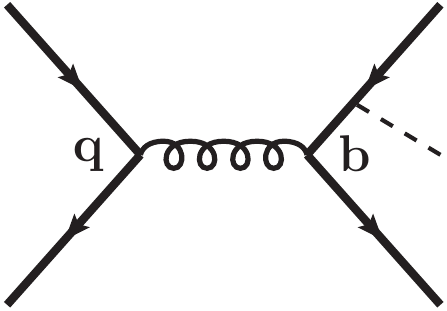}\hspace*{2cm}
    \includegraphics[height=2cm]{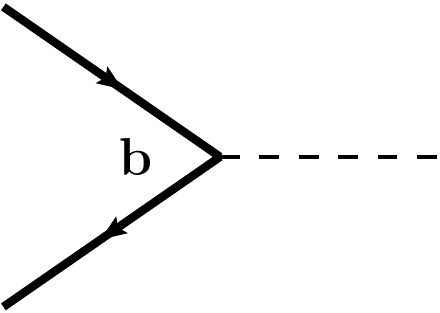}
	\vspace{0.2cm}
  \caption{Typical LO Feynman diagrams contributing to $\bbH$ production in the four-flavour scheme (left, centre) and the five-flavour scheme (right). Incoming particles are shown on the left, and final asymptotic states on the right. In particular, the last diagram represents the LO contribution to the inclusive Higgs-boson production through initial-state bottom-quark fusion mediated by the Yukawa coupling $y_b$. In the five-flavour scheme, this prediction relies on heavy-quark collinear factorisation: the bottom quark is generated through the PDF evolution, and the description is inclusive with respect to the final state-bottom quark radiation.}
  \label{fig:bbhlo}
\end{center}
\end{figure}

\begin{figure}[b]
\begin{center}
    \includegraphics[height=2.15cm]{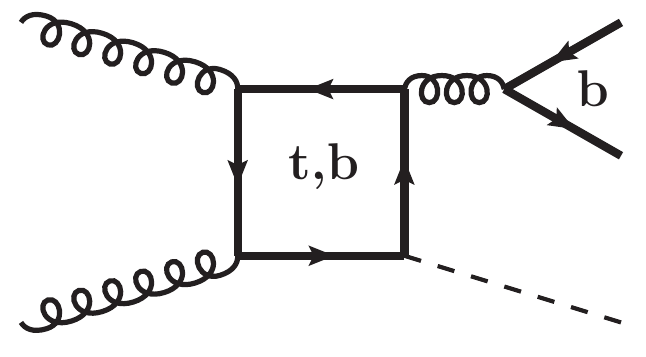}\hspace{1.2cm}
    \includegraphics[height=2.15cm]{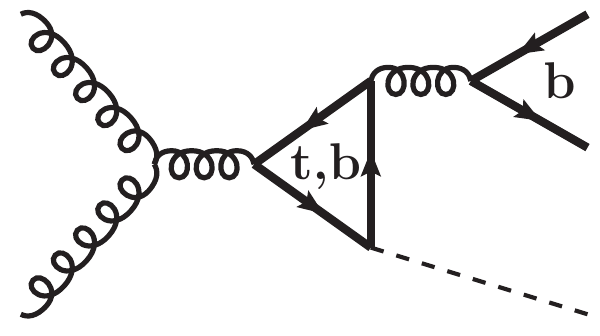}\hspace{1.2cm}
    \includegraphics[height=2.15cm]{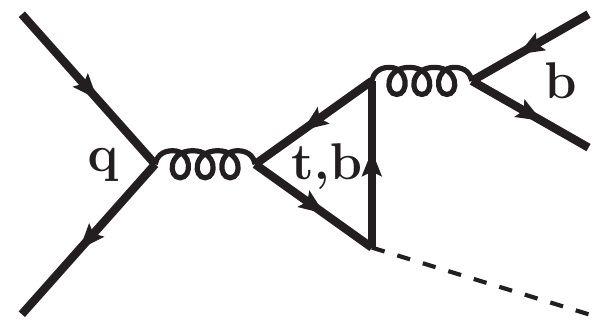}
  \caption{Typical one-loop Feynman diagrams for $\bbH$ production in the four-flavour scheme. The Higgs boson is produced through radiation off internal heavy-quark loops; for the dominant top-quark contribution, the corresponding vertex is governed by the Yukawa coupling $y_t$.}
  \label{fig:bbhyt}
\end{center}
\end{figure}

Higgs boson production in association with bottom quarks ($b\bar bH$)
proceeds via two dominant production mechanisms within the Standard
Model (SM). The typically 
considered $b\bar bH$ process proceeds via tree-level diagrams at
Leading Order (LO) where the Higgs couples to external bottom quarks, see \fig{fig:bbhlo} for example,
which will be referred to as {\it Higgs radiation off bottom quarks}.
Hence, the cross section is proportional to the squared bottom Yukawa coupling ($y_b^2$). The other (in the SM even larger) production 
mechanism is through the {\it loop-induced gluon fusion process}, where the Higgs couples to the quark loop and a radiated gluon splits into 
a bottom-quark pair, see \fig{fig:bbhyt}. The dominant contribution to this process is given by the top-quark loop, and the cross section is therefore proportional 
to the squared top Yukawa coupling ($y_t^2$). Interference effects between the 
two production mechanisms are of order $y_t\,y_b$, but they are bottom-mass suppressed and appear only if the bottom quark is considered in a massive scheme.

We can thus write the \bbH{} production cross section as follows:
\begin{equation}
\begin{split}
    {\rm d} \sigma & = \ybsq{}\,\alpha_s^2\left( \Delta_{\ybsq}^{(0)} + \mathcal O(\alpha_s) \right) +y_t y_b\, \alpha_s^3\left( \Delta_{\ybyt}^{(0)} + \mathcal O(\alpha_s)  \right) + \ytsq{} \,\alpha_s^4\left( \Delta_{\ytsq}^{(0)} + \mathcal O(\alpha_s)  \right)\, ,
\end{split}
\label{eq:hbbxsec2}
\end{equation}
where $\Delta_{X}^{(0)}$ is the LO contribution to each coupling structure. The first term corresponds to Higgs radiation off bottom quarks, the last term
to the loop-induced gluon fusion process, and the central one is their interference. Higher-order corrections, indicated generically
by the $\mathcal O(\alpha_s)$ (and higher) corrections, 
are typically computed separately for each process, with the caveat
that at higher orders in Quantum Chromodynamics (QCD) the two processes mix, giving rise 
to their $y_t y_b$ interference contributions. Notice that the
loop-induced gluon fusion contribution is formally a
Next-to-Next-to-Leading Order (NNLO) correction (relative $\alpha_s^2$)
due to the loop suppression, which however is fully compensated by its $y_t^2/y_b^2$ enhancement compared to the LO $y_b^2$ cross section.
As a result, the cross section of the $y_t^2$ contribution is roughly twice as large as the $y_b^2$ one.
There are further relevant \bbH{} production mechanisms, including
Higgsstrahlung (VH) and Vector Boson Fusion (VBF), 
but their numerical impact is subleading compared to the $y_t^2$ and $y_b^2$ contribution, at least for
the inclusive cross section. We note that, in experimental analyses, these contributions are
already included in the simulations of the VH and VBF processes, and that the corresponding interference
terms are negligible.

For fully inclusive Higgs boson production, the radiation off bottom quarks is about two orders of magnitude smaller than the dominant gluon-fusion cross section and ranks at the same size as Higgs production in association with top quarks ($\ttH{}$) production. Higgs boson production in association with bottom quarks yields therefore a subleading, but yet relevant 
contribution to the inclusive Higgs cross section, especially in the
precision age of the LHC. However, when $b$-tagging is applied in
experimental analyses, the $\bbH{}$ signal yield drops substantially,
as the $b$-jets in this process are typically softer and more forward
than those in $\ttH{}$ production. In addition, $\ttH{}$ events
feature a more distinctive event topology with electroweak (EW) signatures, which allows for more effective background suppression. As a result of the small signal yield and involved experimental signature,
the measurement of the \bbH{} process in the 
SM has not been achieved yet at the LHC.\footnote{Although no evidence
for an observation of $\bbH{}$ production has been reported so far,
dedicated searches have been performed; see, for
example,~\citere{CMS:2024goa}.} In the Minimal Supersymmetric Standard
Model (MSSM) or the Two-Higgs-Doublet
Model (2HDM) of type II, on the other hand, the bottom Yukawa coupling is strongly enhanced for large values of $\tan\beta = v_2/v_1$, where $v_{1,2}$ denote the vacuum expectation values acquired by the gauge-eigenstate Higgs fields, so that this production becomes the
dominant one, i.e.\ even larger than the gluon-fusion mechanism. In addition, \bbH{} production is a major background to di-Higgs searches, where at least 
one of the two Higgs bosons decays to bottom quarks.

The calculation of the $b\bar bH$ cross section can be performed in
two different schemes. In the Four-Flavour Scheme (4FS), the bottom
quarks are massive and thus no bottom Parton Distribution Functions
(PDFs) are taken into account so that the bottom quarks are entirely
generated in the final state starting from light quark-antiquark and
gluon-gluon initial states at LO, see \fig{fig:bbhlo}\,(left,
center). The calculation requires 4FS PDFs and a 4FS strong coupling
$\alpha_s$ in order to avoid artificially large logarithms at higher
orders. The 4FS calculation has been performed at Next-to-Leading
Order (NLO) QCD some time ago \cite{dittmaier:2003ej,dawson:2003kb},
while NNLO QCD results became available only very recently
\cite{Biello:2024pgo}.

In the 4FS, the integration over the transverse momenta of the final-state bottom quarks, however, generates logarithmic contributions in the 
bottom-quark mass that might reduce the perturbative convergence. In order to resum these logarithms, bottom-quark densities need to be introduced by
treating the bottom quark as a massless particle. The
Dokshitzer--Gribov--Lipatov--Altarelli--Parisi (DGLAP) evolution of the PDFs leads to the resummation. This framework defines the Five-Flavour Scheme (5FS) and starts from a $b\bar b$ initial state at LO, see \fig{fig:bbhlo}\,(right), which neglects the off-shellness and transverse momenta of the initial-state bottom quarks
as well as all power corrections in the bottom-quark mass. The first two approximations are resolved by adding higher-order QCD corrections that, order by order, restore the full kinematics of the bottom quarks. However, finite bottom-mass effects cannot be studied in the 5FS, but they can be included
through a combination with the 4FS calculation. The NLO QCD cross section in the 5FS has been obtained some time 
ago \cite{dicus:1998hs,balazs:1998bm}. The NNLO QCD corrections \cite{harlander_2003} can range up to several 10\% (depending on the scale settings)
and they stabilise the scale dependence significantly. More recently,
the 5FS calculation has even been extended to
Next-to-Next-to-Next-to-Leading Order (N$^3$LO) QCD \cite{duhr:2019kwi}, which 
induces a small correction and further reduces the scale uncertainties.

At sufficiently high perturbative order, the two schemes need to approach each other. However, the comparison of the 4FS at NLO and the 5FS at NNLO shows a discrepancy at the level of 20--30\%, so that a combination of both schemes became mandatory for reliable prediction. In order to cope with the logarithmic terms in the 4FS, a factorization scale smaller than the Higgs mass has been introduced (typically $M_H/4$) that reduces the differences between the 4FS and 5FS \cite{campbell2004,Maltoni:2012pa}. The combination of the 4FS and 5FS was first performed by applying the empirical Santander matching \cite{harlander:2011aa}, which is based on a logarithmic weighting between the 4FS and 5FS. This heuristic approach has been replaced by two proper matching procedures some years ago, the \fonll{} approach \cite{forte:2015hba,forte:2016sja} and the \nlonnllpart{} method \cite{Bonvini:2015pxa,Bonvini:2016fgf}. Both approaches rely on a systematic expansion of the 4FS parameters and 5FS PDFs using different methods to merge the two schemes without double-counting of common contributions. The 
combined results show a better agreement with the 5FS at NNLO QCD than with the 4FS at NLO QCD. This situation has changed recently with the computation
of the NNLO QCD corrections in the 4FS \cite{Biello:2024pgo}. At NNLO QCD, 4FS and 5FS predictions agree within 10\% or better, thus solving 
the previous discrepancies between the schemes.

Apart from fixed-order predictions, there has been substantial progress also in the matching of NNLO QCD corrections with parton showers (NNLO+PS) for \bbH{} production recently.
A NNLO+PS calculation in the 5FS has been performed using the \GENEVA{} method~\cite{Gavardi:2025zpf}, while in the \minnlo{} framework~\cite{Biello:2024pgo,Biello:2024vdh} NNLO+PS predictions for both the 4FS and 5FS predictions have been achieved. 
These three implementations provide suitable NNLO event generators for the experimental analyses.

The $y_t^2$-induced contributions to the inclusive \bbH{} cross section (cf.\ \fig{fig:bbhyt}) are not well-defined in the 5FS, since, formally, the rate diverges 
for massless bottom quarks in fixed-order perturbation theory. Only by selecting bottom-flavoured jets ($b$-jets), with an appropriate
definition of the jet flavour (and imposing kinematical selections on them) fixed-order predictions for the $y_t^2$ contribution in the 5FS are infrared safe. So far, the size of the 
$y_t^2$ contribution to the \bbH{} final state has been estimated effectively only at LO in the 5FS, using the inclusive NNLO+PS 
generator for gluon fusion \cite{hamilton:2012rf,hamilton:2013fea,hamilton:2015nsa}. In this case, reshuffling of the bottom momenta onto the mass shell and matching to the parton shower, 
where the bottom quarks are considered as massive particles, renders the selection of bottom quarks in the final state finite, even if the 
underlying perturbative calculation is performed in the 5FS.

In the 4FS, on the other hand, the $y_t^2$ part of the inclusive \bbH{} cross section is finite by construction, since the bottom quarks are treated as massive.
On top of that, there is the additional  $y_t\,y_b$ interference contribution in the 4FS, which vanishes in the 5FS.\footnote{Throughout this contribution, we consider 5FS to correspond to the massless treatment of all power corrections with $m_b=0$. Alternatively, the 5FS can be considered to be the leading-power contribution in the bottom-quark mass, thereby including the dominant interference term of $\mathcal{O}(y_by_tm_b)$, as done in \citere{Dawson:2004sh} for instance.}
The complete NLO QCD corrections to the top-Yukawa induced terms (both $y_t^2$ and $y_b y_t$ interference) 
have been obtained in the Heavy-Top Limit (HTL) \cite{deutschmann:2018avk} and turn out to be substantial. This calculation 
has later been extended in \citere{manzoni:2023qaf} to include the parton-shower matching at NLO (NLO+PS) to study 
\bbH{} production as a background to di-Higgs searches. 
The $y_t\,y_b$ interference amounts to about 10\% \cite{dittmaier:2003ej,dawson:2003kb} compared to the inclusive $y_b^2$ cross section, while the $y_t^2$ contribution is roughly twice as large as the $y_b^2$ one \cite{deutschmann:2018avk}.
This reduces the sensitivity of the $b\bar bH$ process to the bottom Yukawa coupling substantially in the SM.
For heavy Higgs bosons in Beyond the Standard Model (BSM) scenarios with an enhanced bottom-Yukawa coupling, however, the radiation off bottom quarks is the dominant contribution,
while all top-quark Yukawa induced ones become subleading.

Finally, while the discussion up to here focused on QCD corrections, it is worth mentioning that EW corrections for \bbH{} production have also been computed in the 4FS.
In particular, they have been first evaluated for the sole $gg$-induced mechanism~\cite{Zhang:2017mdz}, and then computed for the full production
process~\cite{Pagani:2020rsg}, and they turn out to be rather small (at the few-percent level). 
In the latter case, complete-NLO corrections (including QCD and EW corrections)
have been computed. In this study, it was also shown that other production mechanisms, specifically VH and VBF, 
can have a relevant contribution to the $b\bar b H$ rate when requiring $b$-jets, although their impact can be minimised through
suitable kinematical requirements, as shown in \citere{Grojean:2020ech}. Nevertheless, these are additional \textit{backgrounds} to the $y_b^2$ \textit{signal}, 
further reducing the sensitivity to the bottom Yukawa coupling in \bbH{} final states at the LHC.

In this contribution, we present updated cross section predictions for Higgs boson production in association with bottom quarks in both the 4FS and the 5FS for
the $y_b^2$ contribution and including the $y_b\,y_t$ interference. 
To provide a consistent overview of state-of-the-art \bbH{} predictions at 13.6 TeV centre-of-mass energy, all results are obtained using 
a common computational setup detailed in~\sct{sec:setup}. Updated inclusive cross section predictions at 13.6 TeV are 
reported in~\sct{sec:matchedinclusivenumbers}, which are obtained by an interpolation of existing LHC Higgs Working Group (LHCHWG) data \cite{LHCHiggsCrossSectionWorkingGroup:2016ypw} of the 
matched NLO 4FS and NNLO 5FS predictions from the \nlonnllpart{} calculation at 13 and 14 TeV. 
In~\sct{sec:resummation}, we present the resummed Higgs transverse momentum spectrum in the 5FS 
at Next-to-Next-to-Next-to Leading Logarithmic (N$^3$LL) accuracy for the $y_b^2$ component.
Section~\ref{sec:MCyb} discusses recent developments of Monte Carlo
(MC) event generators for the $y_b^2$ contribution 
and compares NNLO+PS results from \minnlo{} and \GENEVA{} in the 5FS, consideres \minnlo{} predictions for heavy Higgs bosons in BSM scenarios, 
and subsequently examines NNLO+PS predictions from \minnlo{} in the 4FS.
In~\sct{sec:HH}, we analyse the role of \bbH{} production as a background in di-Higgs searches, 
focusing on both $y_b^2$ and $y_t^2$ contributions in phase-space regions relevant for $HH$ studies.

Finally, we discuss a natural extension of the massless calculation
for $\bbtoH$ to describe Higgs-boson production via lighter-parton
fusion ($q\bar q \rightarrow H$). Although these modes are strongly
suppressed by the small Yukawa coupling—proportional to the
light-quark masses—they receive a PDF enhancement, with distinctive
differential effects. We therefore present novel results for Higgs
boson production via light-quark fusion at N$^3$LL' + approximate
Next-to-Next-to-Next-to-Leading Order (aN$^3$LO) and NNLO+PS in~\sct{sec:lightYukawa}, 
with particular emphasis on the sensitivity of the transverse momentum distribution on the light-quark Yukawa couplings.
We summarise our findings and discuss possible future directions for improved \bbH{} predictions in \sct{sec:conclusions}.

\section{Computational setup}\label{sec:setup}
All simulations in this paper are obtained at a centre-of-mass energy of $\sqrt{s}=13.6$ TeV, employing the PDF4LHC21\_40 (LHAPDF ID \texttt{93100}) set for the parton distribution functions for massless-scheme calculations and the set PDF4LHC21\_40\_nf4 (LHAPDF ID \texttt{93500}) for simulations with massive bottom quarks~\cite{PDF4LHCWorkingGroup:2022cjn}. We used the \texttt{LHAPDF} interface~\cite{buckley:2014ana} to utilise the PDF sets in our calculations.

\subsection{Numerical Inputs}
The parametric inputs, including the strong coupling and EW parameters
relevant for the Higgs vacuum expectation value, follow the latest
recommendations from the LHCHWG~\cite{Karlberg:2024zxx} and the
Particle Data Group (PDG) parameters~\cite{ParticleDataGroup:2024cfk}. For completeness, the most relevant inputs for \bbH{} predictions are reported in the following: The predictions in the massive schemes are obtained using
an on-shell value of the bottom-quark mass of $m_b^{\rm OS}=(4.92\pm0.13)$\,GeV for internal and external bottom quark lines.
The Yukawa coupling, on the other hand, is renormalised in the $\MSbar{}$ scheme using an input value of $m_b(m_b)=(4.18\pm0.03)$\,GeV
for the renormalization group running of the Yukawa mass. This scheme choice for the Yukawa coupling is crucial in order to evaluate it at its natural
scale, which is of the order of the Higgs boson mass, and to avoid large logarithmic corrections. The scale settings are discussed in the following subsection.
The following EW input parameters are used: $m_W = 80.379$\,GeV, $\Gamma_W=2.085$\,GeV, $m_Z = 91.1876$\,GeV, $\Gamma_Z=2.4952$\,GeV. Employing complex masses with the electromagnetic coupling set to $\alpha=1/132.3489045$, the corresponding vacuum expectation value is $v=(246.403-3.8060\,i)$\,GeV.
The Higgs boson is considered to be stable in all simulations,
except where its decay into photons or tau leptons is explicitly considered. 
Also in these cases it is treated as on-shell and with zero width.
The strong coupling is set to $\alpha_s(m_Z) = 0.1180$ 
in the 5FS calculations. The corresponding 4FS value is obtained 
by accounting for heavy-quark decoupling effects, as discussed below.
The strong coupling is obtained directly from the corresponding PDF set
for consistency.

\subsection{Running of the bottom Yukawa coupling}
The Yukawa coupling is renormalized in the \MSbar{} scheme in both the 4FS and the 5FS, adapting the following recommendation: We derive the coupling $y_b(\muR)=m_b(\muR)/v$ by 
evolving $\hat{m}_b\equiv m_b(m_b)=4.18$\,GeV to $m_b(\muR)$ via four-loop running, solving the Renormalization Group Equation (RGE) \cite{harlander:2002wh,baikov_2014}. 
This evolution is directly related to the one of the strong coupling, which is evaluated both at $\muR$ and at $\hat{m}_b$. 
In the massless scheme, these values are straightforwardly obtained by evolving the input $\alpha_s(m_Z) = 0.1180$ with four-loop QCD running and $n_f = 5$ active flavours.
In the massive scheme, the procedure follows the approach commonly adopted in modern PDF evolution libraries: starting from $\alpha_s(m_Z) = 0.1180$ we evolve down to $\hat{m}_b$
with $n_f = 5$, where the decoupling relation~\cite{vogt:2004ns} is applied to extract $\alpha_s(\hat{m}_b)$ in the 4FS. 
This value is then used as a boundary condition to evolve to $\alpha_s(\muR)$ with $n_f = 4$ active flavours.

For consistency, scale uncertainties are obtained by varying the scale with respect to the central one using the order of the evolution consistent with the 
calculation, i.e.\ two-loop running for NLO predictions and three-loop running for NNLO calculations, and with the number of active flavours consistent with the flavour scheme under consideration.

Predictions involving the top-quark Yukawa contribution employ the on-shell definition of the coupling. A top-quark mass of $172.5$\,GeV is used throughout the simulations.

\subsection{Scale settings}
A scale choice of the order of the Higgs-boson mass is recommended for the factorization ($\muF$) and renormalization ($\muR$) scales. The precise choices of these scales are specified in this Report before the corresponding predictions are presented. The theory uncertainty is estimated using the standard 7-point variation, i.e.\ by changing the scales by a factor of 2 with the constraint $\frac{1}{2}\leq\muR/\muF\leq 2$ in order to avoid large logarithmic effects. The renormalization scales associated with the Yukawa coupling and the strong coupling are simultaneously rescaled by the same factor. We have performed detailed studies on the impact of this correlation and found that such a correlated variation provides a reliable estimate of the overall renormalization uncertainty.

In earlier 4FS predictions, a dynamical scale—specifically, one quarter of the transverse mass of the \bbH{} system—was often adopted for the Yukawa coupling~\cite{Wiesemann:2014ioa}. Such small scale
resulted in higher 4FS cross sections, which reduced the gap with the 5FS predictions. However, with the inclusion of NNLO QCD corrections in the 4FS, 
such a scale choice is no longer necessary. Moreover, as shown in~\citere{Biello:2024pgo}, comparisons between dynamical and fixed scales at NNLO 
show minimal differences. Consequently, we recommend aligning the scale choices of the Yukawa coupling in the 4FS and 5FS at NNLO QCD accuracy, with a scale of the order of the Higgs-boson 
mass providing a natural and consistent choice.

\section{Updated inclusive cross sections at 13.6 TeV}
\label{sec:matchedinclusivenumbers}
In this section, we report predictions for the inclusive \bbH{}  cross section at 13.6 TeV. 
The best description of the \bbH{} process is obtained by combining the calculations in the massless and massive schemes. The 5FS computation includes 
the resummation of large collinear logarithmic contributions, while the 4FS captures all mass effects order by order in perturbative QCD. 
As already mentioned in the introduction, two main approaches exist that have been applied to obtain the \bbH{} cross section at 
high accuracy: \fonll{}~\cite{forte:2015hba,forte:2016sja,Duhr:2020kzd} and \nlonnllpart{}~\cite{Bonvini:2015pxa,Bonvini:2016fgf}.

FONLL-B~\cite{forte:2015hba,forte:2016sja} and \nlonnllpart{}~\cite{Bonvini:2015pxa,Bonvini:2016fgf} match the NNLO 5FS and the NLO 4FS cross section.
Both approaches yield results that are fully consistent with each other, as shown in~\citere{LHCHiggsCrossSectionWorkingGroup:2016ypw}. 
The novel FONLL-C matching~\cite{Duhr:2020kzd} adds the \fnnnlo{} corrections in the 5FS to FONLL-B. The \nlonnllpart{} combination introduces a resummation scale $\muB$, which enables an estimate of the 
matching uncertainties.
The $y_by_t$  contribution is easily included as a fixed-order non-singular term, since these interference effects are power corrections that 
vanish to all perturbative orders in the small bottom-quark mass limit.

The setup of the calculation follows that is described in~\sct{sec:setup}, with the choice of the following central scales,
\begin{align}
\muF=\frac{1}{4}(m_H+2m_b),\hspace{1cm}\muR=\frac{1}{2}m_H\,,
\end{align}
in order to ensure a good perturbative convergence in the fully-inclusive calculation.

\begin{table}[b]
\begin{center}%
\begin{small}%
\tabcolsep5pt
\begin{tabular}{|c|c|c|c|}%
\hline
$m_h$[GeV] & $\sigma^{}$[fb] & $\Delta_{\left(\mu_{R},\mu_{F}\right)\oplus\mu_{B}}$[\%] & $\Delta_{\mathrm{PDF}\oplus\alpha_s}$[\%]  \\\hline\hline
$125.00$ & $569.1$ & $\pm8.63$ & ${{+3.35}}/{-3.44}$ \\\hline
$125.09$ & $567.9$ & $\pm8.63$ & ${{+3.36}}/{-3.44}$ \\\hline
$125.10$ & $567.7$ & $\pm8.63$ & ${{+3.36}}/{-3.44}$ \\\hline
$125.20$ & $566.3$ & $\pm8.63$ & ${{+3.35}}/{-3.44}$ \\\hline
\end{tabular}%
\end{small}%
\end{center}%
\caption{Total \bbH{} cross sections in the SM for a LHC center-of-mass energy of $\sqrt{s}=13.6$ TeV obtained via linear interpolation of results from 13 TeV and 14 TeV. The results are given with symmetrised uncertainties from the 7-point scale variation combined with the resummation dependency and the parametric uncertainty from PDFs and strong coupling.}
\label{tab:bbH136lin}
\end{table}

\begin{table}[t]
\begin{center}%
\begin{small}%
\tabcolsep5pt
\begin{tabular}{|c|c|c|c|}%
\hline
$m_h$[GeV] & $\sigma^{}$[fb] & $\Delta_{\left(\mu_{R},\mu_{F}\right)\oplus\mu_{B}}$[\%] & $\Delta_{\mathrm{PDF}\oplus\alpha_s}$[\%]  \\\hline\hline
$125.00$ & $568.1$ & $\pm8.64$ & ${{+3.35}}/{-3.44}$ \\\hline
$125.09$ & $566.9$ & $\pm8.64$ & ${{+3.36}}/{-3.44}$ \\\hline
$125.10$ & $566.7$ & $\pm8.64$ & ${{+3.36}}/{-3.45}$ \\\hline
$125.20$ & $565.3$ & $\pm8.64$ & ${{+3.35}}/{-3.45}$ \\\hline
\end{tabular}%
\end{small}%
\end{center}%
\caption{Total \bbH{} cross sections in the SM for a LHC center-of-mass energy of $\sqrt{s}=13.6$ TeV obtained via logarithmic interpolation of results from 13 TeV and 14 TeV.}
\label{tab:bbH136log}
\end{table}

\begin{table}[t]
\begin{center}%
\begin{small}%
\tabcolsep5pt
\begin{tabular}{|c|c|c|c|}%
\hline
$m_h$[GeV] & $\sigma^{}$[fb] & $\Delta_{\left(\mu_{R},\mu_{F}\right)\oplus\mu_{B}}$[\%] & $\Delta_{\mathrm{PDF}\oplus\alpha_s}$[\%]  \\\hline\hline
$125.00$ & $567.7$ & $\pm8.86$ & ${{+3.30}}/{-3.68}$ \\\hline
$125.09$ & $566.4$ & $\pm8.92$ & ${{+3.43}}/{-3.51}$ \\\hline
\end{tabular}%
\end{small}%
\end{center}%
\caption{Total \bbH{} cross sections in the SM for a LHC center-of-mass energy of $\sqrt{s}=13.6$ TeV obtained via linear interpolation of results from 13.5 TeV and 14 TeV.}
\label{tab:bbH136linalt}
\end{table}

In~\tab{tab:bbH136lin}, we report the results of the \nlonnllpart{} combination via a linear interpolation between the existing 13 TeV and 14 TeV results, using the relation
\begin{align}
\sigma(13.6\,\text{TeV})=0.4\sigma(13.0\,\text{TeV})+0.6\sigma(14.0\,\text{TeV})\,.
\end{align}
The cross section is provided together with an overall theoretical uncertainty. This includes the standard 7-point renormalization and factorization scale variation, the resummation scale uncertainty, and the uncertainties associated with the PDFs and the strong coupling. An alternative cross section estimation is presented in~\tab{tab:bbH136log} where the same numbers have been combined using a logarithmic interpolation. In order to test the accuracy of the interpolation, we have also considered linear interpolation of 13.5 TeV and 14 TeV results. Using the latter, we have estimated the cross section in~\tab{tab:bbH136linalt} with the following linear interpolation,
\begin{align}
	\sigma(13.6\,\text{TeV})=0.8\sigma(13.5\,\text{TeV})+0.2\sigma(14.0\,\text{TeV})\,.
\end{align}
We observe a good agreement between the cross section values obtained with different interpolations or different choices of boundary conditions. Due to the good accuracy of the interpolation and the substantial theoretical and parametric uncertainty, dedicated runs at 13.6 TeV are not required.

\section{Transverse momentum resummation at third order}
\label{sec:resummation}
The Higgs transverse momentum spectrum provides a promising approach for extracting the Yukawa coupling from Higgs boson production processes, because its shape is sensitive to the precise value of the coupling.
In particular, one can exploit the pattern of QCD emissions from the incoming quarks and gluons to discriminate between the gluon and various quark channels in the initial state~\cite{Ebert:2016idf}.
That is, the radiation pattern for different initial states yields different shapes for the transverse momentum ($q_T$) spectrum of the recoiling Higgs boson. As a result, a precise measurement of the Higgs $q_T$ spectrum, especially at small $q_T$, allows one to gain sensitivity to the quark Yukawa couplings~\cite{Bishara:2016jga, Soreq:2016rae}.
At small $q_T \ll m_H$, this requires the all-order resummation of logarithms of $q_T/m_H$ that would otherwise spoil the convergence of perturbation theory in this regime.

In this section, we provide recent results for the resummed $q_T$
spectrum for $b\bar{b}\to H$ at N$^3$LL$^{\prime}$ order matched to
fixed NNLO and aN$^3$LO in the 5FS
\cite{Cal:2023mib}.\footnote{Here, we use the logarithmic counting
of~\citere{Berger:2010xi}, where the prime denotes inclusion of higher
order boundary terms.} Previously, the spectrum was calculated to
Next-to-Next-to-Leading Logarithmic (NNLL) + NNLO accuracy~\cite{Harlander:2014hya} using the
Collins--Soper--Sterman (CSS) formalism~\cite{Collins:1981uk,Collins:1981va, Collins:1984kg}. For this prediction, we use Soft-Collinear Effective Theory (SCET)
\cite{Bauer:2000yr,Bauer:2001ct,Bauer:2001yt,Bauer:2002nz,Beneke:2002ph} to resum the logarithms of $q_T/m_H$ which is equivalent to a modern formulation of the CSS formalism.
We employ the rapidity
renormalization group~\cite{Chiu:2012ir} together with the
exponential regulator~\cite{Li:2016axz} for which the ingredients
required for the resummation at N$^3$LL$^{\prime}$ are known.
The singular cross section can be written in a factorised form as
\begin{align} \label{tmd_factorization}
\frac{\mathrm{d} \sigma^\mathrm{sing}}{\mathrm{d} Y \mathrm{d}^2 \qt}
&= \sum_{a,b} H_{ab}(m_H^2; \mu) [B_a\otimes  B_b \otimes S_{ab}]( x_a, x_b, \qt; \mu)
\,,\end{align}
where the kinematic quantities $\omega_{a,b}$ and $x_{a,b}$ are given by
\begin{align}
\omega_{a}=m_H e^{+ Y}, \quad \omega_{b}=m_H e^{-Y} \quad \text{ and} \quad \quad x_{a,b}= \frac{\omega_{a,b}}{E_{cm}}
\,,\end{align}
with $E_{cm}$ denoting the hadronic center-of-mass energy.
The hard function $H_{ab}$ is process dependent and describes physics at the hard scale $\mu\sim m_H$. The beam functions $B_{a,b}$ and the soft function $S_{ab}$ describe collinear and soft radiation at the low scale $\mu \sim q_T$. 
The term $[B_a\otimes  B_b \otimes S_{ab}]$,
\begin{align}
	[B_a\otimes  B_b \otimes S_{ab}]( x_a, x_b, \qt; \mu) &\equiv \int \dd^2 k_a \dd^2 k_b \dd^2 k_s \,\delta^{(2)}(\qt-k_a-k_b-k_s) \nonumber\\
	&\times B_a(x_a,k_a;\mu,\nu/\omega_a) B_b(x_b,k_b;\mu,\nu/\omega_b) S_{ab}(k_s;\mu,\nu)\,,
\end{align}
is usually evaluated in Fourier-conjugate $b_T$ space, as the convolutions in $q_T$ in
\eqn{tmd_factorization} turn into products in $b_T$ space.

\begin{figure}[t!]
	\begin{center}
		\begin{tabular}{cc}
			\includegraphics[width=.55\textwidth, page=1]{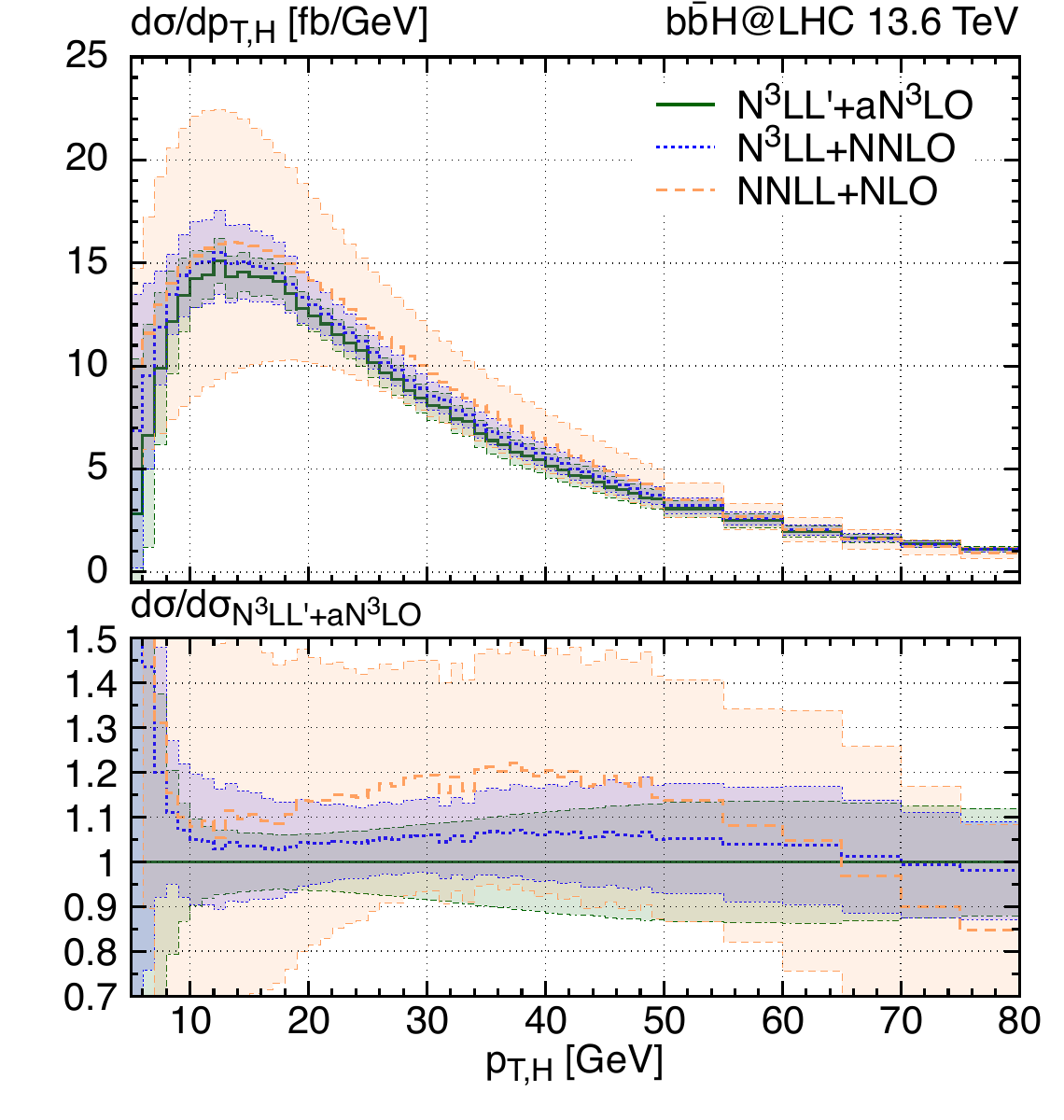}
		\end{tabular}
		\vspace*{1ex}
		\caption{Higgs transverse momentum spectrum in the 5FS predicted by the analytic resummed prediction at $\text{NNLO+NLO}$ (orange, dashed), $\text{N}^3\text{LL+NNLO}$ (blue, dotted) and $\text{N}^3\text{LL$^{\prime}$+aN}^3\text{LO}$ (green, solid) accuracy.\label{fig:resumplot}}
	\end{center}
\end{figure}
To perform the all-order resummation, each function is first evaluated at its own
natural boundary scale(s): $\mu_H$, $(\mu_B, \nu_B)$, and $(\mu_S, \nu_S)$. By
choosing appropriate values for the boundary scales close to their canonical
values, each function is free of large logarithms and can
therefore be evaluated in fixed-order perturbation theory. Next, all functions are evolved from their respective boundary conditions to a common arbitrary point
$(\mu, \nu)$ by solving their coupled system of RGEs. For more details, we refer to \citeres{Ebert:2016gcn,Ebert:2020dfc,Cal:2023mib}.
For the resummation at N$^3$LL$^{\prime}$ we require the N$^3$LO
boundary conditions for the hard function~\cite{Gehrmann:2014vha,
	Ebert:2017uel}, and the beam and soft functions~\cite{Lubbert:2016rku,
	Li:2016ctv, Billis:2019vxg, Luo:2019szz, Ebert:2020yqt}. We also need the 3-loop
noncusp virtuality anomalous dimensions~\cite{Lubbert:2016rku, Moch:2005id, Stewart:2010qs,
	Bruser:2018rad, Billis:2019vxg} and rapidity anomalous
dimension~\cite{Lubbert:2016rku, Li:2016ctv, Vladimirov:2016dll}, as well as
the 4-loop cusp anomalous dimension
$\Gamma_\mathrm{cusp}$~\cite{Korchemsky:1987wg, moch:2004pa, Bruser:2019auj,
	Henn:2019swt, vonManteuffel:2020vjv} and QCD $\beta$
function~\cite{Tarasov:1980au, Larin:1993tp, vanRitbergen:1997va,
	Czakon:2004bu}.

To arrive at a consistent prediction, we need to match the N$^3$LL$^{\prime}$ resummed cross section to a NNLO Higgs+jet prediction for which we rely on an approximation. 
Our aN$^3$LO fixed-order cross section must contain the correct singular terms which are part of $\mathrm{d} \sigma^\mathrm{sing}$.
Further, there are large cancellations between the singular and the non-singular cross sections at large values of $q_T$, which must not be spoiled in the approximation procedure.
To satisfy both requirements,
we developed a general method to decorrelate the singular and non-singular contributions in~\citere{Cal:2023mib} which involves shifting a correlated piece between the singular and the non-singular~\cite{Dehnadi:2022prz}.
After the decorrelation procedure, we perform a Pad\'e-like approximation for the non-singular $\mathcal{O}\left(\alpha_s^3\right)$ coefficient. 

Our numerical results for the
resummed and fixed-order singular contributions are obtained with
\textsc{SCETlib}~\cite{scetlib}.
In \fig{fig:resumplot}, we show the resummed $q_T$ spectrum for $b\bar{b}\to H$ at different resummation orders up to the highest N$^3$LL$^{\prime}+$aN$^3$LO. The bands show the perturbative uncertainty estimate. For a detailed breakdown of the uncertainties, we refer to~\citere{Cal:2023mib}. We observe excellent perturbative convergence, with reduced uncertainties at each higher order.

\section{Monte Carlo simulations at NNLO+PS}\label{sec:MCyb}
This section highlights recent advancements in MC simulations for modelling the $y_b^2$ contribution to \bbH{} production. The first matching with parton showers was carried out in the 5FS scheme in~\citere{Biello:2024vdh} using the \minnlo{} method. More recently, the \GENEVA{} approach has been applied to 
simulate \bbtoH{} production in the 5FS~\cite{Gavardi:2025zpf}. In this section, we present the numerical comparison of the two matching 
methods for this process using the LHCHWG-recommended setup 
detailed in \sct{sec:setup}. In the MC simulations with massless bottom quarks, we consider the Higgs mass as the central value for the factorization ($\muF$) and renormalization ($\muR$) scales. 

We continue by discussing BSM studies using the \minnlo{} generator for an example scenario in the MSSM used as a proof-of-concept. 
The last part of the section is dedicated to the recently developed NNLO+PS MC generator in the 4FS using the \minnlo{} method~\cite{Biello:2024pgo}, 
which provides the first calculation of the NNLO QCD corrections in the massive scheme. In this context, we compare MC predictions from the 
4FS and 5FS \minnlo{} generators.

\subsection{NNLO+PS predictions in the 5FS}\label{sec:5FSNNLOPS}
\subsubsection{\minnlo{} and \GENEVA{}  methods in a nutshell}\label{sec:nutshell}

\minnlo{} \cite{Monni:2019whf,Monni:2020nks} and \GENEVA{} \cite{Alioli:2012fc,Alioli:2013hqa} are two methods that provide a consistent matching
between NNLO QCD corrections and parton showers. They both rely on the
combination of fixed-order calculations and analytic resummation
in the relevant resolution variables to achieve NNLO accuracy.
Both have been proven to work with different types of resolution variables \cite{alioli:2021qbf,Gavardi:2023aco,Ebert:2024zdj,Gavardi:2025zpf}
and they have been successfully applied to obtain NNLO+PS predictions for many processes of colour-singlet production \cite{Lombardi:2020wju,Lombardi:2021rvg,Buonocore:2021fnj,Lombardi:2021wug,Zanoli:2021iyp,Gavardi:2022ixt,Haisch:2022nwz,Lindert:2022qdd,Niggetiedt:2024nmp,Biello:2024vdh,Alioli:2015toa,Alioli:2019qzz,Alioli:2020qrd,alioli:2021qbf,Alioli:2021egp,Alioli:2022dkj,Alioli:2023har,Gavardi:2023aco,Gavardi:2025zpf,Alioli:2025xcu}.
The \minnlo{} approach has also been extended and applied to heavy-quark pair production at NNLO+PS \cite{mazzitelli:2020jio,mazzitelli:2021mmm,Mazzitelli:2023znt}
as well as heavy-quark pair production in association with colour singlets \cite{mazzitelli:2024ura,Biello:2024pgo}.
The two methods follow different philosophies to reach NNLO accuracy, whose fundamentals are presented below.

We start by briefly reviewing the \minnlo{} method for colour-singlet production. We refer to \citeres{Monni:2019whf,Monni:2020nks,Ebert:2024zdj} 
for further details. The \minnlo{} procedure is derived from the differential matching of the resummation 
formula in a suitable jet-resolution variable with the fixed-order prediction at NNLO.
By choosing a specific matching scheme, where the Sudakov is factored out, the 
cross section is cast into a form that has several important features. The structure
mimics the one of the parton shower, NNLO accuracy is achieved inclusively over
first and second radiation, and due to the Sudakov suppression no slicing cut-off
is required, making the procedure very efficient for event generation.
The matching to the parton shower is then performed for the second radiation
by means of the \POWHEG{} method \cite{frixione:2007vw}.

In practice, one starts from an NLO+PS \POWHEG{} event
generator~\cite{alioli:2010xd} for colour-singlet $F$ plus one jet $J$
production, which provides the differential description of the 
second radiation, consistently matched to subsequent emissions
by the shower. Such generator includes the NLO $FJ$ cross section 
(inclusive over the second radiation), which is governed by the following
function:
\begin{equation}
  \bar{B} = \frac{d\sigma_{\scriptscriptstyle\rm
      FJ}^{(1)}}{d\Phi_{\scriptscriptstyle\rm FJ}}\!\left(\mu\right) +
  \frac{d\sigma_{\scriptscriptstyle\rm
      FJ}^{(2)}}{d\Phi_{\scriptscriptstyle\rm FJ}}\!\left(\mu\right)\,.
\end{equation}
This formula can now be extended by means of the \minnlo{} procedure,
which replaces the NLO $FJ$ cross section by the NNLO $F$ cross section
with the overall Sudakov form factor in the jet-resolution variable, mimicking
the first shower emission.
\begin{equation}
  \label{MiNNLOPS_cross_section}
  \bar{B} = e^{-\tilde{S}\left(p_{\scriptscriptstyle\rm T}\right)}
  \left\{\frac{d\sigma_{\scriptscriptstyle\rm
      FJ}^{(1)}}{d\Phi_{\scriptscriptstyle\rm
      FJ}}\!\left(p_{\scriptscriptstyle\rm T}\right) \left[1 +
    \tilde{S}^{(1)}\!\left(p_{\scriptscriptstyle\rm T}\right)\right] +
  \frac{d\sigma_{\scriptscriptstyle\rm
      FJ}^{(2)}}{d\Phi_{\scriptscriptstyle\rm
      FJ}}\!\left(p_{\scriptscriptstyle\rm T}\right) + D^{(\geq
    3)}\!\left(p_{\scriptscriptstyle\rm T}\right)
  F^{\scriptscriptstyle\rm corr}\!\left(\Phi_{\scriptscriptstyle\rm
    FJ}\right)\right\},
\end{equation}
with $e^{-\tilde{S}}$ denoting the Sudakov form factor and 
the factor $F^{\scriptscriptstyle\rm corr}$ spreading the NNLO corrections
in the full phase space.
Note that it is crucial to choose a jet-resolution variable 
consistent with the ordering variable in the parton shower in order not to break
its logarithmic accuracy. For instance, in the case of standard Leading Logarithmic (LL) 
transverse momentum ordered showers (which is the current default, and 
considered throughout here) the transverse 
momentum of the colour singlet fulfils this criterion.
Here, $D^{(\geq 3)}$ contains the relevant terms to reach 
NNLO accuracy for observables inclusive over first and second radiation.
Without these corrections the inclusive observables would only be NLO accurate,
as originally introduced in the \minlo{} approach \cite{hamilton:2012rf,hamilton:2012np}.

The original \minnlo{} formulation was based on transverse momentum resummation \cite{Monni:2019whf,Monni:2020nks}, but it 
can be applied, in principle, to any resolution variable. For instance, it was explicitly derived for N-jettiness in \citere{Ebert:2024zdj},
which however formally breaks the LL accuracy of the shower, due to a mismatch of the resummation and the shower ordering variable. 
Beyond colour-singlet production, the \minnlo{} method has been extended to heavy-quark pair production without \cite{mazzitelli:2020jio,mazzitelli:2021mmm}
and with extra colour singlets in the final state \cite{mazzitelli:2024ura}. This advancement enables NNLO+PS predictions for \bbH{} production
not only in the massless scheme, but also in the massive one, allowing for a complete description of bottom-quark kinematics, as discussed in~\sct{sec:bbH4FS}.

The \GENEVA{} approach also builds on the analytic resummation in 
jet-resolution variables. The matching in both first 
and second radiation in \GENEVA{} are achieved through the 
analytic resummation, opposed to \minnlo{} where the second 
radiation is matched through \POWHEG{}, as discussed before.
\GENEVA{} generates the partonic events with the aim of not distorting
the spectrum in the zero-jet resolution variable (here generically called $r_0$).
Conceptually, the \GENEVA{} method boils down to three steps.
\begin{enumerate}
\item Matching. The $r_0$ spectrum at fixed order, differential over
  the phase space with no final state partons $\Phi_0$, is matched
  additively to its analytic resummation, obtaining
  \begin{equation}
    \frac{d\sigma}{d\Phi_0 \> dr_0} =
    \frac{d\sigma^{\scriptscriptstyle\rm FO}}{d\Phi_0 \> dr_0} +
    \frac{d\sigma^{\scriptscriptstyle\rm res}}{d\Phi_0 \> dr_0} -
    \left. \frac{d\sigma^{\scriptscriptstyle\rm res}}{d\Phi_0 \>
      dr_0}\right|_{\scriptscriptstyle\rm FO}.
  \end{equation}
  The main difference to the \minnlo{} matching is that the 
  matching of the resummation is kept in this additive form, instead 
  of factoring out the overall Sudakov form factor $e^{-\tilde{S}}$ in \eqn{MiNNLOPS_cross_section}.
\item Slicing. A slicing scale $r_0^{\scriptscriptstyle\rm cut}$ is
  introduced to separate events where the radiation is not resolved
  from those where it is, whose distributions are described
  respectively by
  \begin{equation}
    \frac{d\sigma_0}{d\Phi_0}\!\left(r_0^{\scriptscriptstyle\rm
      cut}\right) = \int_0^{r_0^{\scriptscriptstyle\rm cut}} dr_0 \>
    \frac{d\sigma}{d\Phi_0 \> dr_0}
    \quad \mbox{and} \quad
    \frac{d\sigma_{\geq 1}}{d\Phi_0 \> dr_0} = \frac{d\sigma}{d\Phi_0
      \> dr_0} \> \theta\!\left(r_0 - r_0^{\scriptscriptstyle\rm
      cut}\right).
  \end{equation}
  This step is not required in the \minnlo{} approach, where the
  exponential suppression  by the overall Sudakov form
  factor makes the differential cross section numerically integrable
  down to $p_{\scriptscriptstyle\rm T} = 0$ (after regulating the Landau
  pole).
\item Splitting. The ${\cal P}_{0 \to 1}$ function is introduced to
  make the $r_0$ spectrum differential over the $\Phi_1$ phase space
  with one final state parton, thus obtaining
  \begin{equation}
    \frac{d\sigma_{\geq 1}}{d\Phi_1} =
    \left\{\frac{d\sigma^{\scriptscriptstyle\rm FO}}{d\Phi_1} +
    \left[\frac{d\sigma^{\scriptscriptstyle\rm res}}{d\Phi_0 \> dr_0}
      - \left. \frac{d\sigma^{\scriptscriptstyle\rm res}}{d\Phi_0 \>
        dr_0}\right|_{\scriptscriptstyle\rm FO} \right] {\cal P}_{0
      \to 1}\!\left(\Phi_1\right)\right\} \theta\!\left(r_0 -
    r_0^{\scriptscriptstyle\rm cut}\right).
  \end{equation}
  Here, ${\cal P}_{0 \to 1}$ plays the corresponding role as
  $F^{\scriptscriptstyle\rm corr}$ in
  \eqn{MiNNLOPS_cross_section}.
\end{enumerate}
The second emission is then generated by iterating this procedure 
for a one-jet resolution variable $r_1$ on
the $d\sigma_{\geq 1} / d\Phi_1$ differential cross section at NLO. Most
\GENEVA{} implementations use the 0- and 1-jettiness $r_0=\mathcal{T}_0$
and $r_1=\mathcal{T}_1$ as resolution variables. The first implementation
that adopted the colour-singlet transverse momentum
$r_0=q_{\scriptscriptstyle\rm T}$ instead was presented
in~\citere{alioli:2021qbf}. This approach was refined and extended
in~\citere{Gavardi:2025zpf} (also using a $p_{\scriptscriptstyle\rm
  T}$-like observable as 1-jet resolution variable $r_1$), where the \bbtoH{} process was discussed.

A subtlety related to the slicing of the phase space in \GENEVA{} is
that, since the scales appearing in the resummed cumulant and spectrum
are a function of the resolution variable $r_0$, the operations of
setting the scales and integrating the spectrum do not commute. This
leaves us with two main options. Setting the scales in the spectrum (\texttt{spectrum} option)
provides the best theoretical description of the $r_0$ distribution in
the region of small $r_0$ at the price of introducing spurious
subleading contributions in the distributions inclusive over the
radiation. Setting the scales in the cumulant and defining the
spectrum as its derivative (\texttt{cumulant} option), on the other hand, enforces the generated
events to reproduce the exact NNLO inclusive distributions by
construction. The numerical difference between the two choices is discussed in the following section, specifically in \fig{fig:genevaptH}.

Besides treating the resolution variables in different ways, the two
approaches may also produce small differences in the distributions
inclusive over the radiation. The \minnlo{} approach can differ from 
fixed-order NNLO predictions by terms beyond the nominal accuracy,
due to the specific matching scheme, where the Sudakov is factored out.
 The additive approach
adopted by \GENEVA{}, on the other hand, can exactly align 
terms beyond accuracy with the fixed-order differential NNLO cross section.
On the other hand, the presence of the overall Sudakov form factor in
\minnlo{} has the advantage of a better numerical efficiency with a small number of 
negative weights and without having to deal with large cancellations in a
slicing cutoff, as introduced in \GENEVA{}. Distributions sensitive
to QCD emissions 
(which also have lower perturbative accuracy) are instead more heavily
dependent on the specifics of the two approaches, and differences between
the methods can be used to estimate the respective theoretical
uncertainties. Despite these differences, \minnlo{} and \GENEVA{} are generally 
agree within their respective uncertainties, especially for observables inclusive 
over radiation, where any difference is beyond NNLO QCD and should be 
covered by scale uncertainties.

\subsubsection{Numerical comparison of \minnlo{} and \GENEVA{} predictions}
In this section, we present a numerical comparison between two MC generators for \bbtoH{} production in the 5FS in the \minnlo{} and \GENEVA{} frameworks. The settings adopted are described in~\sct{sec:setup}, with the only notable difference being the treatment of the running bottom-quark Yukawa coupling. 
The \minnlo{} predictions follow the setup in \sct{sec:setup} using a four-loop running 
\cite{harlander:2002wh,baikov_2014} to evolve the bottom Yukawa from 
the input to the hard scale, while the \GENEVA{} generator employs a three-loop running throughout derived from SCET. This difference is numerically very small and can be safely neglected in the following comparison. Both the MC generators produce events that are showered using~\PYTHIA{8}~\cite{Bierlich:2022pfr} with a local recoil.

The \minnlo{} generator includes an option to change the scale in the resummed logarithmic terms from the scale $m_X$ by a factor of $\KQ$, where $m_X$ denotes the invariant mass of the final state of the Born process (for $\bbtoH$, this corresponds to the Higgs-boson mass). This scale $Q = \KQ m_X$ controls the matching of resummed and fixed-order contributions within \minnlo{}\footnote{See section 4.2 of~\citere{mazzitelli:2021mmm} for further details.}. It can be applied to assess the matching uncertainties in the \minnlo{} generator. The $\bbtoH$ results are obtained using the default central value of $\KQ = 0.25$, with the alternative choice $\KQ = 0.5$ included in the uncertainty estimate. The overall theoretical uncertainty is assessed via a 14-point scale variation, combining standard renormalization and factorization scale variations with the additional resummation scale variation. In the matching region, the logarithmic contributions are smoothly suppressed towards large transverse momenta using modified logarithms,
\begin{align}
	\ln \frac{Q}{\pt} \rightarrow \frac{1}{p} \ln\left(1+\left(\frac{Q}{\pt}\right)^p\right),
\end{align}
as originally introduced in~\citere{Monni:2019whf}. The free positive parameter $p$ is set to the default value, $p=6$. We have tested alternative parameters for the modified logarithms and found only minor effects, which are well within the \minnlo{} scale uncertainties.

The \GENEVA{} generator uses the cumulant scale choices in order to have a better agreement of inclusive observables with NNLO QCD predictions. Moreover, it applies default theoretical-uncertainty estimation with the included matching uncertainty as discussed in~\citere{Gavardi:2025zpf}.

\begin{figure}[t!]
\begin{center}
\begin{tabular}{cc}
\includegraphics[width=.45\textwidth, page=1]{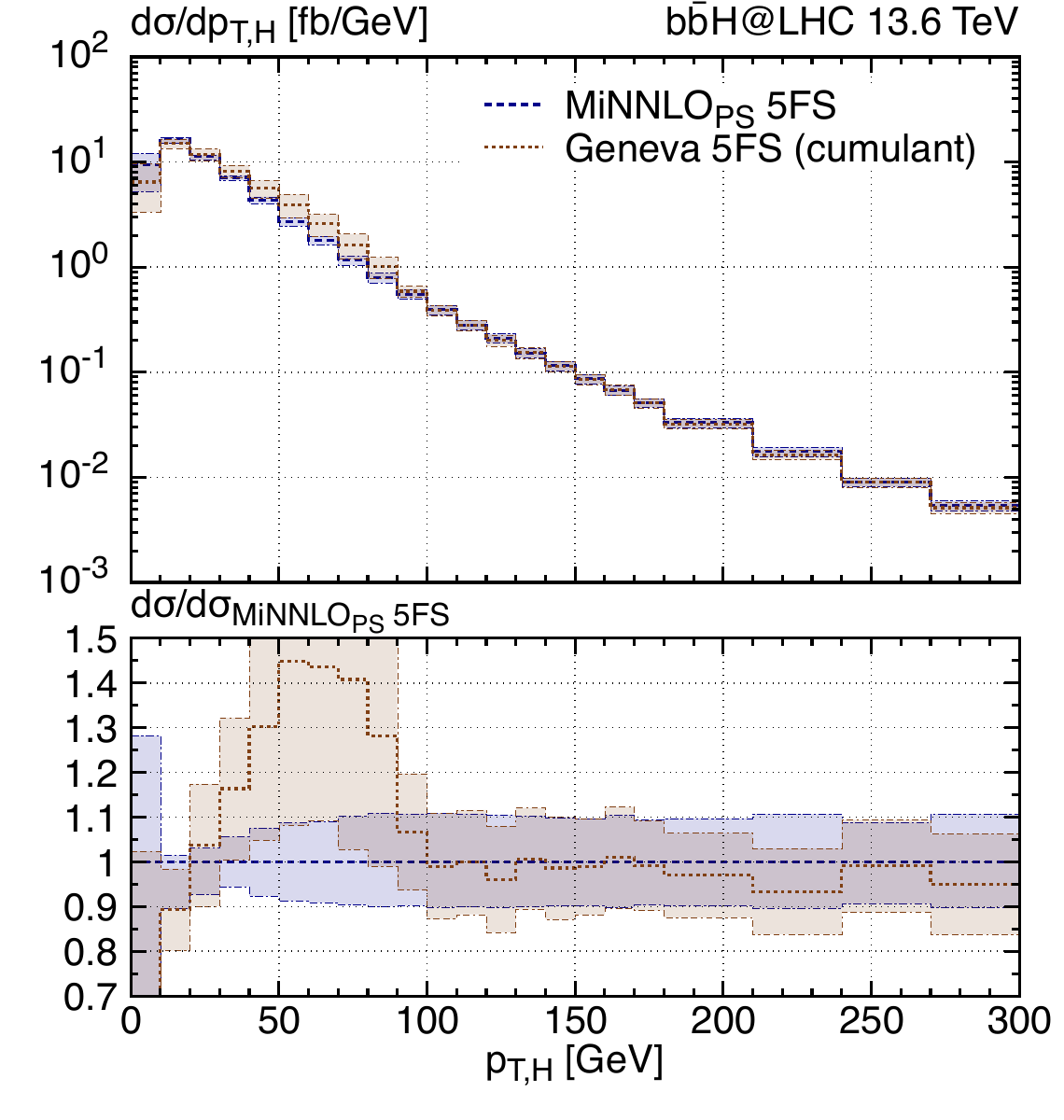}&
\includegraphics[width=.45\textwidth, page=1]{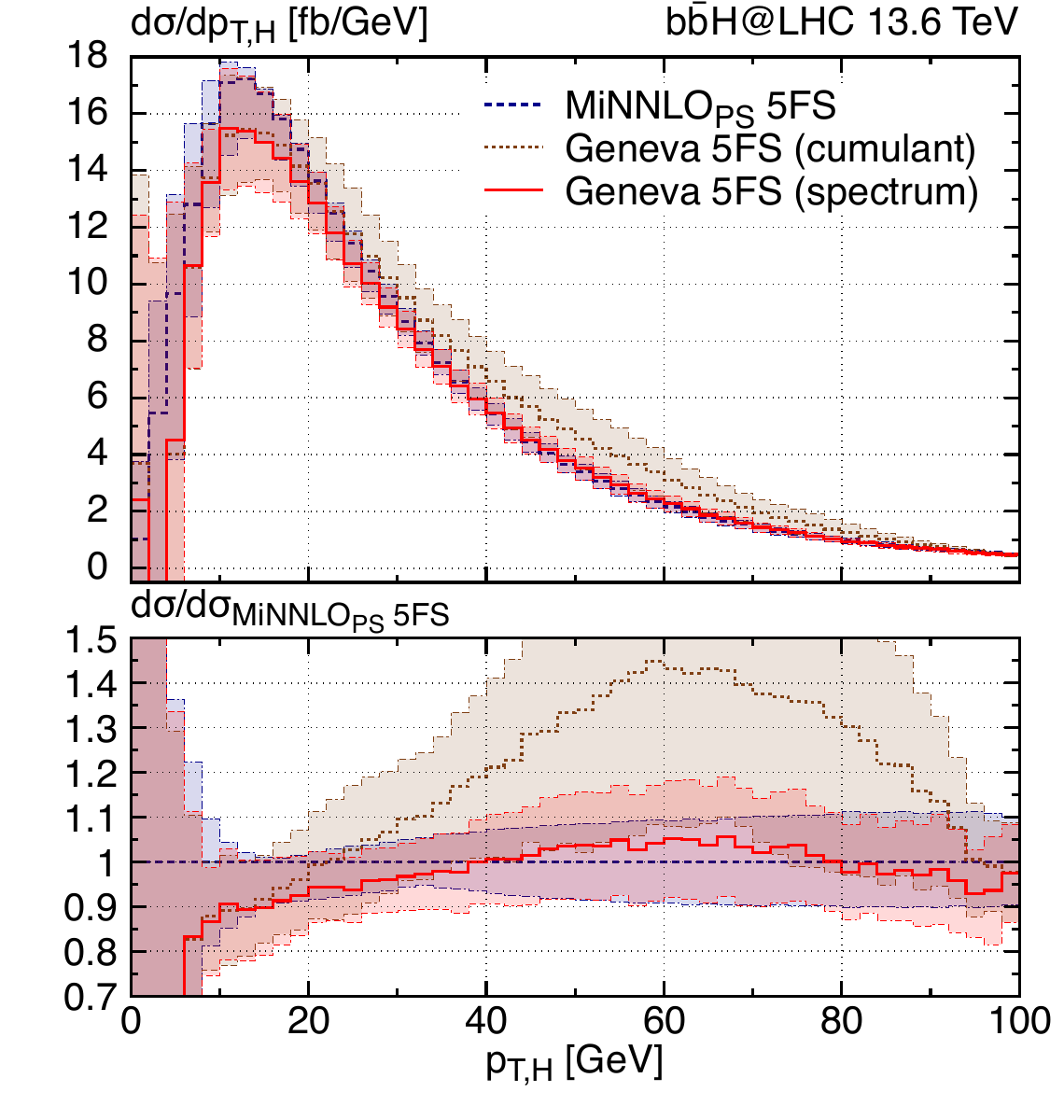}
\end{tabular}
\vspace*{1ex}
\caption{Comparison of \minnlo{} (blue, dashed) and \GENEVA{} (brown, dotted) predictions at NNLO+PS level in the 5FS for the transverse momentum distribution of the Higgs boson, where \GENEVA{} uses their default cumulant scale choice. The zoomed version on the right shows also the alternative spectrum scale choice of the \GENEVA{} generator (red, solid).\label{fig:genevaptH}}
\end{center}
\end{figure}

We start the comparison by focusing on the transverse momentum spectrum of the Higgs boson in~\fig{fig:genevaptH}. The left plot shows the spectrum in a large range. As 
expected, we observe good agreement between \minnlo{} and \GENEVA{} predictions
at large transverse momenta, where both calculations are effectively NLO accurate and 
governed by $H$+jet production. Also in the range of $10-30$ GeV both generators yield relatively 
close results, consistent within their respective uncertainty. However, directly
before and after the peak, and especially in the
$30-100$ GeV range, we observe substantial differences between \minnlo{} and \GENEVA{} with cumulant scale choice, barely covered by scale uncertainties,
which are particularly large for \GENEVA{} in that range. The large difference in the transition region between \GENEVA{} with cumulant scale choices
and \minnlo{} is due to differences in the matching procedures. The analytic resummed prediction, used in the \GENEVA{} framework, underestimates the total cross section. However, this underestimated component must be included to obtain the correct total cross section. Since \GENEVA{} requires it to not affect either the
peak region nor the fixed-order accuracy in the tail,
it is accommodated in the matching procedure by modifying the transition region.
In the \GENEVA{} matching procedure with the cumulant option, this leads to the observed larger scale uncertainties  
in the intermediate range of the $\ptH\sim[30,90]$ GeV.

The right plot of \fig{fig:genevaptH} provides a zoomed-in view of the same observable and includes also the \GENEVA{} prediction obtained using the spectrum scale choices, as described in~\sct{sec:nutshell}, which are more appropriate for the transverse momentum spectrum. 
Moreover, in \GENEVA{}, the scale variation is implemented through variations of the profile scales at the spectrum level, making them particularly well-suited for transverse observables when using the spectrum scale choice. Indeed, the substantial 
scale uncertainties observed for the cumulant scale choice
in the matching region are strongly reduced for the spectrum scale choice, which 
features uncertainties that are of similar size as the \minnlo{} ones.
Most notably, we find that \minnlo{} and \GENEVA{} predictions are in significantly 
better agreement, when the spectrum scale choice is applied in \GENEVA{}.
In fact, the two predictions are in excellent agreement within their respective scale
uncertainties over the entire transverse momentum range.
We stress that at small transverse momenta all 5FS predictions
become essentially unphysical, as power corrections in the 
bottom-quark mass become crucial for $p_T\lesssim m_b$, which are included only 
in 4FS calculations. Indeed, we observe that the scale uncertainty bands are
severely inflated at small transverse momenta, and the predictions can
even turn negative in the first bins, because the bottom PDFs are not valid below
the bottom-quark mass threshold.

\begin{figure}[t!]
\begin{center}
\begin{tabular}{cc}
\includegraphics[width=.45\textwidth, page=1]{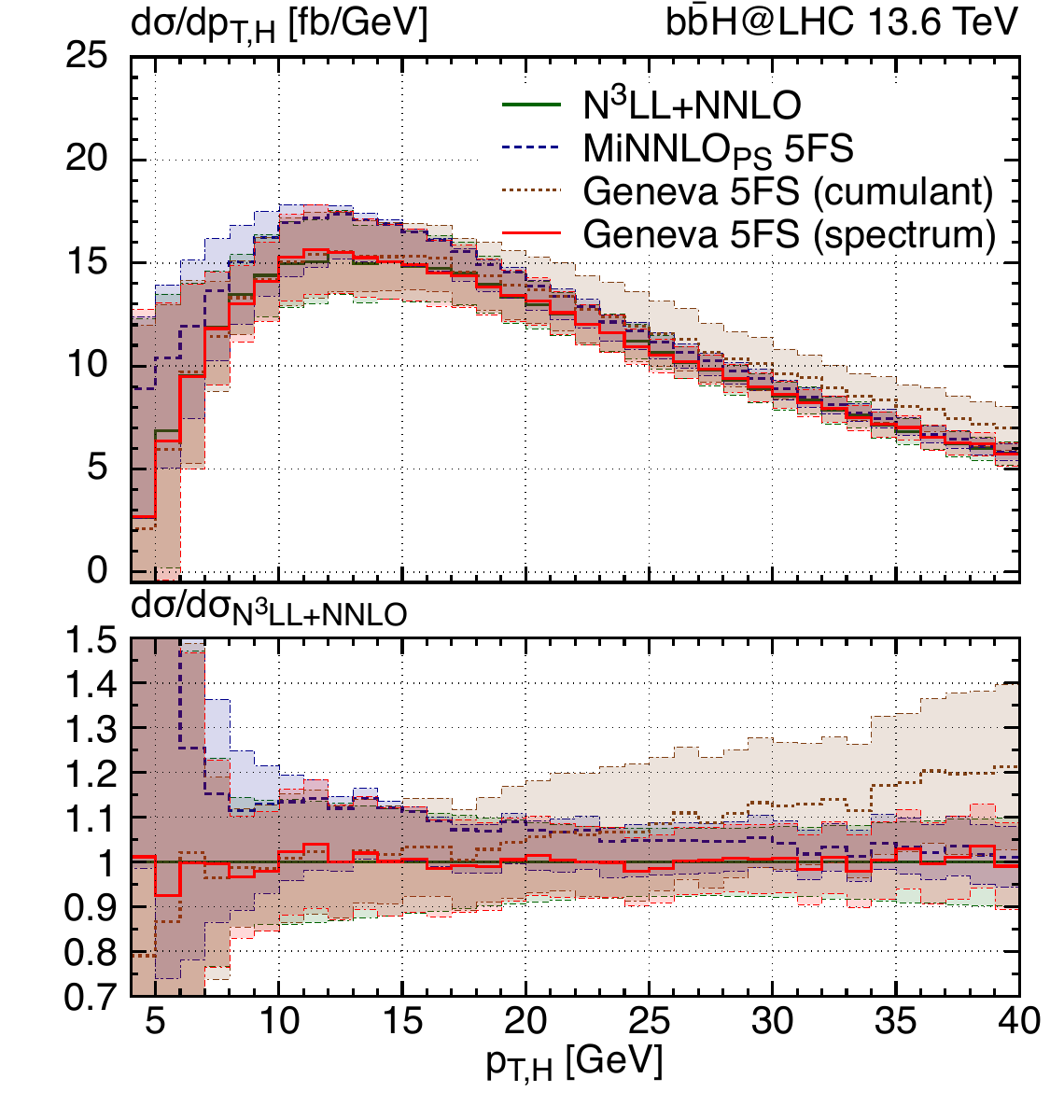}&
\includegraphics[width=.45\textwidth, page=1]{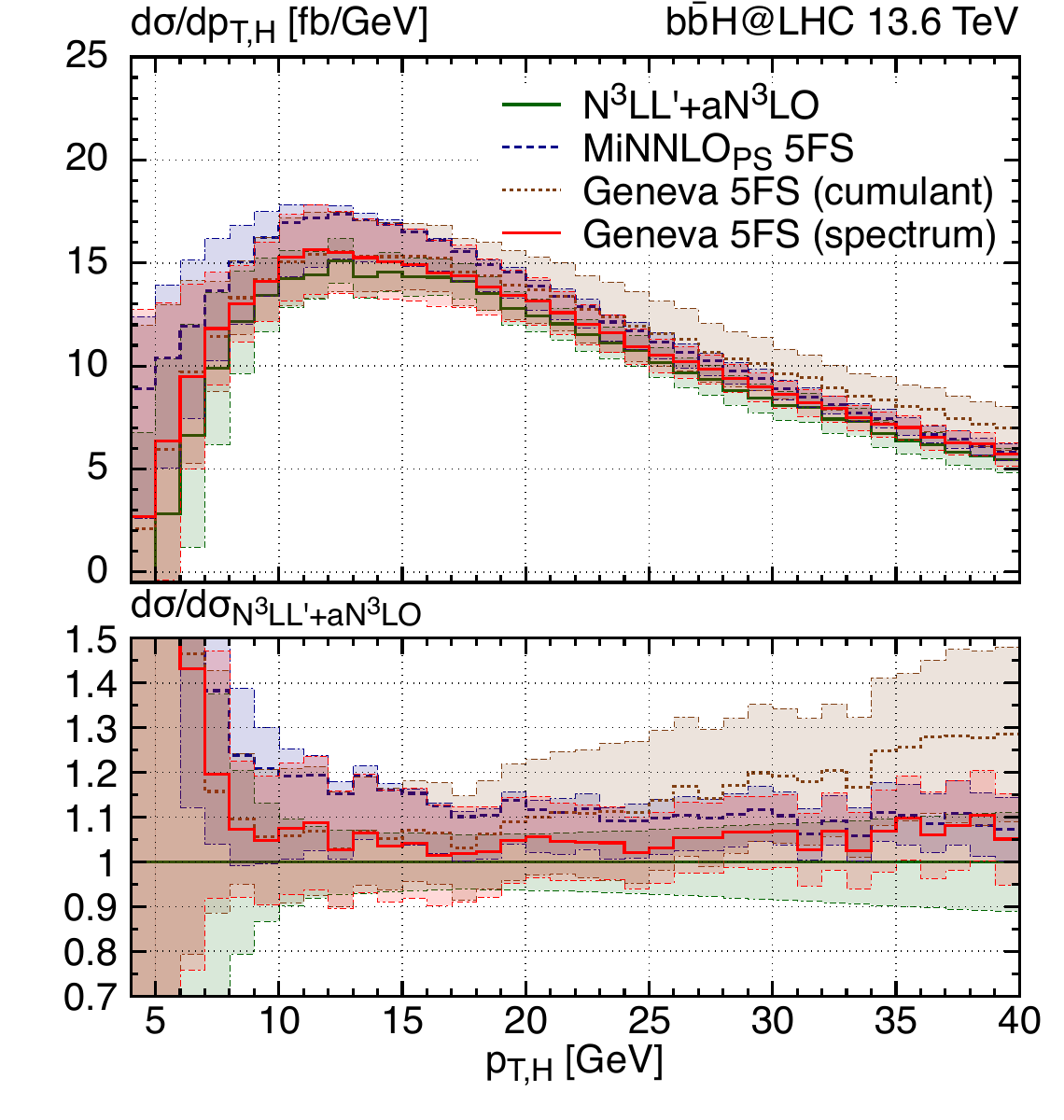}
\end{tabular}
\vspace*{1ex}
\caption{Comparison of the Higgs transverse momentum spectrum predicted by analytic resummed predictions (green, solid) at $\text{N}^3\text{LL+NNLO}$ (left plot) and at $\text{N}^3\text{LL$^{\prime}$+aN}^3\text{LO}$ against the results obtained by the two MC generators: \minnlo{} (blue, dashed) and \GENEVA{} with cumulant (brown, dotted) and spectrum (red, solid) scale choices. All predictions are obtained in the 5FS for the $y_b^2$ contribution. \label{fig:resVSMCs}}
\end{center}
\end{figure}

We compare the transverse momentum spectrum of the Higgs boson with the resummed analytic predictions discussed in~\sct{sec:resummation}. In the left plot of~\fig{fig:resVSMCs}, we show the MC predictions against the $\text{N}^3\text{LL}+\text{NNLO}$ result. By construction, the \GENEVA{} prediction with the spectrum scale choice matches the analytic result exactly before the parton shower, and the shower does not introduce sizable numerical effects due to the local dipole
recoil prescription selected in the shower settings of the calculations. On the other hand, the cumulant scale choice shows good agreement in the resummation region ($p_{\text{T,H}} < 20$\,GeV), but begins to substantially deviate with increasing
uncertainties at higher transverse momenta, as already observed in~\fig{fig:genevaptH}. The \minnlo{} prediction also shows good agreement with $\text{N}^3\text{LL}+\text{NNLO}$ result within its scale uncertainty band, with the largest deviations appearing at very low transverse momenta, well covered by the resummation scale variation.
In that region, $p_T\lesssim m_b$, the massless calculation actually 
breaks down, and finite bottom-quark mass effects become relevant, as pointed out before.
It therefore includes spurious effects induced by the PDF evolution, which turn 
the $\text{N}^3\text{LL}^{\prime}+\text{NNLO}$ cross section negative below $4$\,GeV, where
the figure has been cut for that reason.
In the right plot of~\fig{fig:resVSMCs}, we present the comparison of the same MCs spectra against the $\text{N}^3\text{LL$^{\prime}$+aN}^3\text{LO}$ calculation. The inclusion of higher-order corrections in the analytic spectrum results in a negative shift relative to both the \minnlo{} and \GENEVA{} predictions across the entire distribution, with a mostly flat effect for $p_{\text{T,H}} > 10$,GeV. Compared to the others, the \GENEVA{} prediction with the spectrum scale choice lies closer to the analytic result.

\begin{figure}[t!]
\begin{center}
\begin{tabular}{cc}
\includegraphics[width=.45\textwidth, page=1]{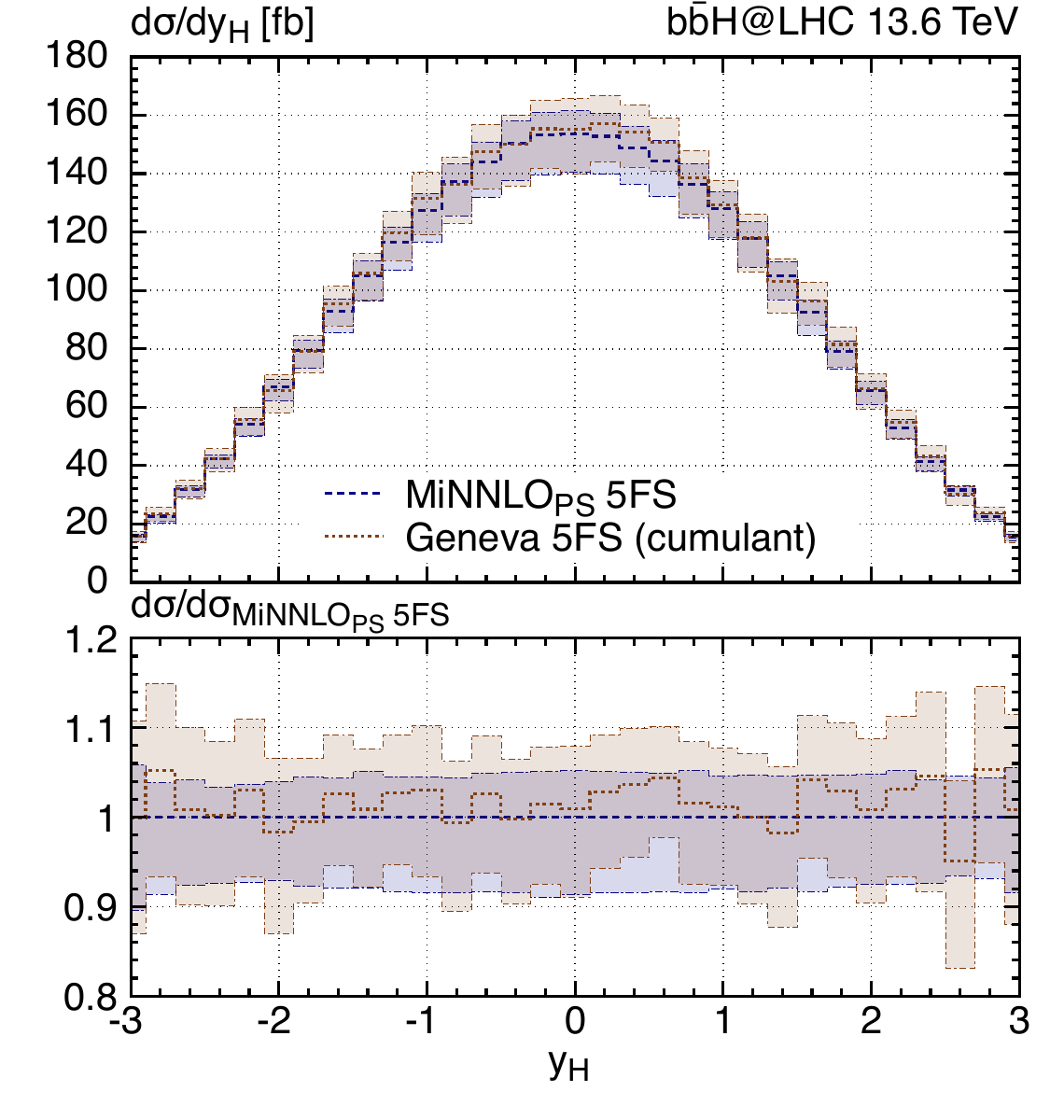}&
\includegraphics[width=.45\textwidth, page=1]{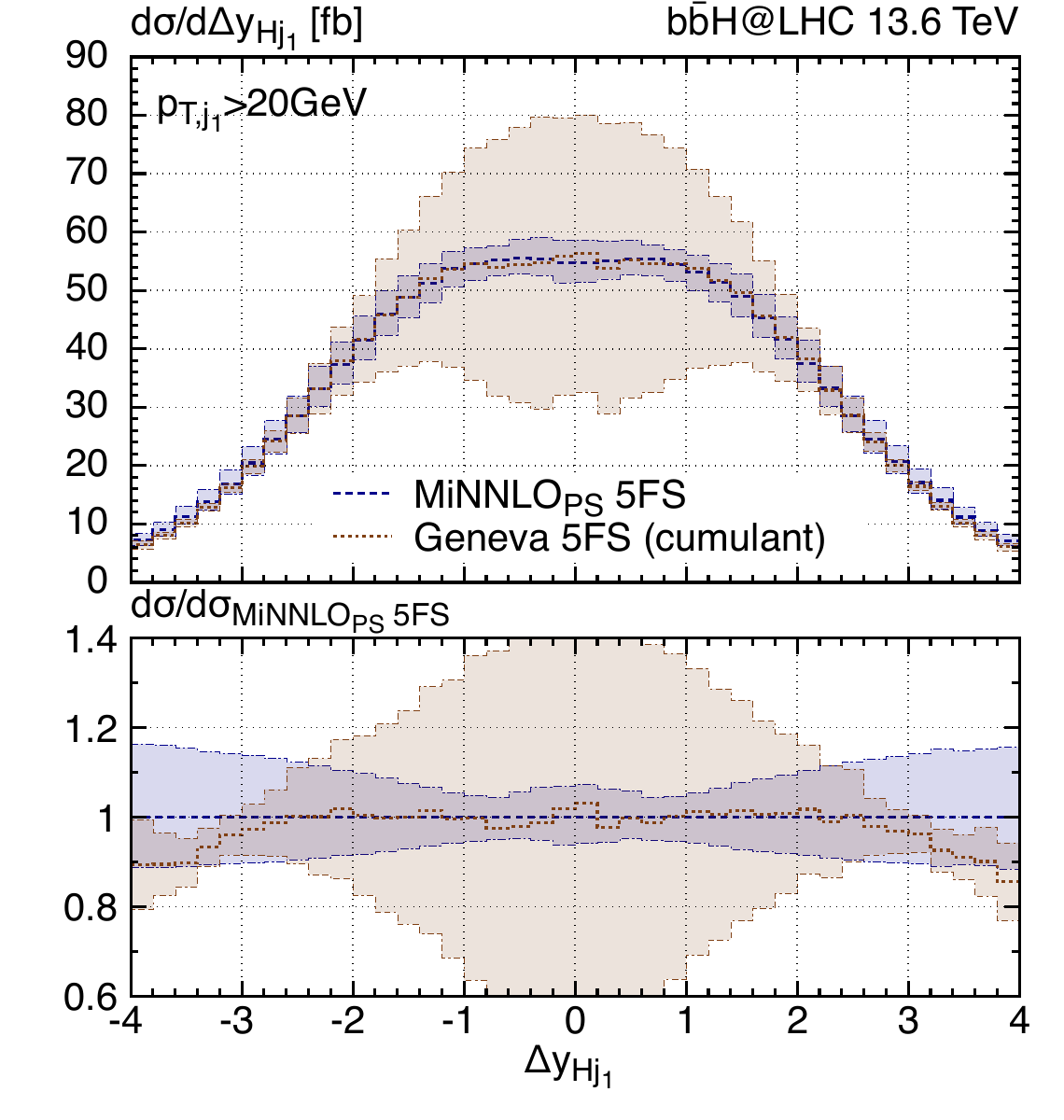}
\end{tabular}
\vspace*{1ex}
\caption{Higgs boson rapidity and difference in rapidity between the scalar and the leading jet as predicted by two MC generators in 5FS interfaced with \PYTHIA{8} using the \minnlo{} (blue, dashed) and \GENEVA{} (brown, dotted) methods. \label{fig:genevay}}
\end{center}
\end{figure}

We conclude the section by presenting a comparison of two angular observables in~\fig{fig:genevay}. The left plot shows the NNLO-accurate Higgs rapidity distribution, displaying excellent agreement between the \minnlo{} prediction and the \GENEVA{} result using the cumulant scale choice, in terms of shape, normalization and size of the scale uncertainties. 
The right plot illustrates an example of an observable that depends on the leading jet. Here, we apply a flavour-blind anti-$k_T$ clustering algorithm with radius $R = 0.4$ and compute the rapidity difference between the Higgs boson and the highest-$p_T$ jet, requiring a minimum jet transverse momentum of 20\,GeV. We observe excellent agreement between the central \minnlo{} and 
\GENEVA{} predictions, with some moderate differences only in the very forward region. However, the associated scale uncertainties differ significantly between the two generators, with \GENEVA{} exhibiting a much larger theoretical uncertainty at central
rapidities.

\subsubsection{Heavy-Higgs for BSM studies in \minnlo{}}

Higgs boson production in association with a bottom-quark pair is of
particular interest in BSM extensions. Indeed, many BSM scenarios predict modifications to the Higgs-bottom quark coupling, which could lead to observable deviations in the production rates and kinematic distributions of the \bbH{} process. In many models, \bbH{} production can become the dominant production mode of exotic Higgs states. The \minnlo{} generator, introduced earlier in this section, is built within the SM framework, but can be easily extended to accommodate BSM predictions. To illustrate its potential, we consider a specific example of how the NNLO+PS generator can be adapted for BSM scenarios. We specifically consider the MSSM~\cite{Fayet:1974pd,Fayet:1976et,Fayet:1977yc,Ovrut:1984uc,Haber:1984rc,Gunion:1984yn}, which corresponds to a Type-II 2HDM~\cite{Branco:2011iw} at LO, but deviates from it at higher orders. The MSSM enforces relations between the Higgs sector and the superpartners of SM particles. Unlike the SM, the MSSM requires two Higgs doublets ($H_u$ and $H_d$) to give mass to both up-type and down-type fermions. An important parameter of the model is the ratio of vacuum expectation values ($v_u$ and $v_d$) of the two SU(2) doublets,
\begin{align}
	\tan\beta=\frac{v_u}{v_d}\,.
\end{align}
The other independent parameter is the CP-even Higgs mixing angle
$\alpha$. This results in five physical Higgs bosons: two CP-even
($h$, $H$), one CP-odd ($A$) and two charged Higgs bosons
($H^{\pm}$). The lightest Higgs boson ($h$) in MSSM can mimic the SM
Higgs, but has different properties depending on the model
parameters. The tree-level mass of $h$ is bounded from above by $m_Z
|\cos 2\beta|$, where $m_Z$ is the Z-boson mass. However, radiative
corrections (mainly from the supersymmetric partners of the bottom and top
quarks) can significantly alter the tree-level prediction, allowing
for $m_h\sim 125$\,GeV (see e.g.~\cite{Slavich:2020zjv}). We perform
the NNLO+PS prediction in the benchmark configuration known as
$M_h^{125}$ scenario~\cite{Bagnaschi:2018ofa}, where all
superparticles are chosen to be so heavy that the presence of these
effects has only a mild impact on the production and decay of the
light MSSM Higgs boson. Thus, the phenomenology of this scenario at
the LHC closely resembles that of a Type-II 2HDM with Higgs couplings
corresponding to the MSSM ones for light scalar states. On the other
hand, for heavy bosons, there are large enhancements in the MSSM
scenario. Loop corrections from Standard Model particles—dominated by the top-quark contribution—as well as from MSSM superpartners modify the effective bottom Yukawa coupling. The $\tan\beta$-enhanced SUSY corrections are resummed to all orders in the effective coupling. For the chosen mass of the SUSY partners, we refer to eq.\,(4) of~\citere{Bagnaschi:2018ofa}. We stress that the NNLO+PS calculation in the 5FS scheme contains only terms proportional to the squared bottom Yukawa coupling. As a result, \bbH{} predictions in the MSSM scenario differ from those in the SM—obtained with an heavy Higgs mass $m_A$—solely by an overall rescaling factor. Using chirality arguments, predictions for pseudo-scalar $b\bar bA$ production can be obtained from the $b\bar b h$ simulation by employing a heavy-boson mass and an effective Yukawa coupling, while neglecting the suppressed non-factorising SUSY-QCD corrections\footnote{The latter have been shown to be numerically small in~\citere{Dittmaier:2014sva}.}. Therefore, in the case of the CP-odd Higgs boson production, we have:
\begin{align}
	\dd \sigma_{b\bar b A}^{\text{MSSM}}(m_A) = \dd \sigma_{b\bar b h}^{\text{SM}}(m_A) \cdot (\tilde g_b^{A})^2\,,	\label{eq:BSMYuk}
\end{align}
with
\begin{align}
	\tilde{g}_b^A = \frac{\tan \beta}{1 + \Delta_b} \left( 1 - \Delta_b \frac{1}{\tan^2\beta} \right)\,.
\end{align}
The parameter $ \Delta_b $ resums higher-order sbottom contributions~\cite{Banks:1987iu,Hall:1993gn,Carena:1994bv,Carena:2000uj,Guasch:2003cv}. In~\citere{Guasch:2003cv} the resummation effects has been studied in a systematic power-counting framework, estimating the uncertainties when keeping only the one-loop effects in the $ \Delta_b $ determination. In order to reduce them, several two-loop calculations have been performed~\cite{Noth:2008tw,Noth:2010jy,Mihaila:2010mp,Crivellin:2012zz} and EW and diagonal contributions~\cite{Ghezzi:2017enb} have been computed recently.  All these corrections have been incorporated into the resummation parameter $ \Delta_b $ for this benchmark, with its numerical value determined using the \texttt{Hdecay} code~\cite{Djouadi:1997yw,Djouadi:2018xqq}. 
Following the current constraints on the $M_h^{125}$ scenario, where the lightest Higgs has a mass of $m_h= 125$\,GeV and is consistent with experimental observations, we consider the case of a CP-odd Higgs boson with a mass of 1.4 TeV and $\tan\beta = 20$, a point in the parameter space which is currently not excluded, as shown in Figure 1 of~\citere{Bagnaschi:2018ofa}. 
We have performed the predictions by running the \minnlo{} 5FS generator with a heavy Higgs-boson mass and adjusted the Yukawa coupling according to~\eqn{eq:BSMYuk}.

\begin{figure}[t!]
\begin{center}
\begin{tabular}{cc}
\includegraphics[width=.45\textwidth, page=1]{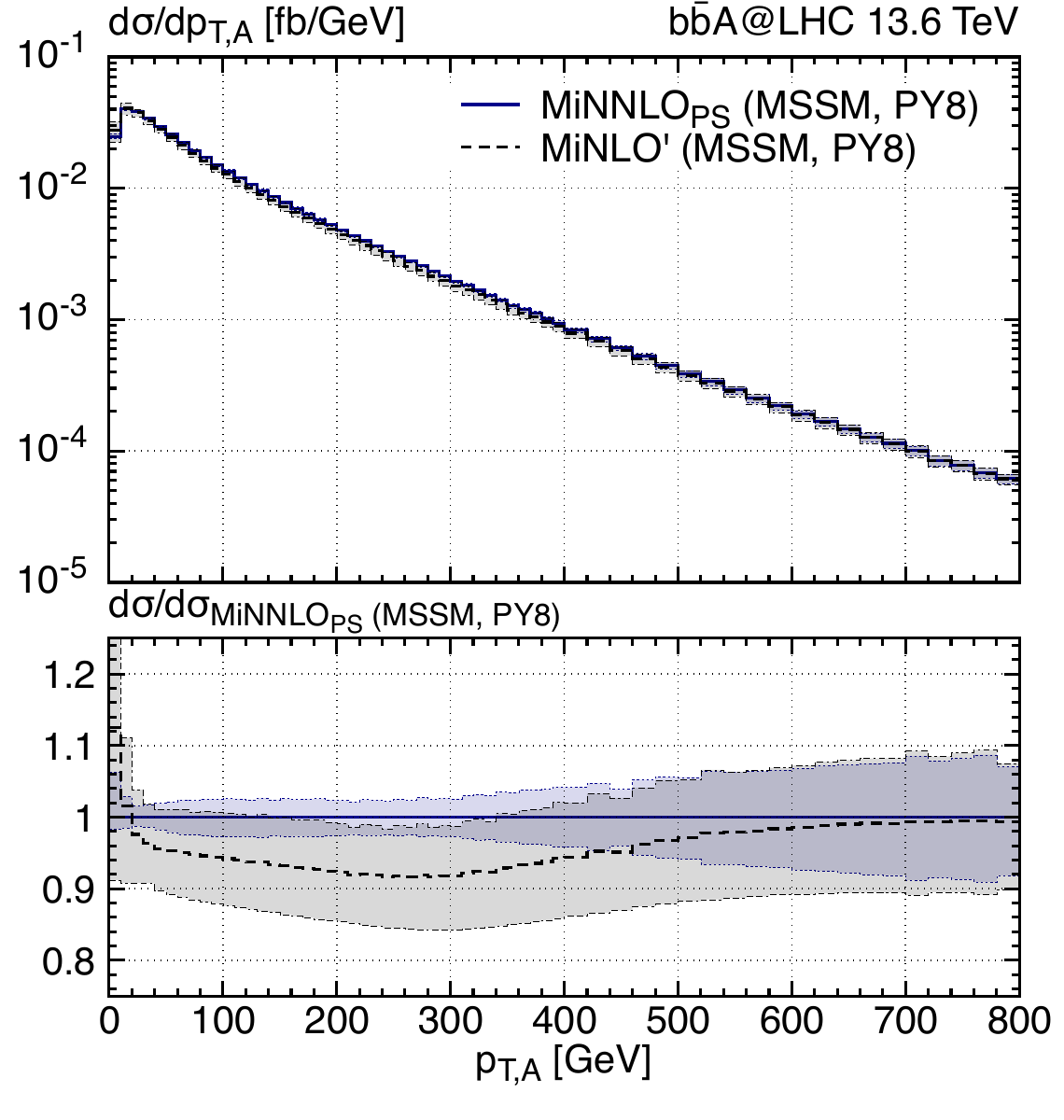}&
\includegraphics[width=.45\textwidth, page=1]{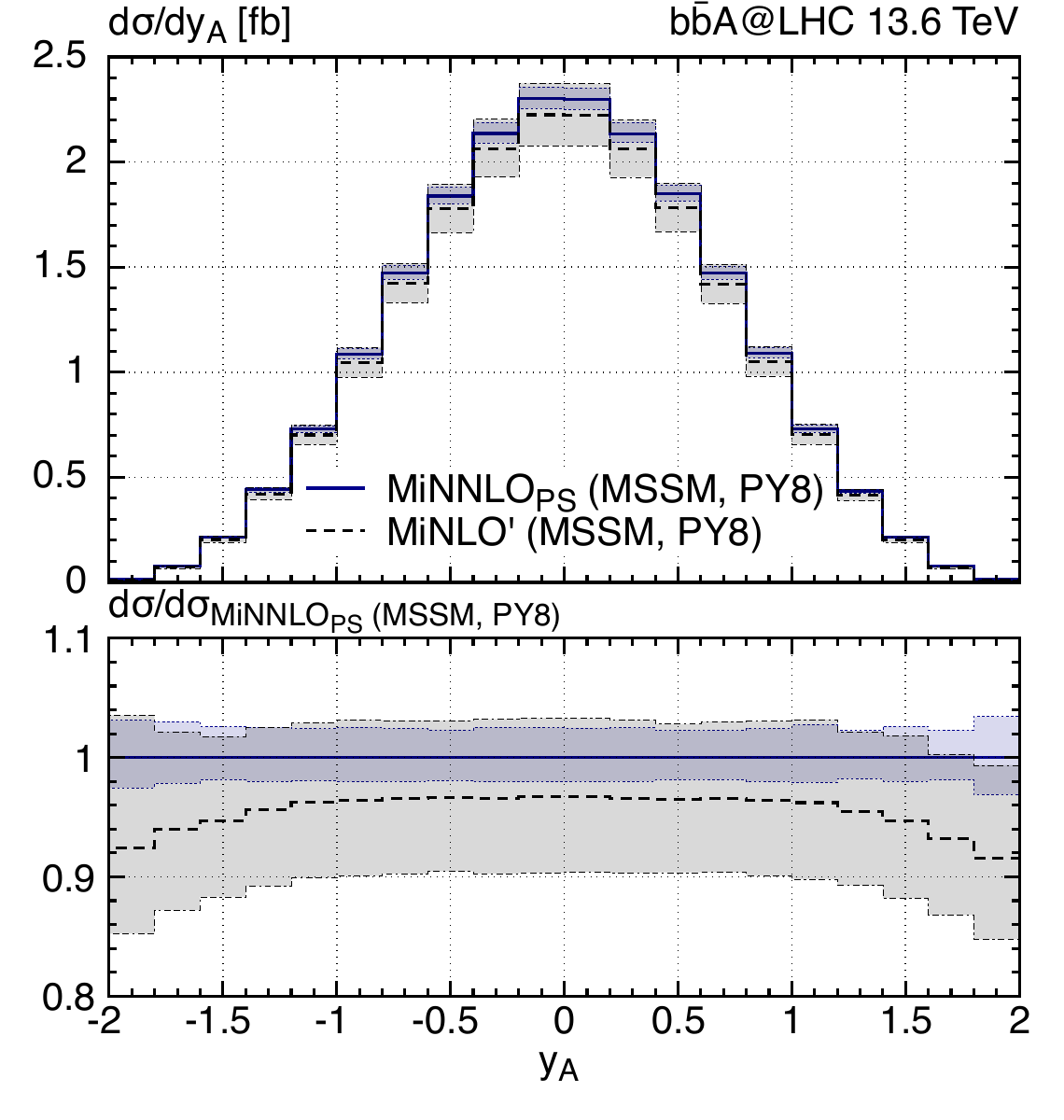}
\end{tabular}
\vspace*{1ex}
\caption{Comparison of \minlo{} and \minnlo{} results for transverse momentum and rapidity distributions of the pseudo-scalar Higgs boson with $m_A=1.4$ TeV and $\tan\beta=20$. \label{fig:MiNLOBSM}}
\end{center}
\end{figure}

\begin{figure}[t!]
\begin{center}
\begin{tabular}{cc}
\includegraphics[width=.45\textwidth, page=1]{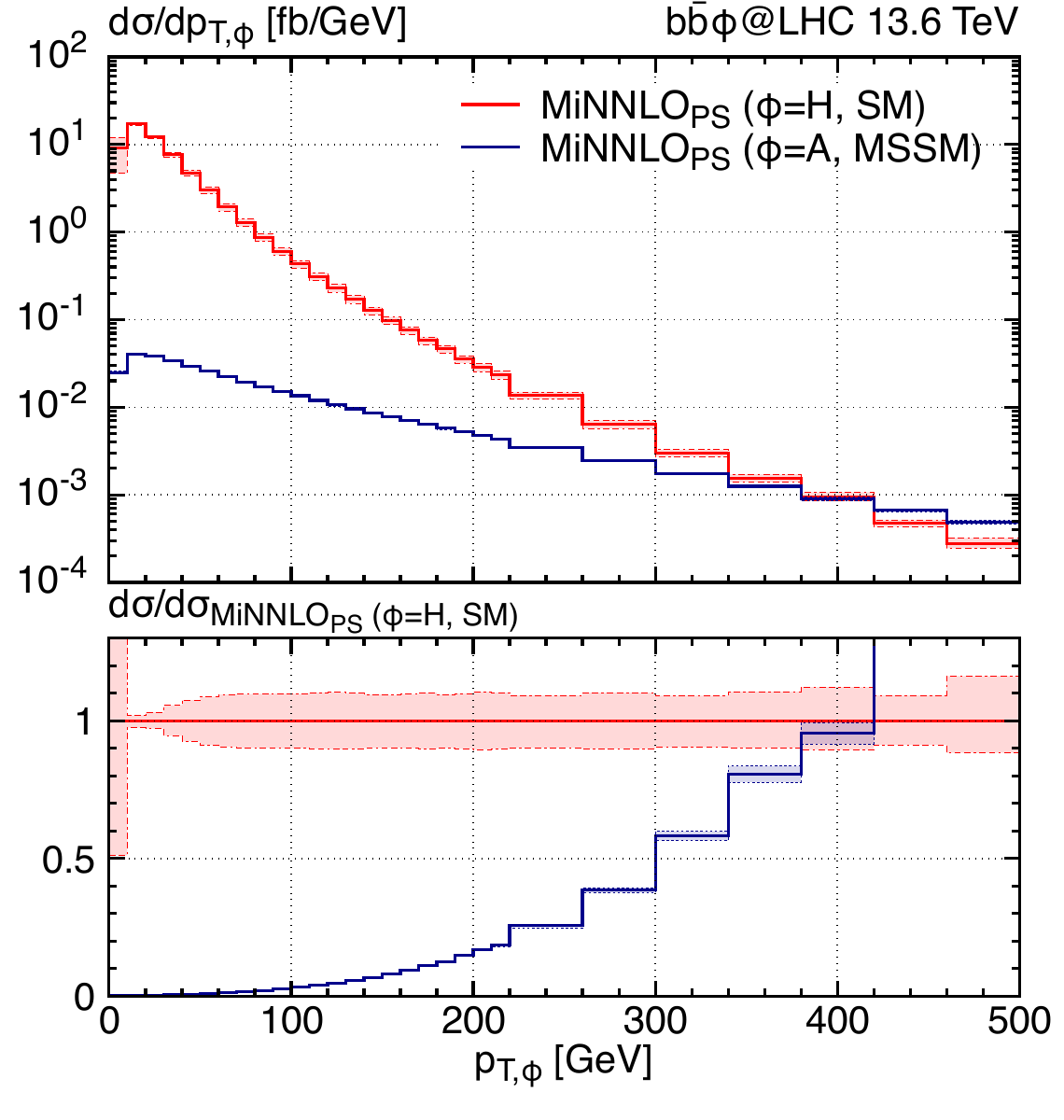}&
\includegraphics[width=.45\textwidth, page=1]{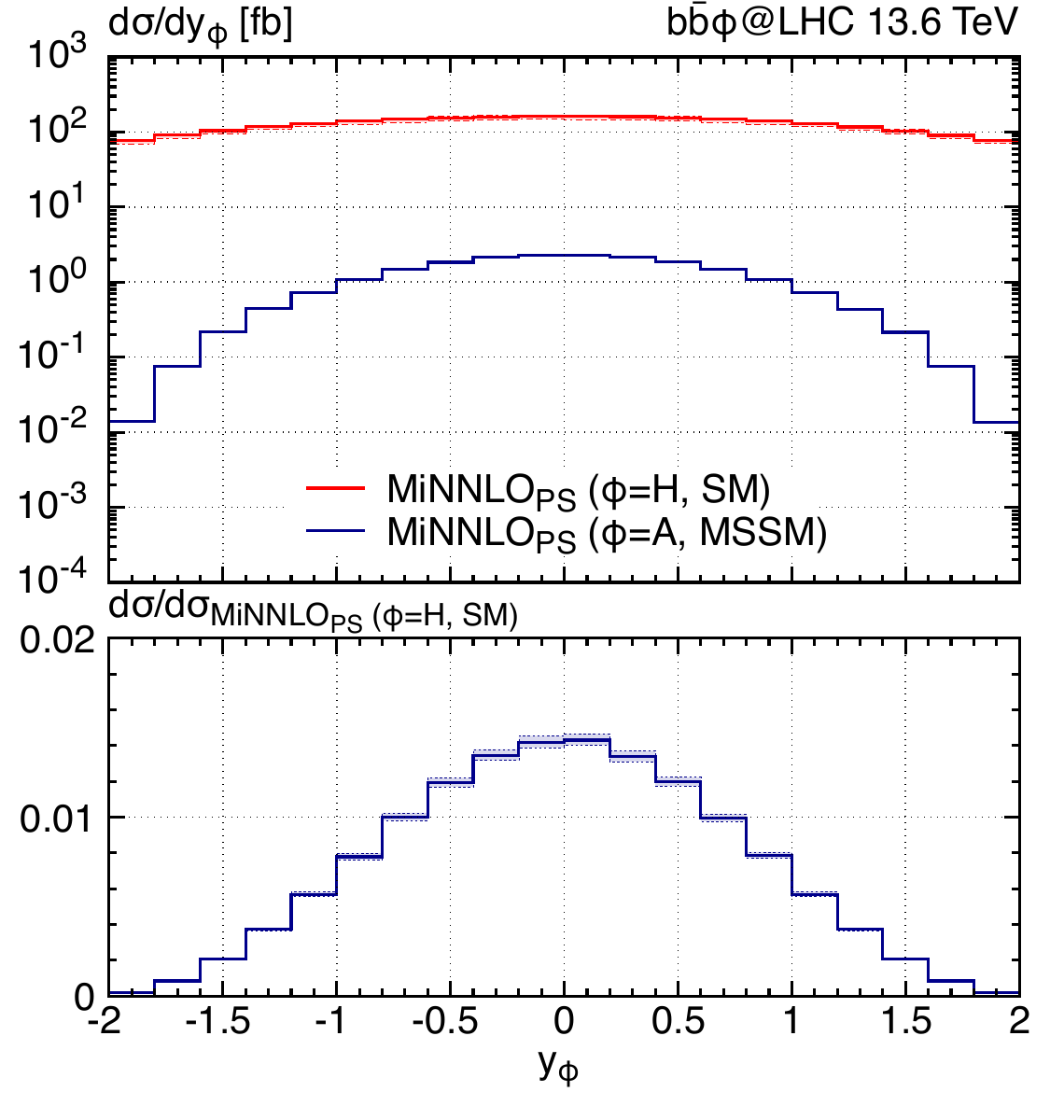}
\end{tabular}
\vspace*{1ex}
\caption{Higgs transverse momentum and rapidity spectra for \minnlo{} results with the heavy pseudo-scalar Higgs $\text{A}$ compared to SM predictions.\label{fig:SMvsBSM}}
\end{center}
\end{figure}

In~\fig{fig:MiNLOBSM}, we present a comparison between \minnlo{}  and \minlo{} predictions. 
The left plot shows the transverse momentum spectrum.
Due to the high mass of the bosonic state, the resummation region extends over a broader range of the transverse momentum spectrum compared to the SM Higgs, with a mass of 125\,GeV. As a result, the NNLO corrections present in \minnlo{} have a larger impact even at intermediate transverse momentum values up to $\sim 600$\,GeV, leading to more accurate predictions and smaller scale uncertainties for \minnlo{} compared to the \minlo{} in a broad transverse momentum range. 
In the right plot of~\fig{fig:MiNLOBSM}, we compare the two predictions for the Higgs rapidity spectrum: NNLO corrections increase the cross section with a flat correction in the central region $|y_A|<1$. Notably, the NNLO effects encoded in \minnlo{} are positive in the case of a 1.4 TeV Higgs boson, compared to the SM case where the authors of~\citere{Biello:2024vdh} have observed a flat negative correction in the Higgs rapidity spectrum.

Finally, we compare the differential behaviour of \minnlo{} predictions for the heavy Higgs state with the SM distributions in~\fig{fig:SMvsBSM}. In the left plot, we show the transverse momentum distribution. 
As expected, the BSM transverse momentum spectrum is substantially harder. With the chosen settings, BSM effects become even larger than the SM prediction for transverse momentum values greater than 400\,GeV. 
In the right plot of~\fig{fig:SMvsBSM}, we compare the rapidity distributions. The heavy pseudo-scalar Higgs receives a larger contribution from central rapidities compared to the SM one, with the total contribution remaining always below 2\% of the SM prediction. We stress that the numerical comparison is highly sensitive to the choice of the MSSM parameters. However, the purpose of this analysis is not to focus on a specific choice of the MSSM parameters, but to demonstrate the potential of the \minnlo{} generator 
for the modelling of a BSM signal of \bbH{} production at NNLO+PS.

\subsection{NNLO+PS predictions in 4FS}\label{sec:bbH4FS}

We now turn to NNLO+PS predictions computed in the massive 4FS for the $y_{b}^{2}$ component of \bbH{} production, i.e.\ the 
corresponding 4FS calculation to the 5FS one presented in the previous section.
Section~\ref{sec:5FSNNLOPS} demonstrated that, in 5FS, the \minnlo{} and \GENEVA{} procedures provide fully exclusive 
NNLO+PS results for the \bbH{} process with massless bottom quarks. While that approach naturally resums collinear logarithms 
through the $b$-PDFs, it cannot account for power corrections in $m_{b}$, which become particularly important at 
small Higgs transverse momenta. Moreover, the 4FS is better suited to model observables that depend on the $b$-jet kinematics.

\subsubsection[\minnlo{} method for $Q\bar Q F$ production]{\minnlo{} method for \boldmath{$Q\bar Q F$} production}\label{sec:bbH4FS}

We consider a fully exclusive NNLO+PS generator for the \bbH{} process in the 4FS \cite{Biello:2024pgo}.  This generator has been constructed using 
the extension of the \minnlo{} method to heavy-quark–associated colour-singlet ($Q\bar{Q}F$) production, developed in \citere{mazzitelli:2024ura}, 
which itself builds on the \minnlo{} approach for $Q\bar{Q}$ production in \citere{mazzitelli:2020jio,mazzitelli:2021mmm}. 
While large logarithmic contributions at small transverse momentum in the $Q\bar{Q}F$ final state have the same 
general structure as for $Q\bar{Q}$ production, the more general kinematics of the $Q\bar{Q}F$ system, compared to the back-to-back configuration
for $Q\bar Q$ production, has to be accounted for in the calculation of the coefficient functions for the resummation. The singular structure is governed
by the factorization theorem, which is expressed in Fourier-conjugate (impact-parameter or $b$) space \cite{Zhu:2012ts,Li:2013mia,Catani:2014qha,Catani:2018mei}:
\begin{align}\label{eq:facformula}
        \frac{\mathd \sigma}{ \mathd^2\vec{\pt}\, \mathd \Phi_{Q\bar{Q}{\rm F}}}&=\sum_{c=q,\bar{q},g}
  \frac{|M^{(0)}_{c\bar{c}}|^2}{2 m_{Q\bar{Q}{\rm F}}^2 }\int\frac{ \mathd^2\vec{b}}{(2\pi)^2} e^{i \vec{b}\cdot
  \vec{\pt} } e^{-S_{c\bar{c}} \left(\frac{b_0}{b}\right)}\sum_{i,j} \Tr({\mathbf H}_{c\bar{c}}{\mathbf \Delta})\,\,
  ({C}_{ci}\otimes f_i) \,({C}_{\bar{c} j}\otimes f_j) \,,
\end{align}
where $|M^{(0)}_{c\bar{c}}|^2$ is the squared LO matrix element,
$m_{Q\bar{Q}{\rm F}}$ the invariant mass of the $Q\bar{Q}F$  system and $b_0=2 e^{-\gamma_\text{E}}$. The factor $e^{-S_{c\bar{c}}}$ represents the same Sudakov radiator that appears in the small-$p_{T}$ resummation for 
colour-singlet production.  In \eqn{eq:facformula}, the sum over $c=q,\bar{q},g$ spans all possible flavour assignments for the incoming partons, with the first parton carrying flavour $c$ and the second parton carrying flavour $\bar{c}$. The collinear coefficient functions
$C_{ij} = C_{ij}\bigl(z,\,p_{1},\,p_{2},\,\vec{b};\,\alpha_{s}(b_{0}/b)\bigr)$ arise from collinear emissions, which include
also the constant terms, and the parton densities $f_{i}$ are evaluated at the soft scale $b_{0}/b$.  The composite factor
$\Tr\bigl({\mathbf H}_{c\bar{c}}\,{\mathbf \Delta}\bigr)\;\bigl(C_{ci}\otimes f_{i}\bigr)\;\bigl(C_{\bar{c}j}\otimes f_{j}\bigr)$
encodes the differences with respect to colour singlet production originating from the more involved colour structure 
and initial-final/final-final interferences for $Q\bar{Q}$ processes. It differs in its explicit form for the $q\bar{q}$ and $gg$ channels and written symbolically here.
 In particular, this factor contains a nontrivial Lorentz structure—omitted here for brevity—that generates azimuthal correlations in the collinear limit \cite{Catani:2010pd,Catani:2014qha}.

All quantities in bold face denote operators in colour space, and the trace $\Tr\bigl({\mathbf H}_{c\bar{c}}\,{\mathbf \Delta}\bigr)$
in \eqn{eq:facformula} runs over colour indices.  The hard function ${\mathbf H}_{c\bar{c}}$ is extracted from the infrared-subtracted amplitudes for $Q\bar{Q}F$ production, 
where the ambiguity in its definition corresponds to a choice of resummation scheme~\cite{Bozzi:2005wk}.  The operator ${\mathbf \Delta}$ encodes quantum interferences arising from soft radiation exchanged at large angles between the initial and final states, as well as among final-state quarks. It is given by ${\mathbf \Delta}={\mathbf V}^\dagger{\mathbf D}{\mathbf V}$, where
\begin{align}
\label{eq:soft}
{\mathbf V} &= {\cal
  P}\exp\left\{-\int_{b_0^2/b^2}^{m_{Q\bar{Q}F}^2}\frac{dq^2}{q^2}{\mathbf
  \Gamma}_t(\Phi_{\rm Q\bar{Q}F};\alpha_s(q))\right\}\,.
\end{align}
The symbol ${\cal P}$ denotes path ordering of the exponential matrix with respect to the integration variable $q^{2}$, arranging scales from left to right in increasing order.  The anomalous dimension $\mathbf{\Gamma}_{t}$ governs the effect of real soft radiation emitted at large angles.  Meanwhile, the azimuthal operator $\mathbf{D} \;\equiv\; \mathbf{D}\bigl(\Phi_{Q\bar{Q}F},\,\vec{b},\,\alpha_{s}\bigr)$
encodes azimuthal correlations of the $Q\bar{Q}F$ system in the small-$p_{T}$ limit.  Upon averaging over the azimuthal angle $\phi$, it satisfies
$\bigl[\mathbf{D}\bigr]_{\phi} \;=\; \mathbf{1}$.
Starting from the $Q\bar Q F$ resummation formula, the procedure to construct a \minnlo{} improved $\bar{B}$ function system is the same as that of the colour-singlet case.
All technical details of this derivation are provided in~\citeres{mazzitelli:2020jio,mazzitelli:2021mmm,mazzitelli:2024ura,Biello:2024pgo}.

In the case of \bbH{} production in the 4FS, the two-loop virtual corrections, required for NNLO accuracy,
are not known in exact form with finite $m_b$. Instead, the two-loop amplitude is approximated in the NNLO+PS
prediction by a small-mass expansion $m_b$, i.e.\ all logarithmically enhanced terms and constant terms are retained,
while dropping terms that are power-suppressed in $m_b$. 
This massification procedure, developed in \citeres{Mitov:2006xs,Wang:2023qbf}, allows us to capture the dominant 
virtual corrections in the small-$m_b$ limit. The massless full-colour two-loop amplitudes, which provide the constant
terms, are taken from~\citere{Badger:2024mir}. Different from the studies in \citere{Biello:2024pgo}, which relied on 
the leading-colour approximation~\cite{badger:2021ega}, here we perform a phenomenological analysis using events 
generated with the full-colour library of~\citere{Badger:2024mir} for the two-loop constant terms in the massification. 
The NNLO calculation is thus complete in full colour, up to power-suppressed terms in the bottom-quark mass, which are neglected
only in the two-loop amplitude.

\subsubsection{Results and flavour-scheme comparison}
The setup for the phenomenological results presented here---including input parameters and PDFs---has been outlined in \sct{sec:setup}. In addition, we specify the renormalization and factorization scale choices adopted in the \minnlo{} simulations for the Born couplings. The Yukawa coupling is evaluated in the $\overline{\text{MS}}$ scheme at a scale of the Higgs mass
\begin{align}
  \mu_R^{(0),y_b}= m_H,
\end{align}
while the Born strong couplings are computed at a dynamical scale scale:
\begin{align}
  \mu_R^{(0),\alpha_s}= H_T/4, \quad \text{with} \quad H_T = M_{T,b}+M_{T,\bar b}+ M_{T,H}\,,
\end{align}
where $M_{T,X}=\sqrt{M_X^2+p_{T,X}^2}$ is the transverse mass of particle $X$.

We discuss the new theoretical predictions obtained at NNLO+PS in the 4FS for Higgs radiation off bottom quarks, and we compare them to the NNLO+PS results
in the 5FS. We start by considering predictions of integrated cross sections in \tab{tab:NNLO4FS_xs}. Apart from the fully inclusive cross section, we consider rates with identified $b$-jets, including a $b$-jet veto, the inclusive one- and two-$b$-jet rates. We define $b$-jets following the 
standard experimental definition, namely as anti-$k_T$ jets that 
contain at least one bottom quark/$B$ hadron. 
We apply a jet radius of $R=0.4$, a minimum transverse momentum of $30$\,GeV and a 
maximum absolute pseudo-rapidity of $2.4$. Note that we have also considered a flavour-aware definition of $b$-jets using the 
{\it Interleaved Flavour Neutralization} (IFN) algorithm (with $\alpha = 2$ and $\omega = 1$)~\cite{Caola:2023wpj}, 
which enables an infrared- and collinear-safe definition of anti-$k_T$ $b$-jets also for fixed-order predictions in the massless scheme. 
Interestingly, we found that all numerical IFN results for the $b\bar b H$ process presented here differ by less than 
a few percent from the standard experimental definition, which is why we refrain from showing the IFN results and focus 
on the standard $b$-jet definition that is straightforward to apply in experimental analyses.

\begin{table}[b]
  \vspace*{0.3ex}
  \begin{center}
	   \renewcommand{\arraystretch}{1.6}
    \begin{tabular}{|c||c|c|c|c|}
    \hline
    \makecell[c]{\shortstack{\rule{0pt}{2ex}Fiducial region}} &  
    \makecell[c]{\shortstack{\rule{0pt}{2ex}NLO$_{\rm PS}$ \\ 5FS} } & 
    \makecell[c]{\shortstack{\rule{0pt}{2ex}NLO$_{\rm PS}$ \\ 4FS} }  & 
    \makecell[c]{\shortstack{\rule{0pt}{2ex}\minnlo{} \\ 5FS} } &  
    \makecell[c]{\shortstack{\rule{0pt}{2ex}\minnlo{} \\ 4FS} } \\
    \hline \hline
	    inclusive & $725.4(2)_{-10\%}^{+11\%}$  fb& $ 389.0(1)_{-20\%}^{+24\%}$ fb& $ 574.8(4)_{-8.0\%}^{+4.5\%}$  fb& $ 519.6(3)_{-15\%}^{+19\%}$ fb\\
     \hline
        H $+$ $0\,b\text{-jets}$\;  & $619.5(9)_{-10\%}^{+11\%}$  fb& $ 312.1(4)_{-20\%}^{+23\%}$  fb& $ 458.1(3)_{-7.8\%}^{+3.3\%}$ fb&$ 420.2(2)_{-15\%}^{+21\%}$  fb\\
        \hline
	    \;H $+\geq$$1\,b\text{-jets}$ & $105.9(1)_{-10\%}^{+10\%}$  fb& $ 76.91(6)_{-20\%}^{+26\%}$  fb& $ 116.7(1)_{-8.7\%}^{+9.4\%}$ fb& $ 99.38(2)_{-12\%}^{+7.5\%}$ fb\\
      \hline
	    \;H $+\geq$$2\,b\text{-jets}$ & $5.992(3)_{-10\%}^{+10\%} $  fb& $ 5.116(2)_{-23\%}^{+32\%}$  fb& $ 8.621(3)_{-10\%}^{+11\%}$ fb& $ 7.154(6)_{-10\%}^{+1.7\%}$  fb\\
       \hline
    \end{tabular}
  \end{center}
  \vspace{-1em}
  \caption{
	Predictions for cross section rates in fb of different {\sc Powheg} and \minnlo{} generators in the 4FS and 5FS for the $y_b^2$ contribution. 
	\label{tab:NNLO4FS_xs}}
\end{table}

Looking at the total inclusive cross section in \tab{tab:NNLO4FS_xs}, NNLO corrections in the 4FS increase the rate by about 30\% relative to NLO. 
This highlights the importance of NNLO accuracy for precise predictions in the massive scheme. 
The \minnlo{} result in the 4FS shows a reduced dependence on the scale choices as compared to the NLO+PS prediction in the 4FS. 
Also the 5FS predictions, which correspond to the $b\bar{b} \to H+X$ process, receive sizable, but negative NNLO corrections
of about $-20$\%. Scale uncertainties are smaller in the 5FS, but they are significantly underestimated considering the fact
that NLO and NNLO predictions do not agree within their respective uncertainties.
At NLO, there is a severe discrepancy between the 4FS and 5FS cross sections, with 
the 5FS one being about 80\% larger.
By contrast, the NNLO+PS predictions in the two schemes 
agree well with each other within their respective scale uncertainties,
due to the negative NNLO correction in the 5FS and the sizable positive 
correction in the 4FS. As a result, the long-standing tension between the 4FS and 5FS 
inclusive cross section is resolved once NNLO accuracy is included.

Considering the fiducial cross sections in the presence of $b$-jets in \tab{tab:NNLO4FS_xs}, we observe that about 80\% of the events 
are produced in the $0$-$b$-jet bin. This result is consistent between the 4FS and 5FS NNLO+PS predictions. By contrast, at NLO+PS 
the 5FS predictions a fraction of 85\% of events with no $b$-jets. Requiring at least one $b$-jet leads to a significant reduction in the cross section by a factor 
of approximately 5, i.e.\ about $20$\% of the events have one or more $b$-jets. Requiring a second $b$-jet further suppresses the rate by roughly 
an additional order of magnitude, with about 1.4\% of the \bbH{} events having two or more $b$-jets.
These relative $b$-jet acceptances are in remarkable agreement between the 4FS and 5FS NNLO+PS generators (in fact, also
compared to the 4FS NLO+PS generator), while
the 5FS NLO+PS predictions being quite different. It is reassuring to observe this high level of agreement, despite the fact that 
in the 5FS the accuracy effectively reduces by one order for each required $b$-jet, whereas the 4FS predictions remain genuinely 
NNLO accurate in all considered $b$-jet categories.
Note that for the \minnlo{} 4FS predictions the inclusive cross section 
and $0$-$b$-jet bin feature larger uncertainties that the more exclusive $2$-$b$-jet selection,
which could be related to a lower sensitivity to logarithmic corrections in the bottom-quark mass
when two hard $b$-jets are required.
Also for the absolute cross section numbers at NNLO+PS in the 4FS and 5FS, the inclusion of higher-order corrections 
via \minnlo{} improves the overall consistency between the two schemes considerably, especially in the $b$-jet-vetoed cross sections,
which are a factor of two apart in the NLO-accurate predictions. Similarly, the one(two)-$b$-jet cross section at NNLO+PS, which is only NLO-accurate (LO-accurate) in the 5FS, 
is in good agreement with the 4FS one, as soon as the \minnlo{} corrections are included.

\begin{figure}[t!]
\begin{center}
\begin{tabular}{cc}
\includegraphics[width=.45\textwidth, page=1]{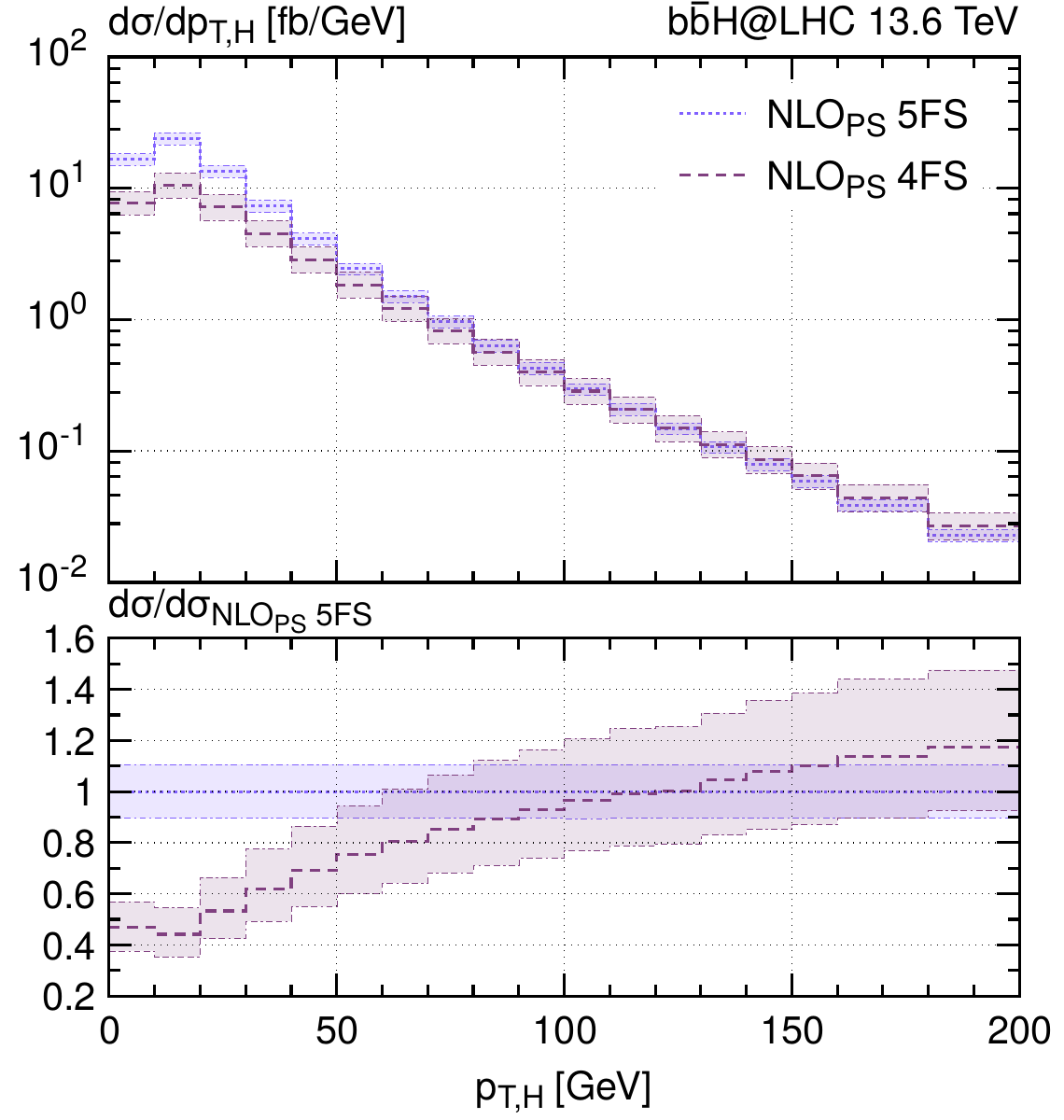}&
\includegraphics[width=.45\textwidth, page=1]{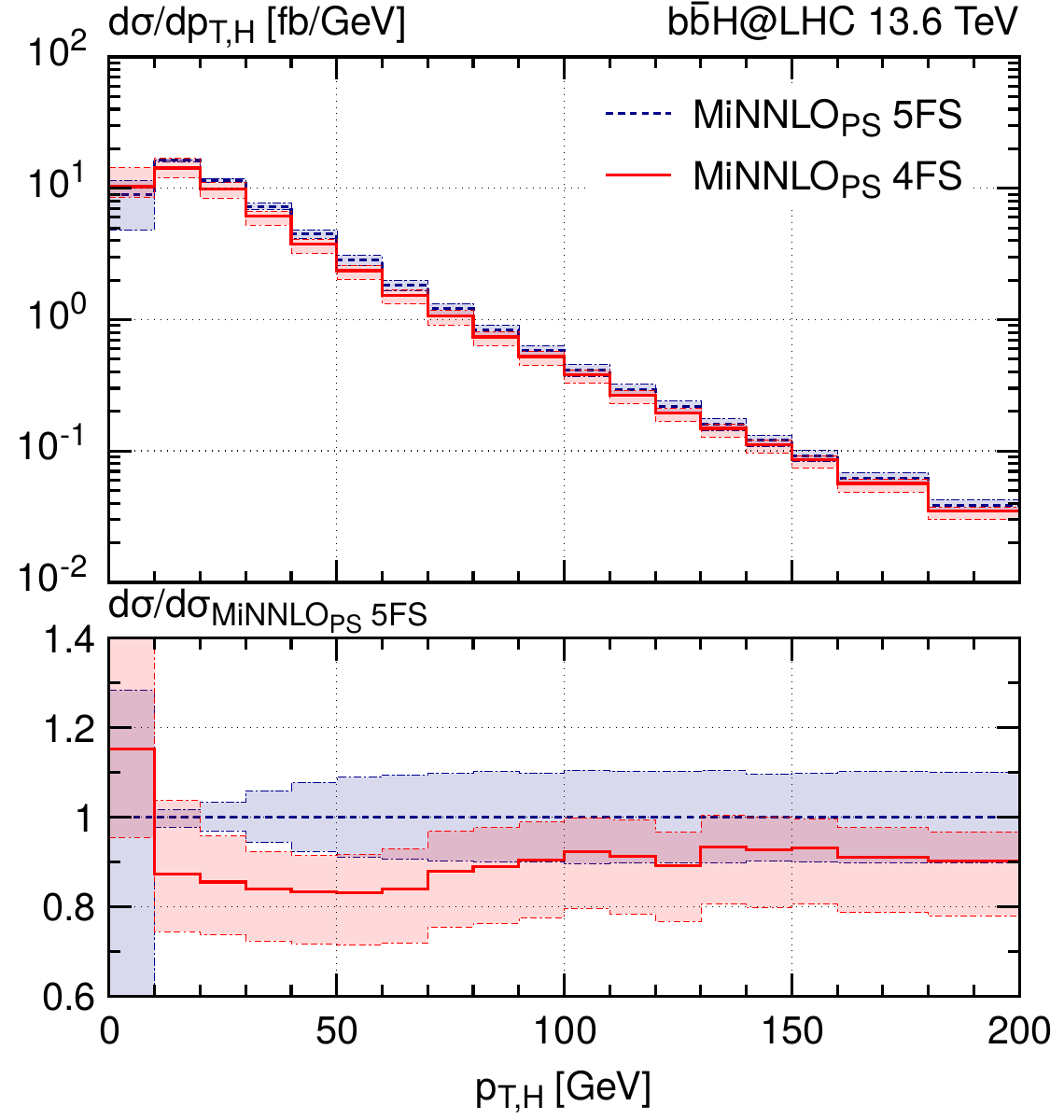}
\end{tabular}
\vspace*{1ex}
\caption{Comparison between different flavour scheme choices for the Higgs transverse momentum spectrum at NLO+PS (left) and NNLO+PS level (right) for the $y_b^2$ contribution. \label{fig:4fsA}}
\end{center}
\end{figure}

\begin{figure}[t!]
\begin{center}
\begin{tabular}{cc}
\includegraphics[width=.45\textwidth, page=1]{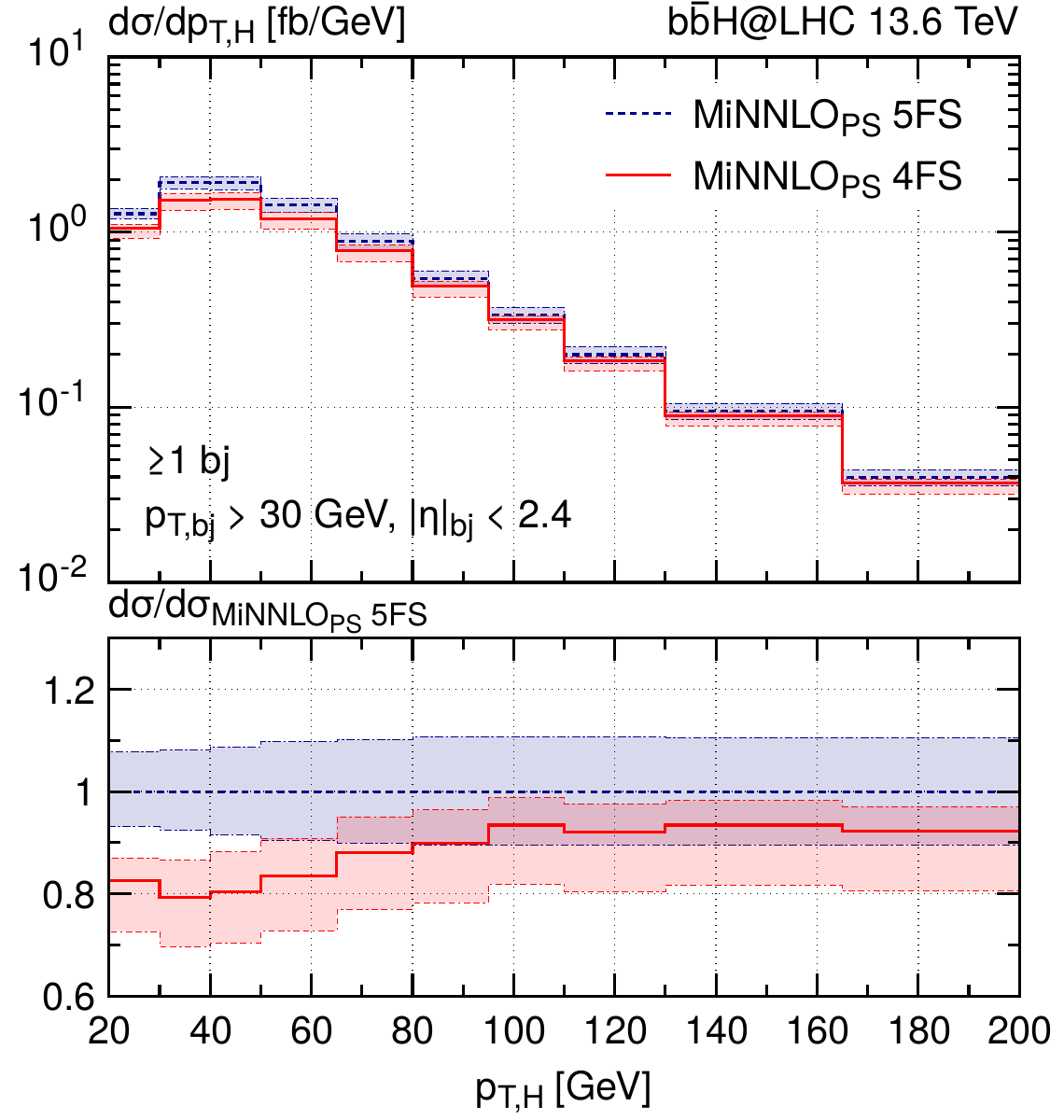}&
\includegraphics[width=.45\textwidth, page=1]{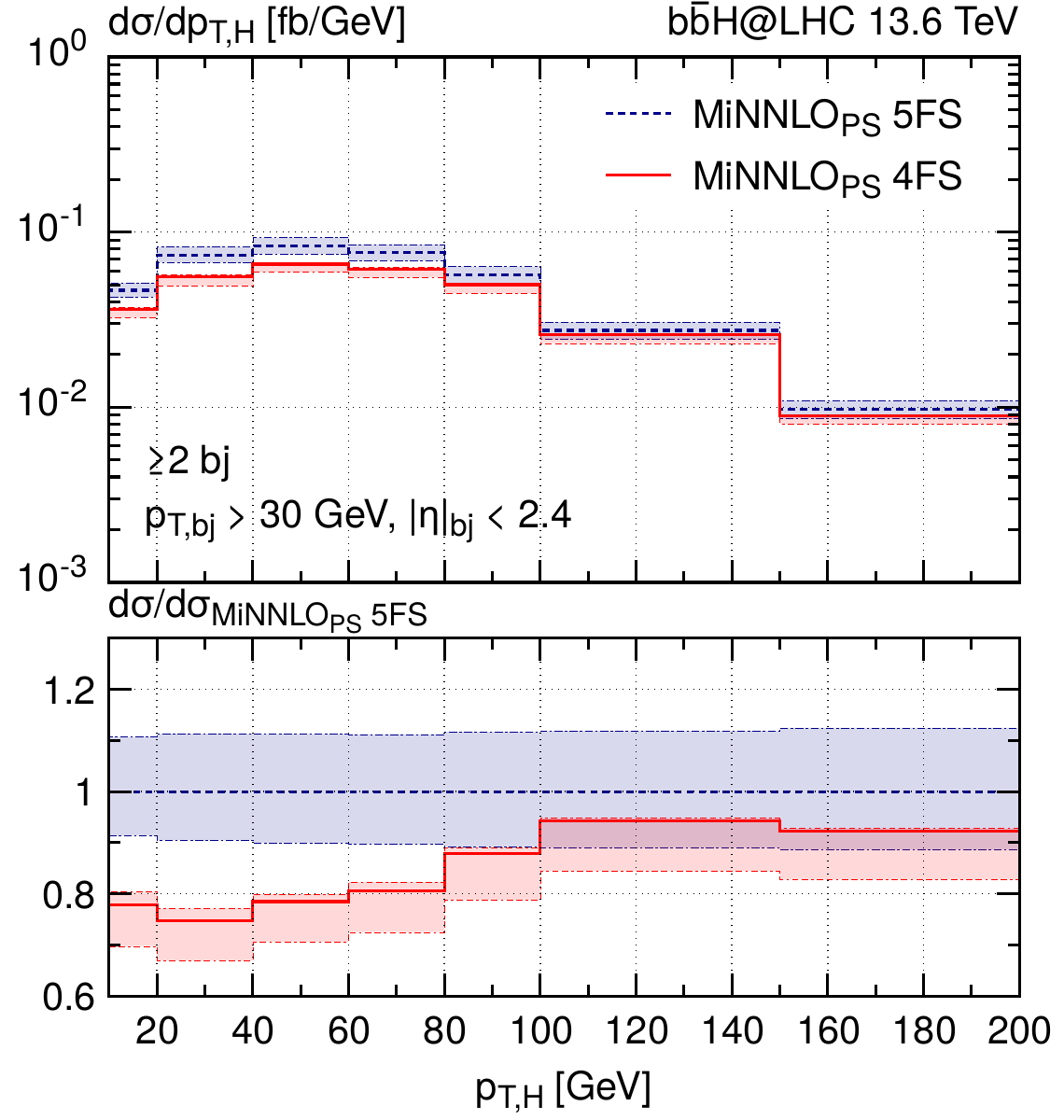}
\end{tabular}
\vspace*{1ex}
\caption{Higgs transverse momentum spectrum with at least one (left plot) or two (right plot) $b$-jets identified using the experimental-like algorithm applied to the events generated by \minnlo{} in the massless (blue, dashed) and massive (red, solid) schemes for the $y_b^2$ contribution.\label{fig:4fsB}}
\end{center}
\end{figure}

Next, we compare differential distributions in 4FS and 5FS at NLO+PS and NNLO+PS in \fig{fig:4fsA} and \ref{fig:4fsB}. 
We show the Higgs transverse momentum distribution in \fig{fig:4fsA}. The left plot shows the NLO+PS predictions and the right plot the NNLO+PS ones. 
At NLO+PS, the 4FS and 5FS predictions differ significantly at low $p_{T,H}$, featuring an entirely different shape. 
At NNLO+PS, this discrepancy is significantly reduced: \minnlo{} matching brings the two predictions into much better 
agreement across the full $p_{T,H}$ range. Disregarding the 10\% difference in the normalization that was already observed for 
the total inclusive cross section, the shape of the two distributions is very similar, with the exception of the first bin,
where the 5FS provides an invalid description, as discussed before. To be more precise, for $p_{T,H} \gtrsim 70$\,GeV the ratio of the 
two predictions is very flat, while for $p_{T,H} \lesssim 70$\,GeV we observe shape variations of about 5\%.
In all cases, 4FS and 5FS predictions are consistent within uncertainties.

Figure~\ref{fig:4fsB} shows the Higgs transverse momentum distribution in events with at least one (left) or two (right) $b$-jets identified. For the inclusive one-$b$-jet selection, the differences between the schemes are moderate and largely flat, except in the small $p_{T,H}$ region, where they reach up to 20\%, slightly beyond the respective uncertainties. 
In the two-$b$-jet category, the differences at small $p_{T,H}$ increase, with the 4FS prediction being systematically lower than the 5FS one. This behaviour reflects the more accurate treatment of bottom-quark kinematics in the massive scheme, which becomes particularly relevant when two $b$-jets are required to be hard and well separated. These configurations
are effectively described only at LO+PS accuracy in the 5FS.

These results underline the importance of the 4FS computation for observables involving identified $b$-jets. Due to the partonic structure of the 5FS calculation, it lacks accuracy in the hard matrix-element description of $b$-jet multiplicities compared to the 4FS, which in all observables involving up to two $b$-jets retains the full NNLO+PS accuracy. 
As a result, the 4FS provides a more reliable and accurate prediction, especially in regions sensitive to the kinematics and multiplicity of $b$-jets. This makes the 4FS prediction highly relevant for comparisons with experimental measurements involving $b$-tagged final states.


\section{Modelling of the \boldmath{$b\bar bH$} background in \boldmath{$HH$} searches}\label{sec:HH}

\subsection[Comparison of different \bbH{} generators for $y_t^2$ and $y_b^2$ contributions]{Comparison of different \boldmath{\bbH{}} generators for \boldmath{$y_t^2$} and \boldmath{$y_b^2$} contributions}

In this section, we study \bbH{} production as a background to di-Higgs searches at the LHC. To this end, we 
consider MC predictions for both the $y_t^2$ and  $y_b^2$  contributions to the \bbH{} cross section
with fiducial selections relevant to enhance a $HH$ signal at the LHC~\cite{HDBS-2021-18,CMS-HIG-22-001}.\footnote{The $y_t y_b$ interference contribution is 
about $-10\%$ at NLO+PS, which is less than the respective scale uncertainties of the $y_b^2$ and  $y_t^2$ cross section. It therefore does not play a relevant
role for the considerations in this section and can be safely neglected.}
Our study closely follows the work presented in \citere{manzoni:2023qaf}, with the main differences being the updated 
collider energy and input parameters, as described in \sct{sec:setup}, as well as novel predictions for the $y_b^2$ component
computed at NNLO+PS in the 4FS using the \minnlo{} generator \cite{Biello:2024pgo}. The $y_t^2$ predictions 
are provided in the 4FS at NLO QCD accuracy matched to the \PYTHIA{8}~\cite{Bierlich:2022pfr} parton shower. 
The results are obtained automatically with {\sc MadGraph5\_aMC@NLO}~\cite{Alwall:2014hca,Frederix:2018nkq} using the NLO 
calculation of \citere{deutschmann:2018avk} (in particular of the virtual amplitude in the HTL).

We compare the $y_t^2$ predictions to those from an inclusive 
gluon-fusion {\sc NNLOPS} simulation~\cite{hamilton:2012rf,hamilton:2013fea,hamilton:2015nsa} in the 5FS\footnote{
The so-called {\sc NNLOPS} simulation~\cite{hamilton:2012rf,hamilton:2013fea,hamilton:2015nsa} is based on
an a posteriori reweighting of \minlo{} events to reach NNLO-accuracy for Born-level observables, i.e. those
related to the colour-singlet final state. By contrast, the previously discussed \minnlo{} method
incorporates the additional terms required to achieve NNLO accuracy directly in the event generation.}, 
generated using {\sc Powheg\_Box\_v2}~\cite{nason:2004rx,frixione:2007vw,alioli:2010xd}. 
This setup achieves NNLO accuracy for inclusive Higgs boson observables, but only provides LO accuracy for final states 
in which the Higgs boson is produced in association with two additional $b$-jets. In the {\sc NNLOPS} calculation, the 
renormalization and factorization scales are set to $\mu_R = \mu_F = m_H/2$, the PDF4LHC21 parton distribution functions 
are used~\cite{PDF4LHCWorkingGroup:2022cjn}, and \PYTHIA{8} is employed both for parton showering and for simulating 
the decay of the Higgs boson into two photons, which are kept stable in the showering process.

The $HH$ signal region is inspired by the $HH \to b\bar{b}\gamma\gamma$ final state~\cite{HDBS-2021-10,CMS-HIG-19-018}. Specifically, we require two $b$-tagged anti-$k_T$~\cite{Cacciari:2005hq,Cacciari:2008gp,Cacciari:2011ma} jets with
a radius of $R=0.4$ satisfying
\begin{equation}
\label{eq:ptjcuts}
p_T(j) > 25\,\GeV \qquad {\rm and} \qquad |\eta(j)| < 2.5\,,
\end{equation}
with their invariant mass constrained to be compatible with the Higgs boson mass:
\begin{equation}
    \label{eq:mbjcut}
    80\,\GeV < m(b_1,b_2)< 140\,\GeV\,.
\end{equation}
Additionally, we require two photons—produced in the simulation via Higgs decays through \PYTHIA{8}—to pass the following selection criteria:\footnote{Note that, since
we employ the Higgs decay in the zero-width approximation, the Higgs boson is on-shell and the invariant mass of the two photons is exactly equal to Higgs mass
$m(\gamma_1, \gamma_2)=m_H$, which trivially fulfils the first condition and simplifies the other ones.}
\begin{equation}
\label{eq:photoncuts}
    105\,\GeV < m(\gamma_1, \gamma_2) < 160\,\GeV, \quad |\eta(\gamma_i)|< 2.37, \quad
    \frac{p_T(\gamma_1)}{m(\gamma_1, \gamma_2)} > 0.35, \quad \frac{p_T(\gamma_2)}{m(\gamma_1, \gamma_2)} > 0.25\,.
\end{equation}
The presence of two $b$-jets and two photons satisfying the above requirements defines the \emph{fiducial} region. This selection can be further refined in the context of $HH$ analyses by imposing a cut on the variable
\begin{equation}
\label{eq:mbbggs}
    \mbbggs = m_{2b2\gamma} - m(b_1,b_2) - m(\gamma_1,\gamma_2) + 2m_H \,.
\end{equation}
This observable is used in $HH$ searches~\cite{HDBS-2021-10,HDBS-2019-27,CMS-HIG-19-018,ATLAS:2025hhd} to define selection cuts that enhance the sensitivity
to an anomalous trilinear Higgs coupling. 
Here, we consider two scenarios 
in which the fiducial selection includes the requirements $\mbbggs < 500$\,GeV and $\mbbggs < 350$\,GeV, respectively. 

\begin{table}[b]
  \vspace*{0.3ex}
  \begin{center}
	   \renewcommand{\arraystretch}{1.2}
\begin{tabular}{|c|c|c|c|c|}
  \hline 
	Fiducial region  &   \makecell[c]{\shortstack{\rule{0pt}{2ex} $y_t^2$ NLO   \\ (NLO+PS)} } & \makecell[c]{\shortstack{\rule{0pt}{2ex} $y_t^2$ LO  \\ (ggF NNLOPS)} }  & \makecell[c]{\shortstack{\rule{0pt}{2ex} $y_t^2$ LO  $\cancel{g\to b\bar{b}}$  \\ (ggF NNLOPS)} } & \makecell[c]{\shortstack{\rule{0pt}{2ex} $y_b^2$ NNLO   \\ (\bbH{} \minnlo{})} }  \\
  \hline \hline
   \begin{tabular}{@{}c@{}} No cut \end{tabular}
	   & $1696^{+62\%} _{-35\%}$   ab& $7574_{-8\%}^{+8\%}$  ab& $2774_{-7\%}^{+7\%}$  ab & $1180_{-15\%}^{+19\%}$  ab\\
  \hline
   \begin{tabular}{@{}c@{}} Fid. cuts \end{tabular}
	   & $18 ^{+55\%} _{-33\%}$  ab& $31.6_{-18\%}^{+18\%}$  ab& $20.5_{-18\%}^{+18\%}$  ab& $5.6_{-10\%}^{+2.8\%}$  ab\\
   \hline
  \begin{tabular}{@{}c@{}} Fid. cuts +\\ $\mbbggs<500\,\GeV$  \end{tabular}
	& $ 12 ^{+57\%} _{-33\%}$  ab  & $23.3_{-20\%}^{+20\%}$  ab& $15.4_{-20\%}^{+20\%}$  ab & $5.5_{-10\%}^{+2.8\%}$ ab \\
   \hline
    \begin{tabular}{@{}c@{}} Fid. cuts +\\ $\mbbggs<350\,\GeV$  \end{tabular}
   & $5.5 ^{+60\%} _{-34\%}$  ab & $11.5_{-20\%}^{+20\%}$  ab& $7.76_{-20\%}^{+20\%}$  ab & $4.8_{-10\%}^{+3.1\%}$  ab\\
   \hline
\end{tabular}
  \end{center}
  \vspace{-1em}
  \caption{
Cross sections in ab for the $y_t^2$ and $y_b^2$ contributions to $pp \to b \bar b H$ with $H\to \gamma\gamma$ decay at $\sqrt{s}=13.6$ TeV.\label{tab:XS-fid}}
\end{table}

We present inclusive and fiducial cross section yields in \tab{tab:XS-fid}.
We start by comparing the numerical results for \bbH{} $y_t^2$ component, computed at NLO in the 4FS using {\sc MadGraph5\_aMC@NLO} (first column), with the 
expected rate obtained from the {\sc NNLOPS} simulation, with (second column) and without $g\rightarrow b\bar{b}$ splitting included in the parton shower (third column). 
All rates include the Higgs branching ratio into photons, $\textrm{BR}(H \to \gamma\gamma) = 0.227\%$~\cite{LHCHiggsCrossSectionWorkingGroup:2016ypw}. 
Uncertainties due to renormalization and factorization scale variations are quoted, where each scale is independently varied by a factor of two around the 
central value following the 9-point prescription. The theoretical uncertainty in this section is estimated from the envelope of the nine scale variations, following~\citere{manzoni:2023qaf}, in contrast to the other sections where the standard 7-point prescription with the constraint $1/2 \leq \muR/\muF \leq 2$ is employed. Considering 7-point variations has a very small effect on the scale-uncertainty band for the total rates, and we have decided to align the choices with the previous HH-background studies~\cite{manzoni:2023qaf}. Additional theoretical uncertainties, such as those associated with variations of the shower starting scale, 
the choice of parton shower algorithm, or modifications to the MC@NLO matching procedure~\cite{frixione:2002ik,frederix:2020trv} were examined 
in \citere{manzoni:2023qaf} and found to be subleading. 
{\sc NNLOPS} uncertainties are evaluated via multiple sources, accounting for the modelling of jet multiplicity, the Higgs boson $\ptH$, $\ptHjj$, and $\mjj$~\cite{LHCHiggsCrossSectionWorkingGroup:2016ypw,Liu:2013hba,stewart:2013faa,Boughezal:2013oha,Stewart:2011cf,Gangal:2013nxa}, as discussed in \citere{ATLAS:2022tnm}. Parton shower uncertanties are also not shown for the {\sc NNLOPS} prediction, but they are expected to be sizeable: a comparison with an alternative parton shower ({\sc Herwig7}~\cite{Bellm:2015jjp}) yields differences in the fiducial $b\bar{b}H$ ($y_t^2$) cross section ranging from 3\% up to 17\% (depending on the kinematic region).

Significant differences are observed between the {\sc MadGraph5\_aMC@NLO} predictions and 
the {\sc NNLOPS} ones, with the {\sc NNLOPS} fiducial cross section up to twice as large (depending
on the kinematical selections) when the $g\rightarrow b\bar{b}$ splittings are included in the parton shower. 
As discussed in \citere{manzoni:2023qaf}, the parton shower induces a double counting of 
$gg\to H$ + 2~$b$-jets contributions, by dressing the inclusive production of the Higgs boson with hard $b$-jet
radiation, which are also included at the hard matrix-element level in the {\sc NNLOPS} calculation. 
Those hard configurations are formally outside the validity range of a parton shower, which is based on soft/collinear
factorization. This mismodelling created by the {\sc NNLOPS} sample is currently covered by a 
conservative 100\% uncertainty in $HH$ analyses~\cite{HDBS-2021-10,ATLAS:2025hhd}. Given
that \bbH{} production is the major background, reducing this uncertainty through more
accurate simulations for the $y_t^2$ contribution to the \bbH{} background will be 
instrumental for future $HH$ measurements.

We also quote the rates for the \bbH{} $y_b^2$ component in \tab{tab:XS-fid} (last column), which are evaluated at NNLO+PS 
accuracy with the \minnlo{} generator introduced in \sct{sec:bbH4FS}. 
In all fiducial $HH$ categories, the $y_b^2$ and $y_t^2$ contributions to the \bbH{} cross section are of the same order 
as the $HH$ signal cross section~\cite{manzoni:2023qaf} (not included in the table), highlighting the importance of their 
accurate modelling to obtain a precise measurement of the $HH$ process. 
The $y_t^2$ term is larger than the $y_b^2$ contribution to the \bbH{} cross section for the inclusive cross section (by about 50\%) 
and in the fiducial region without $\mbbggs$ cut (by a factor of about three). With kinematical selections on $\mbbggs$, however, the relative contribution of the 
$y_b^2$ component increases. For the fiducial region with $\mbbggs<500$\,GeV, the $y_t^2$ term is larger by roughly a factor of 
two, while for $\mbbggs<350$\,GeV $y_b^2$ and $y_t^2$ contributions are already comparably large. 
The latter is the fiducial region most sensitive to the trilinear Higgs coupling. Hence, the accurate modelling 
of both, $y_b^2$ and $y_t^2$ components, is important to fully control the \bbH{} background.

\begin{figure}[t!]
\begin{center}
\begin{tabular}{cc}
\includegraphics[width=.45\textwidth, page=1]{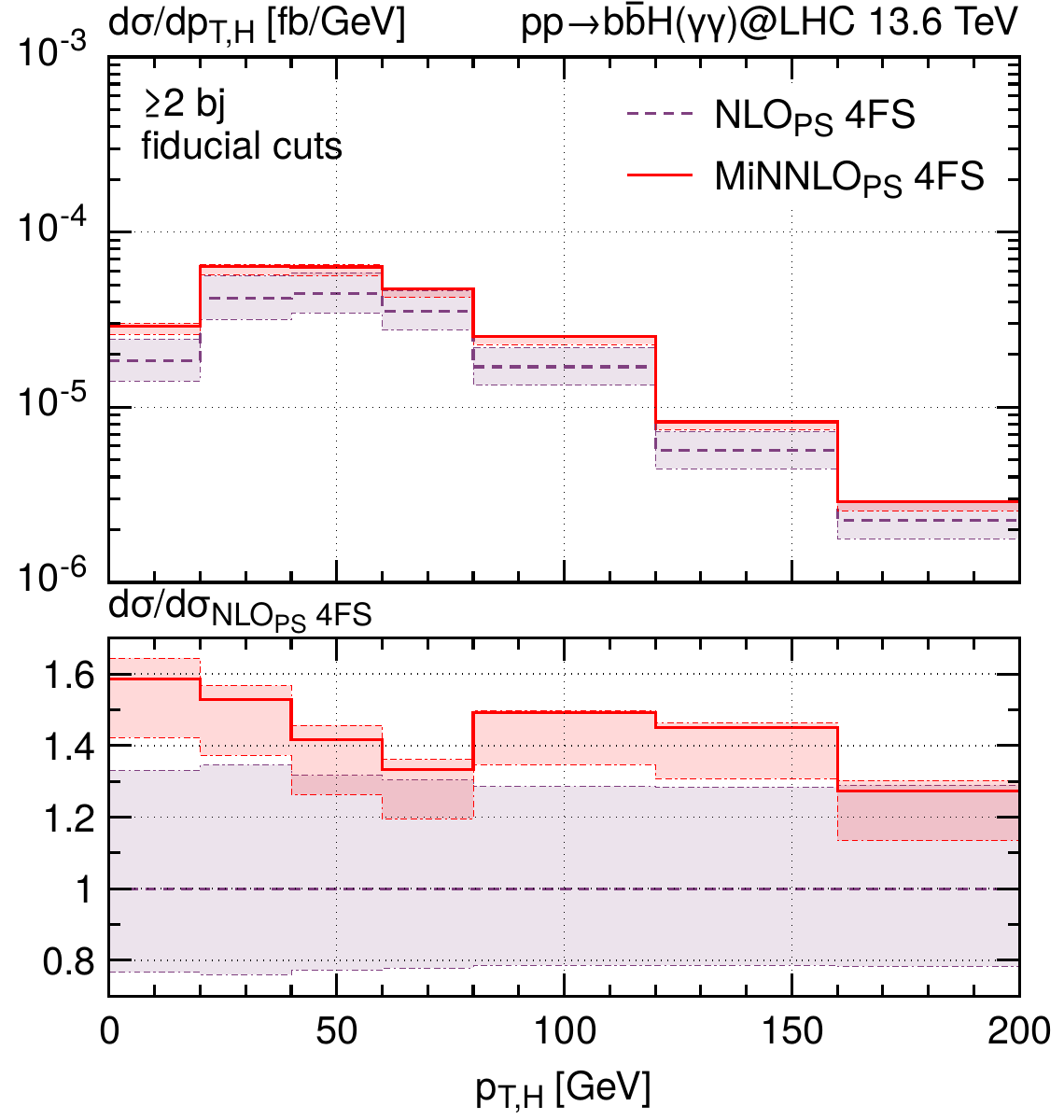}&
\includegraphics[width=.45\textwidth, page=1]{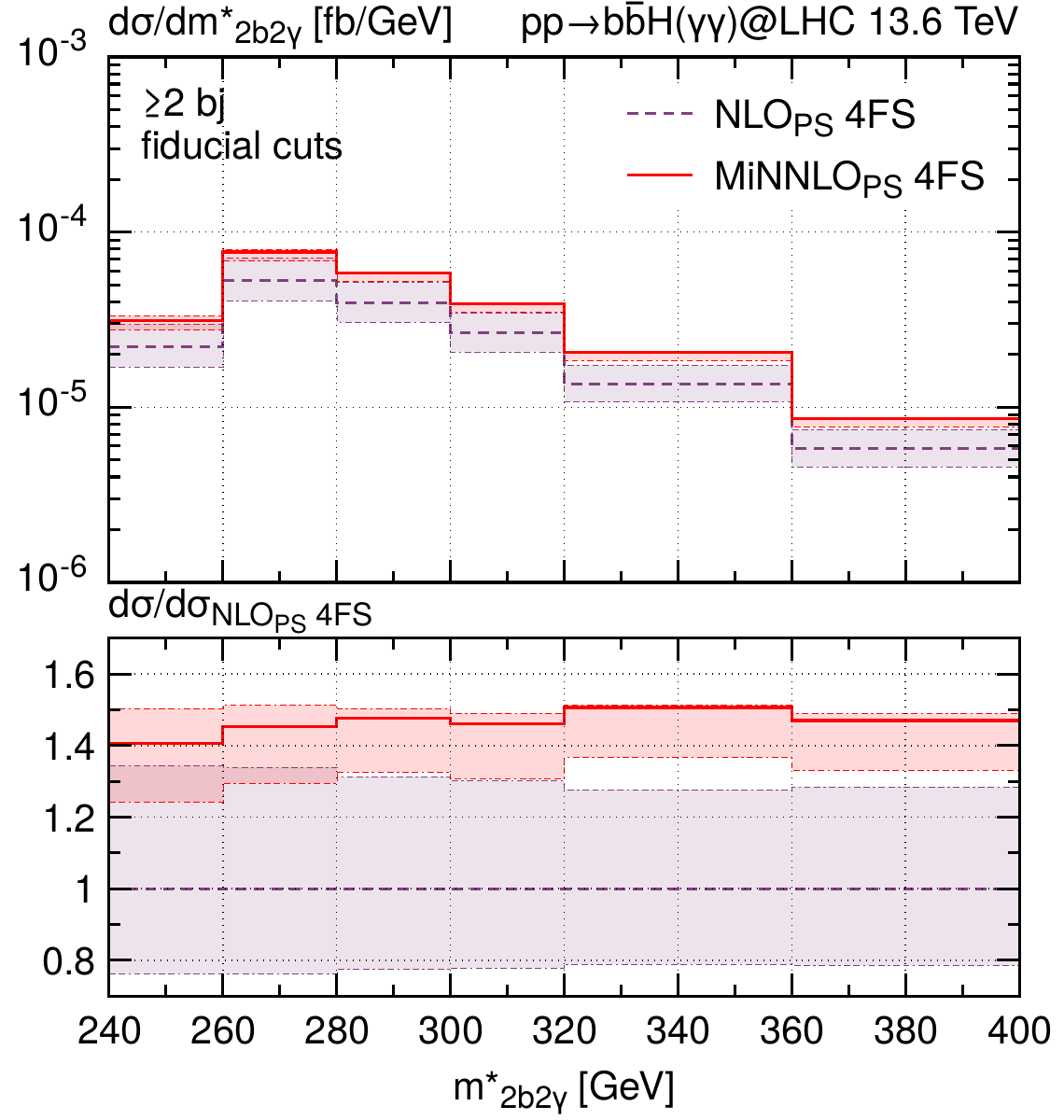}
\end{tabular}
\vspace*{1ex}
\caption{Transverse momentum spectrum of the Higgs boson and invariant mass $\mbbggs$ as defined in~\eqn{eq:mbbggs} in the fiducial volume described by the selections~(\ref{eq:ptjcuts}-\ref{eq:photoncuts}), comparing NLO+PS (purple, dashed) and NNLO+PS (red, solid) 4FS predictions for the $y_b^2$ contribution.\label{fig:4fsFID}}
\end{center}
\end{figure}

\begin{figure}[t!]
\begin{center}
\begin{tabular}{cc}
\includegraphics[width=.45\textwidth, page=1]{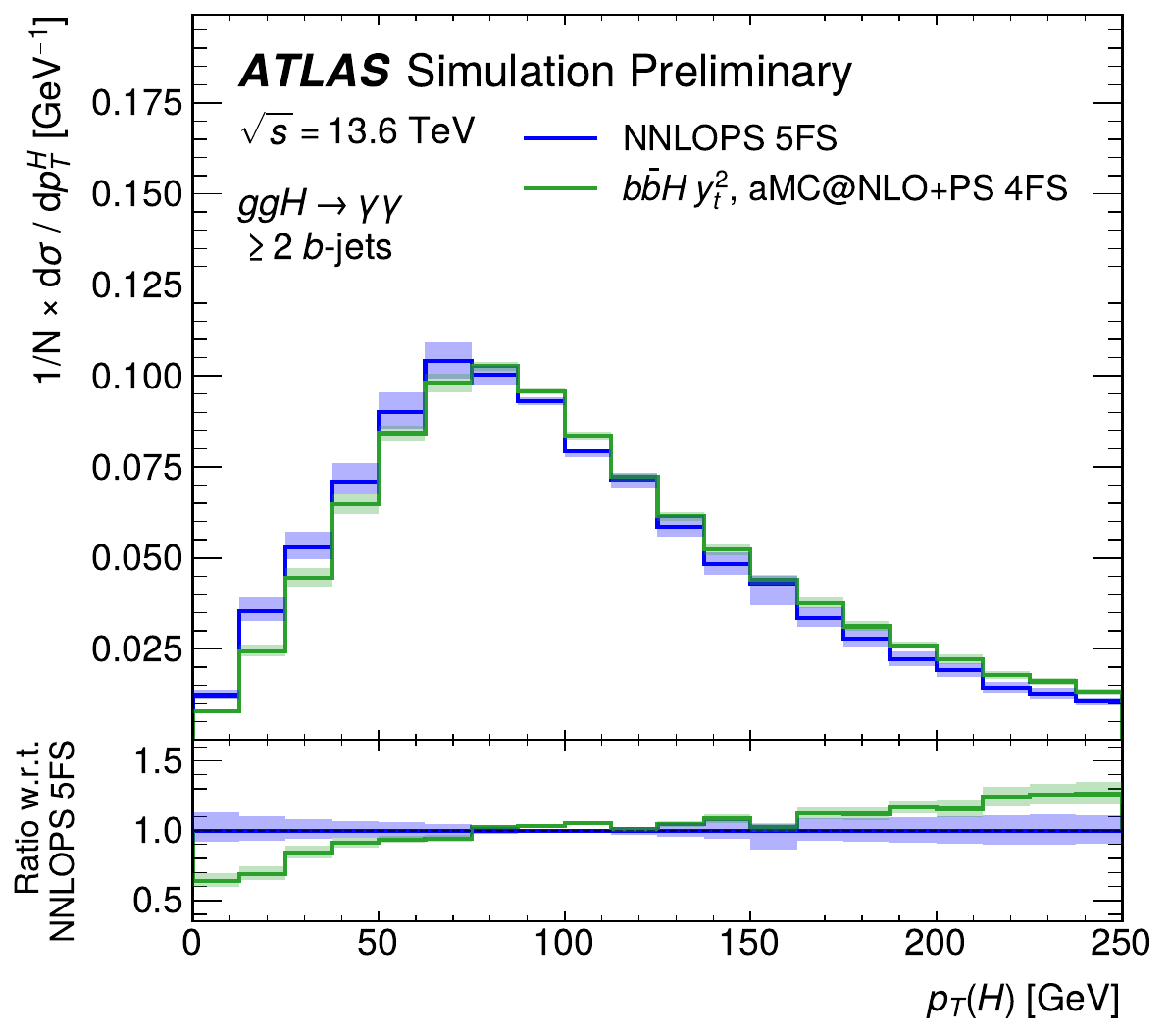}&
\includegraphics[width=.44\textwidth, page=1]{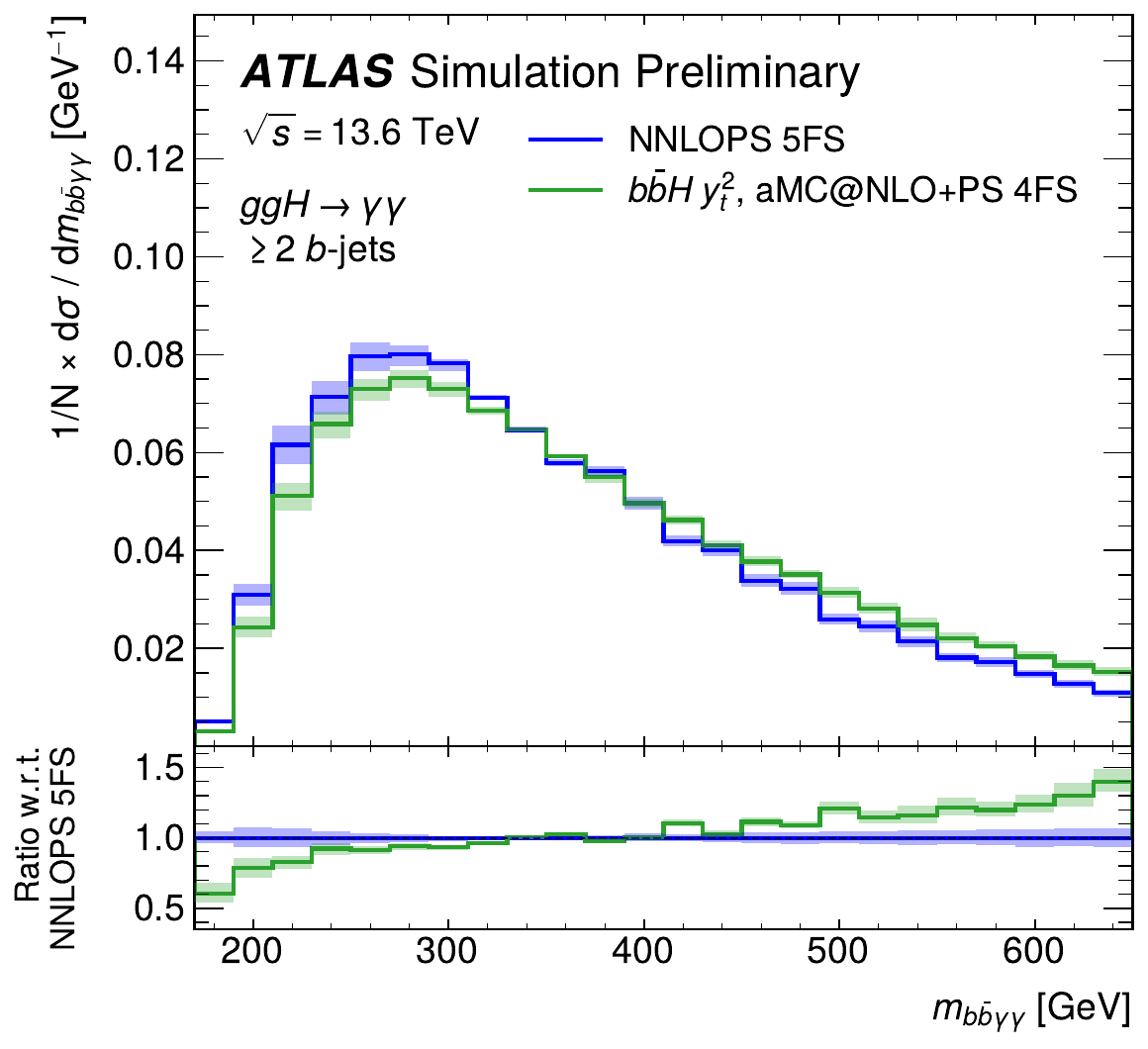}
\end{tabular}
\vspace*{1ex}
\caption{Distributions of the Higgs boson $\pt$, and invariant mass $\mbbggs$ as defined in~\eqn{eq:mbbggs} for the $y_t^2$ contribution predicted by {\sc MadGraph5\_aMC@NLO} and {\sc NNLOPS} samples after requiring at least 2 $b$-jets in the final state as described by \eqn{eq:ptjcuts}. All distributions are normalised to unity. The blue and green shaded bands represent the scale uncertainties (shape only) of the {\sc NNLOPS} and {\sc MadGraph5\_aMC@NLO} samples, respectively. The lower panel displays the ratio of the two predictions.~\cite{atlaspub}\label{fig:4fsNNLOPS}}
\end{center}
\end{figure}

We continue by considering differential distributions in the $HH$ signal region in \fig{fig:4fsFID} and \ref{fig:4fsNNLOPS}, including 
the Higgs transverse momentum spectrum (left figures) and the $\mbbggs$ distribution (right figures).
For the $y_b^2$ component in \fig{fig:4fsFID} we compare 4FS predictions at NLO+PS and NNLO+PS. We find that the inclusion 
of NNLO corrections increases the cross section 40--60\% 
and leads to a significant reduction in the scale uncertainties. Although the NNLO correction in $\mbbggs$ is rather flat,
it is clear that for a precise description of the $y_b^2$ cross section of $\bbH{}$ production 
NNLO-accurate predictions in the 4FS are indispensable.
In \fig{fig:4fsNNLOPS}~\cite{atlaspub} we compare the $y_t^2$ predictions from the {\sc MadGraph5\_aMC@NLO} and {\sc NNLOPS} samples, 
after applying only the $b$-jet selection defined in \eqn{eq:ptjcuts}. 
The distributions are normalised to unity in order to highlight that the 4FS NLO+PS {\sc MadGraph5\_aMC@NLO} generator
predicts a significantly harder spectrum than the 5FS LO+PS prediction from the  {\sc NNLOPS} generator.
By normalising to unity, these shape effects are disentangled from the even larger rate differences with respect to {\sc NNLOPS}, discussed before.

\subsection[Combining $y_t^2$ \bbH{} component and inclusive Higgs boson production in the 4FS]{Combining \boldmath{$y_t^2$} \bbH{} component and inclusive Higgs boson production in the 4FS}

In analyses focusing on final states containing $b$-jets, it is essential to accurately model the contributions 
from light-flavour and charm jets that may enter the event selection due to mistagging. 
To facilitate the use of the {\sc MadGraph5\_aMC@NLO} sample in experimental studies, 
while maintaining a realistic description of all jet flavour components, 
the $y_t^2$ contribution to the \bbH{} process at NLO+PS 
is supplemented with the inclusive Higgs boson production sample (including higher-order light-jet multiplicities) 
provided by the {\sc NNLOPS} generator, consistently within the 4FS.
The {\sc NNLOPS} simulation includes the inclusive description of the $gg \to H$ process at NNLO, but also describes 
one accompanied light-flavour and charm jets at NLO and two of them at LO, which are not represented in 
the {\sc MadGraph5\_aMC@NLO} sample for \bbH{} production.

\begin{table}[b]
\begin{center}%
\begin{small}%
\tabcolsep5pt
\begin{tabular}{|c|c|c|c|}%
\hline
$\sqrt{s}$ & $\sigma^{}$ & $\Delta_{\left(\mu_{R},\mu_{F}\right)}$ & $\Delta_{\mathrm{PDF}}$  \\\hline\hline
$13.0$\,TeV & $667$ fb & $+62\% / -35\%$ & $\pm 1.7\%$ \\\hline
$13.6$\,TeV  & $723$ fb & $+62\% / -36\%$ & $\pm 1.7\%$ \\\hline
$14.0$\,TeV & $763$ fb & $+62\% / -36\%$ & $\pm 1.7\%$ \\\hline
\end{tabular}%
\end{small}%
\end{center}%
\caption{Total \bbH{} cross sections for the $y_t^2$ contribution in the SM at LHC center-of-mass energies of $\sqrt{s} = 13$~TeV, $13.6$~TeV, and $14$~TeV. The cross sections are evaluated for a Higgs mass of $m_h = 125$~GeV and the PDF4LHC21\_40\_nf4 (LHAPDF ID \texttt{93500}) set is employed for the parton distribution functions. Variations in the cross section due to changes in $m_h$ between $125$~GeV and $125.20$~GeV are negligible compared to the scale uncertainty.}
\label{tab:bbHytxsec}
\end{table}

To prevent double counting of events featuring $gg \to H$ production with two $b$-jets, the {\sc NNLOPS} sample is filtered to exclude any event that contains 
{\it at least} one bottom quark, regardless whether it originates from the hard scattering or from parton showering. 
Removing events with a single bottom quark from the {\sc NNLOPS} sample ensures complete orthogonality of configurations between the two samples. 
This excludes $g  b \to b H$ contributions, which are already accounted for in the {\sc MadGraph5\_aMC@NLO} prediction via initial-state $g \to b\bar{b}$ splittings. Thus, by effectively removing the $b\bar{b}H$ component from the {\sc NNLOPS} sample, it is turned 
into a 4FS description of Higgs boson production accompanied by 
only light jets. The filtered {\sc NNLOPS} sample can then be combined with the {\sc MadGraph5\_aMC@NLO} one for 
$b\bar{b}H$ production at NLO+PS by simply adding the two contributions, obtaining a full description of Higgs boson production with light- and $b$-jet
final final states within the 4FS, where inclusive Higgs observables are described at NNLO+PS and $H$+$b$-jet final states at NLO+PS. 
Although this method combines a 4FS calculation with one that was originally obtained in the 5FS, by dropping consistently all bottom-quark contributions
in the NNLOPS prediction, it is turned effectively into a 4FS calculation.

One should bear in mind, however, that for full consistency also the PDFs and strong
coupling need to be updated to 4FS ones within this calculation. In order to exploit the existing NNLOPS samples, this is not done here for practical reasons.
Still, even with this caveat, the method outlined above provides a coherent and practical approach for estimating the $gg \to H$+$b$-jets contribution at analysis level. 
This strategy has already been used in similar contexts, such as the modelling of $Z$ boson production with two bottom quarks in \citere{bagnaschi:2018dnh}, and the treatment of the $t\bar{t}+b\bar{b}$ process in \citere{PhysRevD.93.014019}.  
As discussed in~\citere{atlaspub}, the filtering efficiency, defined as the fraction of {\sc NNLOPS} events surviving the bottom quark veto, is found to be 94\%. The filtered {\sc NNLOPS} sample is then normalised to the inclusive $gg \to H$ cross section at N$^3$LO after subtracting the $y_t^2$ \bbH{} component as listed in~\tab{tab:bbHytxsec}. The comparison of nominal and filtered {\sc NNLOPS} samples is discussed in~\citere{atlaspub}.

\section{Light-quark Yukawa contributions to Higgs boson production}\label{sec:lightYukawa}

The cross section proportional to the bottom-quark Yukawa coupling yields 
the largest (and most relevant) quark-induced contribution to Higgs boson production 
after the top quark. However, Higgs boson production mechanisms induced by the lighter 
quarks (charm, strange, up, down) play an important role to constrain 
the light-quark Yukawa couplings. 
Indeed, while the bottom-quark Yukawa coupling is strongly constrained through measurements of Higgs decays, no stringent bounds exist for lighter quarks~\cite{Kagan:2014ila}. 
 In particular, the charm Yukawa coupling is only weakly constrained, with an observed upper limit of less than 8.5 times the SM prediction based on analyses of Higgs decay products in Higgsstrahlung events~\cite{Atlas:2022ers}. 
Light-quark Yukawa contributions typically have an important
impact at small momentum-transfers, for instance, at small transverse momentum
of the Higgs boson. Therefore, \citere{Bishara:2016jga} suggested
to access the bottom and charm Yukawa couplings through precise measurements
of the Higgs transverse momentum distribution.

The Higgs boson production mechanisms induced by light-quark Yukawa couplings correspond
precisely to the ones of \bbH{} production, with one contribution coming from 
loop-induced gluon fusion, while another coming from the 
tree-level production of the Higgs boson in association with light quarks. Again different
prescription of treating the quarks as massless or massive can be considered. However,
for the lighter quark flavours, the treatment as massless objects is more suitable
in general. As for \bbH{} production, the 
typical tree-level production mode in the massless scheme 
is the quark-fusion process $q\bar q\to H$ with $q\in\{b,c,s,u,d\}$, 
and we write the cross section as follows:
\begin{align}
\label{eq:light-cs}
\sigma_{q\bar q \rightarrow H}(\bar \kappa_q^2)=\sigma_{b\bar b \rightarrow H}+\bar \kappa_c^2 \bar \sigma_{c\bar c \rightarrow H}+\bar \kappa_s^2 \bar \sigma_{s\bar s \rightarrow H}+\bar \kappa_d^2 \bar \sigma_{d\bar d \rightarrow H}+\bar \kappa_u^2 \bar \sigma_{u\bar u \rightarrow H}\,.
\end{align}
All interference effects at higher orders are mass suppressed, and they
vanish in the massless treatment with $ n_f = 5 $. 
We have introduced the coupling $\bar\kappa_q=y_q(m_H)/y_b(m_H)$, which denotes the ratio of the 
Yukawa coupling of a light quark to the SM Yukawa coupling of the bottom quark, both evaluated in the \MSbar{} scheme at the scale $m_H$, and the 
cross section with the corresponding Yukawa coupling stripped off, 
$\bar\sigma_{q\bar q\to H} =\sigma_{q\bar q\to H}\cdot y_b^2(m_H)/y_q^2(m_H)$.
In other words, the cross sections in \eqn{eq:light-cs} are obtained 
using the appropriate quark PDF, while assigning the Yukawa interaction 
to the strength of the bottom-quark coupling. This is done here to obtain cross sections of
comparable size.

A value of $ \bar\kappa_q = 1 $ corresponds to a Yukawa coupling for quark $ q $ equal in strength to that of the bottom quark. Thus, the normalised couplings take the following values in the SM:  
$ \bar \kappa_b = 1 $, $ \bar \kappa_c \simeq 2.3 \cdot 10^{-1} $, $ \bar \kappa_s \simeq 1.9 \cdot 10^{-2} $, $ \bar \kappa_d \simeq 2.7 \cdot 10^{-3} $, $ \bar \kappa_u \simeq 4.4 \cdot 10^{-4} $.
The numerical values are obtained by using the PDG values of the quark
masses in the $\overline{\text{MS}}$
scheme~\cite{ParticleDataGroup:2024cfk} as boundary conditions and
evolving them to the hard scale $m_H$ using four-loop running. The
High-Luminosity LHC prospects~\cite{Cepeda:2019klc} show that the
light-quark Yukawa coupling can be constrained to $|\bar \kappa_c|\leq
3.34$, $|\bar \kappa_d|\leq 0.39$, $|\bar \kappa_u|\leq 0.33$ at 95\%
Confidence Level when using kinematical distributions, in particular the transverse momentum spectrum. In~\citere{Cepeda:2019klc} other methods for constraining light quark Yukawa couplings, such as exclusive decays, have been discussed, leading to an global expectation of $|\bar \kappa_c|\leq 0.52$, $|\bar \kappa_s|\leq 0.31$, $|\bar \kappa_d|\leq 0.34$, $|\bar \kappa_u|\leq 0.33$.

We present results for Higgs boson production induced by light-quark Yukawa interactions
based on two different calculations:
the NNLO+PS calculation for $b\bar b \to H$ production with \minnlo{} 
in \citere{Biello:2024vdh}, which has been extended to all light-quark fusion processes;
and the resummed 
N$^3$LL$^{\prime}$+aN$^3$LO predictions in \citere{Cal:2023mib},
which have been obtained not only for $b\bar{b}\to H$, but also 
for $c\bar{c}\to H$ and $s\bar{s}\to H$ production in that work.
The latter calculation has also been used to construct the 
\textsc{Geneva} NNLO+PS generator for the $c\bar{c}H$ process in~\citere{Gavardi:2025zpf}, which will not be studied here.

In the following, we consider the Higgs transverse momentum spectrum, as one of the 
observables most sensitive to light-quark Yukawa interactions, especially at small 
transverse momentum, and present results at both NNLO+PS level from the 
\minnlo{} generator and for the N$^3$LL$^{\prime}$+aN$^3$LO resummed spectrum.
We further apply the \minnlo{}  simulation to a detailed phenomenological study at 
NNLO+PS level in the diphoton channel $q\bar q\rightarrow H\rightarrow \gamma\gamma$, where we include the Higgs decay to two photons through
the \PYTHIA{8} parton shower.

\subsection{Sensitivity of the Higgs $p_T$ spectrum to light-quark Yukawa couplings}
\begin{figure}[t!]
\begin{center}
\begin{tabular}{cc}
	\includegraphics[width=.45\textwidth, page=1]{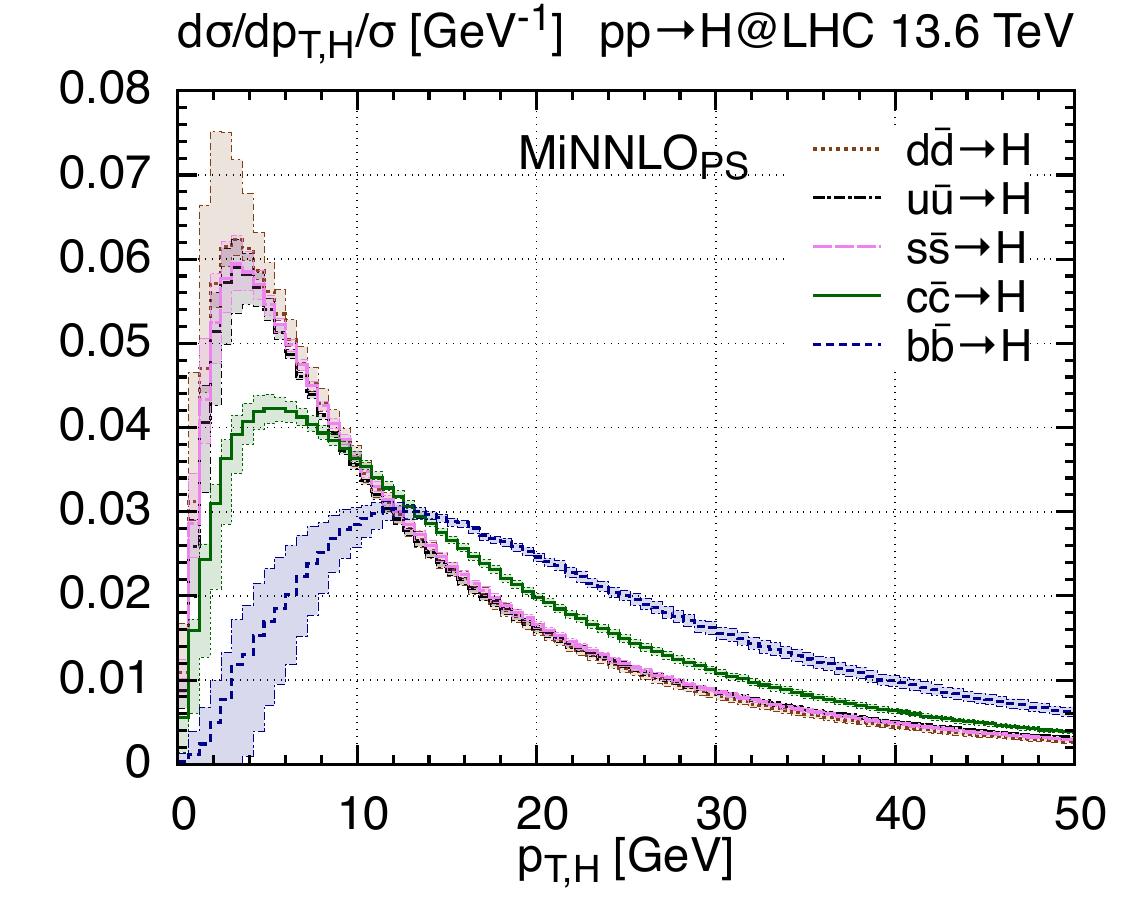}&
	\includegraphics[width=.45\textwidth, page=1]{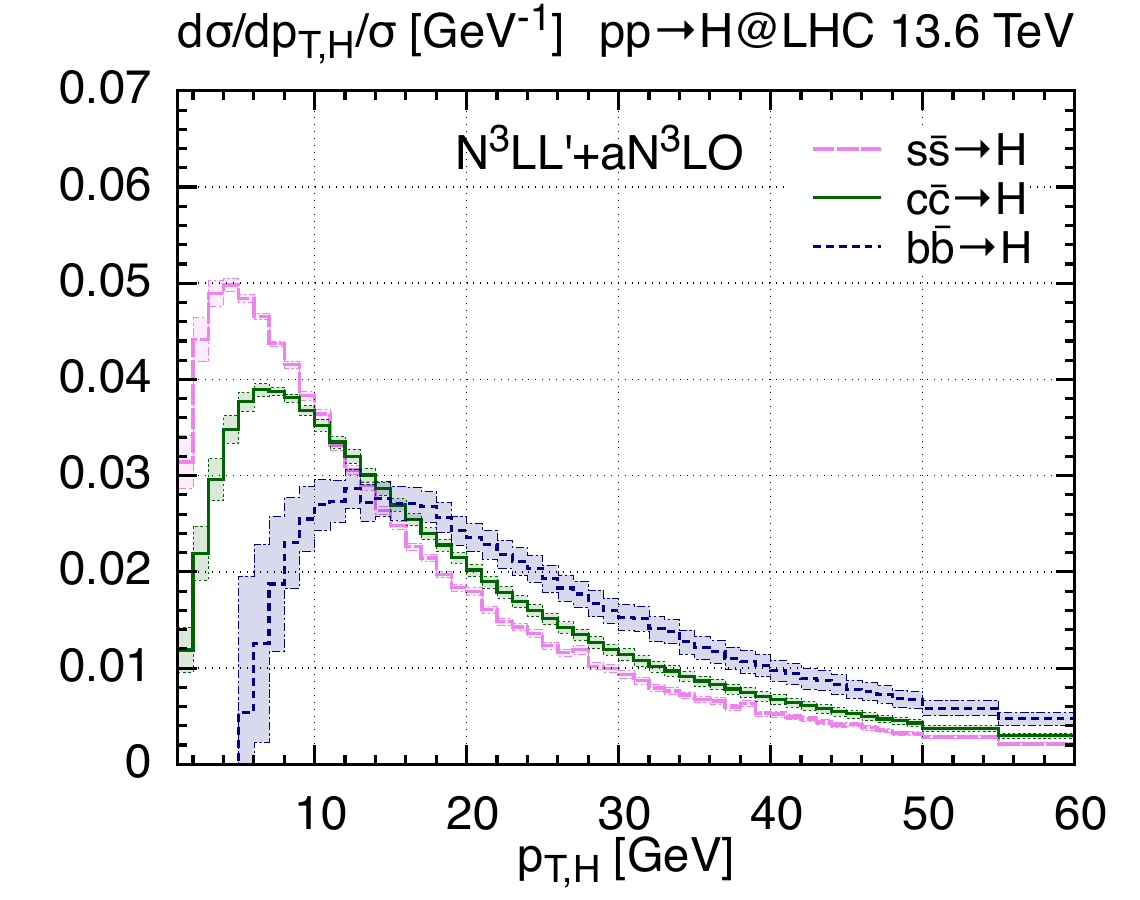}
\end{tabular}
\vspace*{1ex}
\caption{Transverse momentum spectra of the Higgs boson obtained with \minnlo{} (left) and at N$^3$LL$^{\prime}$+aN$^3$LO (right) for the different quark-fusion channels: bottom (blue, dashed), charm (green, solid), strange (pink, long-dashed), and down (brown, dotted), and up (black, dotted-dashed). The distributions are 
normalised to their respective total cross sections.
\label{fig:lightpTHzoom}}
\end{center}
\end{figure}

We begin with the transverse momentum spectrum of the Higgs boson. In~\fig{fig:lightpTHzoom}, we show the normalised distribution for the five different flavour channels obtained from the \minnlo{} generator (in the left plot) and for the second-/third-generation quark channels at N$^3$LL$^{\prime}$+aN$^3$LO (in the right plot).\footnote{
We note that the N$^3$LL$^{\prime}$+aN$^3$LO spectra are normalised to the N$^3$LO cross sections obtained from \texttt{n3loxs}~\cite{Baglio:2022wzu}, which uses the results from \citeres{duhr:2019kwi,Duhr:2020kzd}. The code was modified to obtain the N$^3$LO cross sections for  $c\bar c H$ and  $s\bar s H$.}
By normalising to the total cross section, we remove the dependence on the Yukawa coupling, making the comparison sensitive only to differences in the parton distributions
in each channel. We observe that the position of the peak varies significantly with the initial-state flavour, with light-quark channels exhibiting the softest spectrum
and the bottom quarks the hardest one. The comparatively hard spectrum in the 
bottom-quark channel is a consequence of the bottom-quark PDF being 
generated perturbatively and predominantly from gluon splittings.

Although the small masses of the first- and second-generation quarks make their contributions challenging to detect experimentally, the shape of the Higgs transverse momentum distribution provides a potential handle to disentangle them from the dominant gluon-fusion channel, which features a much harder spectrum, peaked at larger values,
and therefore distributes the cross section towards large transverse momentum values.
In particular, the softer spectrum associated with the charm-quark contribution, combined with the intermediate size of its SM Yukawa coupling, makes this observable promising for extracting the charm Yukawa coupling. Additionally, it can 
 be used to put bounds on the Yukawa couplings of the lighter quarks.

We further comment on the analytically resummed predictions in the right
plot of \fig{fig:lightpTHzoom}.
At NNLL+NLO accuracy, the spectra of the different flavour channels already exhibit different shapes, as observed here for \nnnres{} results with a substantial reduction of the scale uncertainty. The uncertainties for the bottom-quark channel are noticeably larger compared to the other channels.
In fact, the relative uncertainties for $b\bar b \to H$ at
a given order are of similar size as those for $s\bar s \to H$ at one lower order.
The main difference between the channels is the relative size of the PDF luminosities.
As already pointed out in~\citere{Cal:2023mib} for $b\bar b \to H$, the $b\bar b$ Born channel is numerically suppressed by the small b-quark PDFs, the
gluon-induced PDF channels, which start at one higher order, play a much more prominent role and explain the observed pattern of uncertainties for the different cases. We note that the $b\bar b \to H$ cross section in the 
right panel of~\fig{fig:lightpTHzoom} is only shown for $p_T> 4\, \GeV$,
since with a bottom-quark mass of $m_b \sim 4.18\, \GeV$ the assumption of 
$m_q \ll p_T$, where the factorization theorem is valid, no longer applies 
for $p_T\lesssim 4\, \GeV$.

\subsection{Simulations with diphoton signature in \minnlo{}}

In the \minnlo{} generator, the Higgs boson is treated as an on-shell asymptotic state. 
Its decay into photons can be modelled through \PYTHIA{8} using the 
zero-width approximation, assuming a branching ratio of
 ${\rm BR}(H \to \gamma\gamma) = 0.227\%$~\cite{LHCHiggsCrossSectionWorkingGroup:2016ypw}.
Inspired by a fiducial region accessible by both the ATLAS and CMS detectors, we apply the following constraints on the rapidities and transverse momenta of the two photons:
\begin{equation}
|y(\gamma_i)|< 2.37, \quad
\frac{p_T(\gamma_1)}{m(\gamma_1, \gamma_2)} > 0.35,\quad \frac{p_T(\gamma_2)}{m(\gamma_1, \gamma_2)} > 0.25\,. \label{eq:aafidmycuts}
\end{equation}
Here, $ \gamma_1 $ denotes the hardest photon, i.e.\ the one with the largest transverse momentum. We employ the selection of \eqn{eq:aafidmycuts} as our fiducial selections 
and reconstruct the Higgs boson momentum from the photon pair.

To cross-check the \minnlo{} calculation with a fixed-order prediction, we have extended the public code \SuSHi{}~\cite{Harlander:2012pb,Harlander:2003ai},
which computes the NNLO QCD cross section for  $b\bar b\to H$ production,
to support the light-quark fusion processes at NNLO in QCD. 
In~\tab{tab:qqH_xs} we present the total inclusive cross section for
the different light-quark Higgs boson production channels from 
\SuSHi{} and \minnlo{} according using the setup in~\sct{sec:setup}. 
We recall that the values are obtained by assuming 
the bottom-quark Yukawa coupling in all production modes, 
as described in \eqn{eq:light-cs}, and must be rescaled by the ratio of appropriate 
squared Yukawa couplings to obtain the SM prediction. 

\setcellgapes{4pt}
\begin{table}[t!]
  \vspace*{0.3ex}
  \makegapedcells
  \begin{center}
	   \renewcommand{\arraystretch}{1.3}
    \begin{tabular}{|c||c|c||c|}
    \hline
    \makecell[c]{Flavour channel} & \makecell[c]{\shortstack{\SuSHi{}\\[0.1cm]$\bar \sigma_{q\bar q \rightarrow H}^{\rm inclusive}$} } & \makecell[c]{\shortstack{\minnlo{}\\[0.05cm]$\bar \sigma_{q\bar q \rightarrow H}^{\rm inclusive}$} }&  \makecell[c]{\shortstack{\minnlo{}\\[0.05cm]$\bar \sigma_{q\bar q \rightarrow H\rightarrow\gamma \gamma}^{\rm fiducial}$ }}  \\
     \hline \hline
	    $d \bar d \rightarrow H$ $(y_d\rightarrow y_b)$ & $11.46(9)_{-1.1\%}^{+0.5\%}$ pb & $11.442(3)_{-2.4\%}^{+2.8\%}$ pb & $11.420(5)_{-2.4\%}^{+2.8\%}$ fb \\
     \hline
	    $u \bar u \rightarrow H$ $(y_u\rightarrow y_b)$ & $16.46(1)_{-1.1\%}^{+0.6\%}$ pb & $16.182(5)_{-2.3\%}^{+2.7\%}$ pb &  $13.169(8)_{-2.1\%}^{+2.5\%}$ fb\\
      \hline
	    $s \bar s \rightarrow H$ $(y_s\rightarrow y_b)$ & $4.454(3)_{-1.4\%}^{+1.0\%}$ pb & $4.676(1)_{-1.8\%}^{+3.7\%}$ pb & $6.215(3)_{-1.7\%}^{+3.5\%}$ fb\\
       \hline
       $c \bar c \rightarrow H$ $(y_c\rightarrow y_b)$ & $1.849(1)_{-2.9\%}^{+1.5\%}$ pb & $1.778(6)_{-0.9\%}^{+2.3\%}$ pb &  $2.399(1)_{-1.0\%}^{+2.3\%}$ fb\\
        \hline
        $b \bar b \rightarrow H$ &  $0.585(7)_{-9.2\%}^{+7.0\%}$ pb &  $0.5757(4)_{-8.0\%}^{+4.5\%}$ pb & $0.8089(8)_{-8.2\%}^{+4.7\%}$ fb\\
        \hline
    \end{tabular}
  \end{center}
  \vspace{-1em}
  \caption{
	 Comparison of \SuSHi{} cross section numbers against the integrated \minnlo{} results for Higgs boson production via light-quark fusion, with all Yukawa couplings set 
	 to the strength of the bottom-quark Yukawa. The last column presents the NNLO+PS results for $H\rightarrow \gamma\gamma$ production within the fiducial selection defined in~\eqn{eq:aafidmycuts}. \label{tab:qqH_xs}}
\end{table}

We observe that the cross sections are enhanced by the respective quark 
PDF, with the light quarks receiving the strongest enhancement, which reduces
as the quark mass increases.
There is a good agreement between the fixed-order NNLO and \minnlo{} cross 
sections, particularly in the down- and up-quark channels. Scale uncertainties are significantly reduced for the lighter quark channels. Most notably 
bottom-quark fusion includes larger uncertainties due to the less accurate bottom 
PDFs, as discussed before.

In the last column of~\tab{tab:qqH_xs} we report the integrated cross sections 
from the NNLO+PS simulation in the fiducial region defined in \eqn{eq:aafidmycuts}. 
The three orders of magnitude difference compared to the fully inclusive 
case originates from the Higgs branching into photons. 
Among the initial-state quarks, the up quark shows the lowest efficiency in passing the fiducial selection, with only about 36\% of diphoton events surviving the kinematical selections, compared to approximately 44\% for the down quark and 59\% for the charm quark.

\begin{figure}[t!]
\begin{center}
\begin{tabular}{cc}
\includegraphics[width=.45\textwidth, page=1]{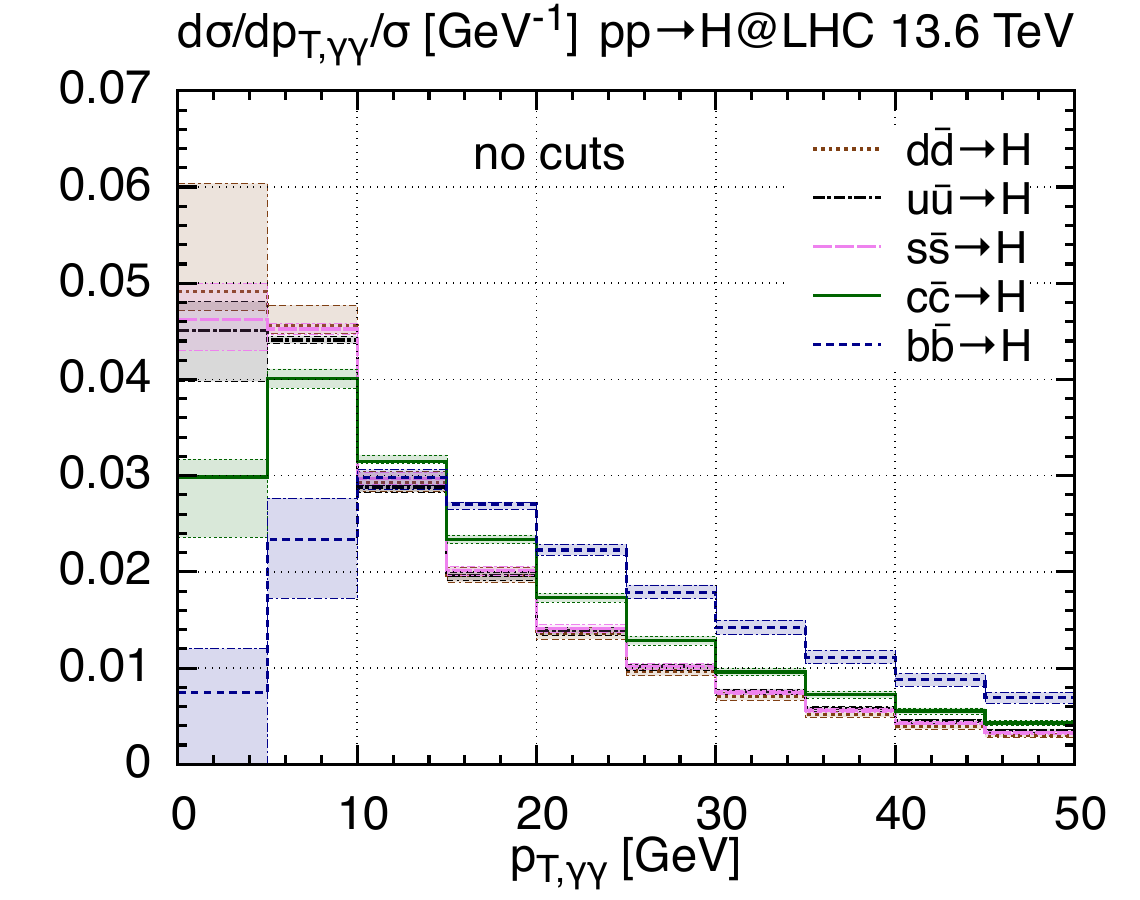}&
\includegraphics[width=.45\textwidth, page=1]{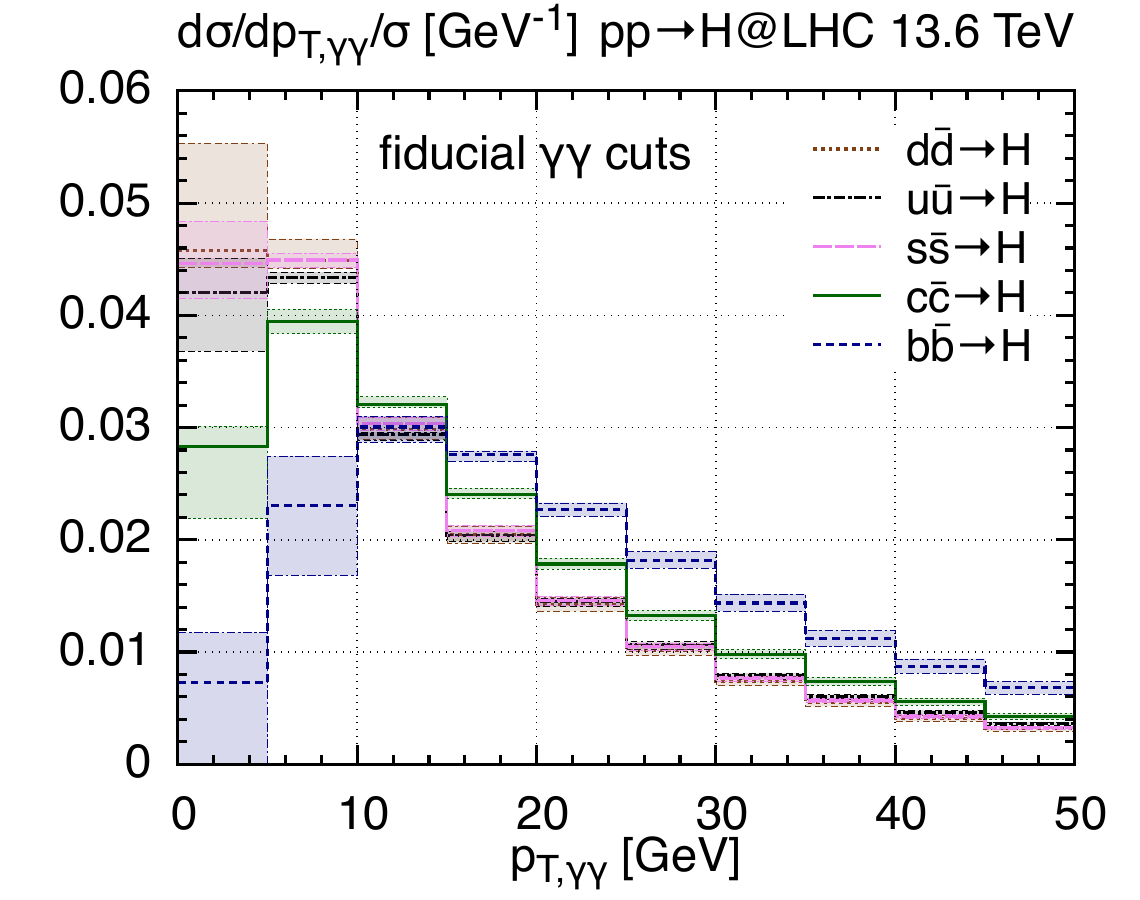}
\end{tabular}
\vspace*{1ex}
\caption{Transverse momentum spectra of the Higgs boson without kinematical selections (left) and within the fiducial region defined in~\eqn{eq:aafidmycuts} (right) for the different quark-fusion channels: bottom (blue, dashed), charm (green, solid), strange (pink, long-dashed), and down (brown, dotted), and up (black, dotted-dashed). The distributions are 
normalised to their respective integrated cross sections.
\label{fig:lightpTH}}
\end{center}
\end{figure}

Finally, we discuss the impact of the fiducial cuts on 
differential distributions of the Higgs boson in the various quark channels.
In the left (right) plot of \fig{fig:lightpTH}, we show the transverse momentum spectrum of the Higgs boson, reconstructed from the diphoton pair in the inclusive (fiducial) 
phase space. The distributions are normalised to their respective integrated  
cross sections.
We observe that the shape of the different distributions changes 
relatively mildly when fiducial cuts on the photons are included. Moreover,
the peak positions stay in the same place. As a result, extrapolations 
from the fiducial to the inclusive phase-space in experimental measurements
should have a rather minor effect on the shape of the distribution.

We note that alternative channels sensitive to light-quark Yukawa couplings have been experimentally investigated recently, such as the reinterpretation of $WW\gamma$ analysis~\cite{CMS:2023rcv} and the Higgs decay rate into four leptons~\cite{CMS:2025xkn}. Novel approaches to probe the light-quark Yukawa couplings have also been discussed recently in~\citeres{Balzani:2023jas,Alasfar:2022vqw,Alasfar:2019pmn}.

\section{Conclusions}\label{sec:conclusions}
We have reported on new predictions for \bbH{} production at the LHC at a centre-of-mass energy of 13.6 TeV. All cross sections are computed following the recommendations of the LHCHWG.

First, we have updated the predictions for the 
total inclusive cross section by interpolating the 13 TeV and 14 TeV cross section
numbers from the 4FS and 5FS matched \nlonnllpart{} calculation. 
So far, the matching of the different flavour schemes has been performed 
only at the fully inclusive level for \bbH{} production.
Matched predictions of differential state-of-the-art \bbH{} 
predictions in 4FS and 5FS could, in principle, be obtained by 
extending differential matching approaches, such as those 
discussed in \citeres{Gauld:2021zmq,guzzi:2024can}, to higher orders.

As far as differential calculations for \bbH{} production are concerned, 
this report focusses on new state-of-the-art predictions in either scheme.
Specifically, novel results for the Higgs transverse momentum distribution 
have been discussed based on the analytic resummation 
at $\text{N}^3\text{LL}^{\prime}+\text{aN}^3\text{LO}$ accuracy. 
In the future, this calculation could be extended by the matching with the 
exact $\text{N}^3\text{LO}$ corrections for Higgs boson production via 
bottom-quark fusion.

We then presented the first numerical comparison of the recently developed 
\minnlo{} and \GENEVA{} generators for \bbH{} production in the massless scheme
(i.e.\ the $b\bar b \to H$ process) at NNLO+PS accuracy. 
We find their predictions to be compatible within the respective theoretical uncertainties 
for the observables studied, showing good agreement also with the analytically resummed 
result for the Higgs transverse momentum spectrum. 
We also included a proof-of-concept study of the \minnlo{} generator for BSM 
applications involving heavy-Higgs boson production with an 
enhanced bottom-quark Yukawa coupling, considering a specific MSSM scenario
for a still allowed parameter setting.

Also the NNLO corrections to \bbH{} production in the massive scheme 
have recently been computed, including parton shower matching, in the \minnlo{} 
framework. We have studied the sizeable NNLO effects, and we compared 4FS 
and 5FS \minnlo{} predictions. For the first time, 4FS and 5FS predictions are finally
in full agreement, when the cross section in both schemes is computed to NNLO
in QCD. Thus, the inclusion of NNLO corrections in the massive scheme, finally
resolves the long-standing 4FS--5FS discrepancy and 
discussion about the {\it appropriate} flavour scheme for the \bbH{} process.
As a result, the corresponding theoretical uncertainty in the modelling of the 
\bbH{} signal cross section is substantially reduced.

Apart from single-Higgs boson production, the \bbH{} process is an important background to
$HH$ searches in all channels involving at least one Higgs boson decaying into two
bottom quarks. 
We have discussed the impact of 4FS predictions in the fiducial phase-space region
relevant for $HH$ studies. For the $y_b^2$ contribution, we have employed the
previously introduced \minnlo{} generator. Also in this case, the inclusion of the NNLO
QCD corrections leads to a strong reduction of the scale uncertainties.
The $y_t^2$ contribution, on the other hand,
which is even larger than the $y_b^2$ one, is modelled from the
NLO+PS generator developed within {\sc MadGraph5\_aMC@NLO}. 
So far, the $y_t^2$ contribution was modelled in the 5FS by selecting events 
from the \textsc{NNLOPS} generator for inclusive Higgs boson production in 
gluon fusion, which is effectively only LO accurate for \bbH{} final state, and
was assigned with a $100$\% uncertainty.
Since also Higgs plus light jet contributions can enter the $HH$ selection 
through mistagging, we have suggested a consistent approach within the 4FS
to combine those contributions from NNLOPS generator (removing all \bbH{} final states)
with the more accurate NLO+PS calculation for \bbH{} production.
As an important outlook for $HH$ studies, the $y_t^2$ \bbH{} component in the HTL approximation could in the future be obtained at NNLO+PS using the existing \minnlo{} framework, with the main limitation being the absence of the required two-loop amplitudes. Moreover, the NLO calculation of the $y_t^2$ \bbH{} contribution beyond the HTL approximation,
which requires the corresponding exact two-loop virtual amplitude, would be highly valuable to further reduce the theoretical uncertainty.

In the last section of the report, we have discussed light-quark Yukawa contributions
to Higgs-boson production. The production mechanism corresponds exactly to 
the one of \bbH{} production in the massless scheme, i.e.\ \qqtoH{}. We find that 
the shape of the transverse momentum spectrum is very sensitive to the different 
initial state quarks, due to PDF effects. We have also considered 
\qqtoH{} production with the Higgs decaying into two photons, 
making use of the possibility to decay the Higgs boson through \PYTHIA{8}
within the \minnlo{} generator. To further improve the theoretical modeling
of \ccH{} final states, which can be important to extract the charm-quark Yukawa coupling, 
the NNLO(+PS) calculation of \ccH{} production in the massive scheme (i.e.\ a three 
flavour scheme) could be considered within the \minnlo{} framework.

\textbf{Citation policy. }The present work consists of contributions from different collaborations. If some of the results are employed for scientific publications, together with this work, one should cite the corresponding work in~\citeres{deutschmann:2018avk,manzoni:2023qaf,Cal:2023mib,Biello:2024vdh,Biello:2024pgo,Gavardi:2025zpf,atlaspub}.

\textbf{Data Availability Statement.} All data used to produce the plots are available upon request by contacting the LHCHWG bbH/bH theory conveners via email.

\textbf{Acknowledgements. }This work has been carried out within the
LHCHWG, as a contribution to the Yellow Report 5. We
thank all the Working Group 3 conveners for coordinating the activities, in particular Tatjana Lenz. CB is grateful to Tommaso Giani and Jan Lukas Sp\"ah for discussions on Yukawa interactions via light-quark fusion. We have used the Max Planck Computing and Data Facility (MPCDF) in Garching to carry out the \minnlo{} simulations presented here. 
AG, RvK and FK have received funding from the European Research Council (ERC) under the European Union's Horizon 2020 research and innovation programme (Grant agreement 101002090 COLORFREE).
MZ~acknowledges the financial support by the MUR (Italy), with
funds of the European Union (NextGenerationEU), through the PRIN2022
grant 2022EZ3S3F.

\bibliography{bbh}
\bibliographystyle{JHEP}

\end{document}